\documentclass{aa}  

\usepackage{natbib}
\bibpunct{(}{)}{;}{a}{}{,}
\usepackage{orcidlink}
\usepackage{hyperref}
\usepackage{xcolor}
\usepackage{rotating}
\usepackage{amsmath}
\usepackage{ulem}
\usepackage{siunitx}

\usepackage{ulem}

\usepackage[document]{} 
\usepackage{graphicx}

\usepackage{txfonts}

\begin{document}

    \title{Estimation of the magnetic field strength from ALMA dust polarization in the protocluster G327.29}{}
    
   \author{A. Koley$^{1}$\thanks{Fondecyt Postdoctoral fellow}\orcidlink{0000-0003-2713-0211}, P. Sanhueza$^{2}$\orcidlink{0000-0002-7125-7685}, A. M. Stutz$^{1}$\orcidlink{0000-0003-2300-8200}, J. Liu$^{3,4}$\orcidlink{0000-0002-4774-2998},  P. Saha$^{5,6}$\orcidlink{0000-0002-0028-1354}, F. A. Olguin$^{6,7}$\orcidlink{0000-0002-8250-6827}, A. Ginsburg$^{8}$\orcidlink{0000-0001-6431-9633}, N. Sandoval-Garrido$^{1}$\orcidlink{0000-0001-9600-2796}, N. Castro-Toledo$^{1}$\orcidlink{0009-0005-6363-5104} \\
    \vspace{5mm}
    (Affiliations can be found after the references)}
    \authorrunning{Koley et al.}{}
          \institute{}
    \vspace{5mm}

   \date{Received xxxxxxx; accepted xxxxxx}{}

 \abstract 
     {Magnetic fields and turbulence may play crucial roles in the evolution of molecular clouds and ultimately in the formation of dense cores and stars. Despite being studied in many molecular clouds, the exact role of magnetic fields and turbulence in star formation is still poorly understood. We report the high-resolution plane of sky magnetic field ($B_{\text{pos}}$) morphology towards the high-mass star-forming region G327.29, obtained with the 12-metre configuration of the Atacama Large Millimeter/sub-millimeter Array (ALMA) telescope. From our analysis, we obtain a complex $B_{\text{pos}}$ morphology in which the magnetic field orientation is uniformly distributed across the entire range from $-$90 to $+$90 deg. The observed area is composed of one filament and one dense central clump, both of which harbour multiple dense cores. The total magnetic field strengths ($B_{\text{tot}}$) in these regions are 1.7 $\pm$ 0.8 and 3.6 $\pm$ 1.5 mG at a number density ($n$) of 1.5 $\pm$ 0.3 $\times$ 10$^{6}$ and 2.3 $\pm$ 0.7  $\times$ 10$^{6}$ cm$^{-3}$, derived using the Davis–Chandrasekhar–Fermi (DCF) angular dispersion function (ADF) method. The corresponding $B_{\rm tot}$ obtained using the Skalidis--Tassis (ST) method are $0.6 \pm 0.3$ and $1.1 \pm 0.5$ mG, approximately three times lower. The virial parameters ($\alpha_{\text{vir}}$) in these regions are 2.8 $\pm$ 1.9 and 0.5 $\pm$ 0.4, suggesting that the regions may be gravitationally bound or unbound after accounting the errors. The ratio of turbulent to magnetic energy ($\sim$ 0.17 to 0.34) indicates that the magnetic field is dynamically more important than turbulence. The relative influence of turbulence and magnetic fields on core dynamics appears to depend on how $B_{\text{tot}}$ scales with gas density ($\rho$) in the densest regions. In summary, this work, enabled by high-resolution ALMA observations, presents a comprehensive analysis of the relative roles of magnetic fields, turbulence, and gravity in regulating high-mass star formation in G327.29.}
   

   \maketitle
%


\section{Introduction}\label{section_1}
Most high-mass stars form within cold, dense clusters known as protoclusters \citep{2017A&A...601A..60C,2022A&A...662A...8M,2025A&A...696A.150S}. Magnetic fields, turbulence, and gravity can play crucial roles in the evolution of these protoclusters, ultimately shaping the formation of dense cores and stars within them \citep{2016A&A...590A...2S,2018MNRAS.473.4890S}. However, the precise interplay between these factors that drives the fragmentation process remains uncertain \citep{2021MNRAS.501.4825K,2021ApJ...915L..10S,2024MNRAS.529.2220R,2025A&A...702A.133K,2025A&A...696A..11V,2025ApJ...994..233Y,2025arXiv250810128C,2025ApJ...980...87S,2026arXiv260113473K}. On parsec scales, magnetic fields often display ordered structures, typically aligned with the long axes of low-density elongated gas filaments, whereas in high-density filaments, the field lines tend to be oriented perpendicular to the filament’s main axis \citep{2016A&A...586A.135P,2016A&A...586A.138P,2016A&A...586A.136P,2017A&A...607A...2S}. 
On small (core) scales, the magnetic field geometry becomes more complex and depends on the relative strengths of the magnetic and turbulent forces. When the magnetic field dominates over turbulence, cores generally exhibit disc-shaped morphologies, and the pinching effect of gravity produces the characteristic hourglass-shaped field morphology \citep{2009Sci...324.1408G,2024ApJ...972L...6S}. Conversely, when turbulence is the dominant factor, the field structure becomes chaotic \citep{2004Ap&SS.292..225C,2017ApJ...847...92H}. If the strengths of the two effects are comparable, increasingly intricate field patterns can emerge. Consequently, the relative influence of magnetic fields, turbulence, and gravity during the different stages of cloud evolution remains poorly constrained. Some observational studies of high-mass star-forming cores suggest that magnetic fields are dynamically more significant than turbulence \citep{2009Sci...324.1408G,2016A&A...593L..14F,2023ApJ...951...68C}, while others find the opposite \citep{2010ApJ...724L.113B,2013ApJ...772...69G,2020ApJ...905..158W,2025ApJ...980...87S}. Therefore, additional observational investigations are essential for a more complete understanding of the star formation process. Dust polarization currently provides the most powerful means of probing magnetic field morphologies in molecular clouds.

\begin{figure*}
        \centering 
        \includegraphics[width=7.3in,height=2.0in,angle=0]{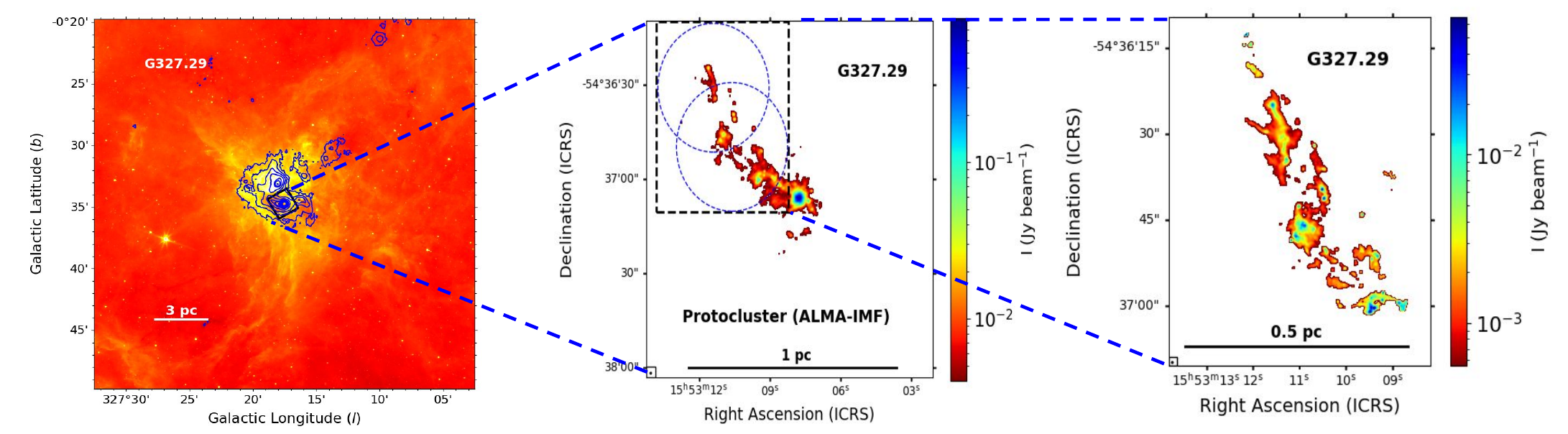}
        \caption{\textit{Left}: \textit{Spitzer} (GLIMPSE) 8 $\mu$m emission towards the G327.29 star-forming region \citep{2003PASP..115..953B}. The contours represent the ATLASGAL 870 $\mu$m emission obtained from \cite{2009A&A...504..415S}. Contour levels range from 0.45 to 34 Jy beam$^{-1}$  with a 19.2$''$ beam. \textit{Middle}: Map of the 1.2 mm continuum emission towards the central clump (protocluster) from the ALMA-IMF large programme  \citep{2022A&A...662A...8M}. The two dashed blue circles mark the areas observed for polarization with the ALMA 12m array. \textit{Right}: 1.2 mm continuum emission map obtained in this work by mosaicking the two pointings shown in the middle panel. }
        \label{fig:fig1}
\end{figure*}

G327.29, located at a distance of approximately 2.5 kpc, is a massive star-forming region with a central 870 $\mu$m clump (hereafter, protocluster) of about 5000 \(\textup{M}_\odot\). The region hosts multiple dense cores \citep{2022A&A...662A...8M,2024A&A...690A..33L}. Based on Lyman-$\alpha$ emission and the flux ratio between the 1.3 and 3.0 mm continua, this protocluster is classified as young \citep{2022A&A...662A...8M,2024ApJS..274...15G}. Figure \ref{fig:fig1} presents a colour image of the \textit{Spitzer} (GLIMPSE) 8 $\mu$m emission of the G327.29 region \citep{2003PASP..115..953B}, the APEX Telescope Large Area Survey of the Galaxy (ATLASGAL) 870 $\mu$m emission from \cite{2009A&A...504..415S}, the 1.2 mm continuum emission of the central 870 $\mu$m clump from the Atacama Large Millimeter/sub-millimeter Array (ALMA) IMF large programme, and the 1.2 mm continuum area where we conducted polarization observations using ALMA’s 12m configuration. The observed region is slightly blueshifted relative to the systemic velocity ($V_{\text{LSR}}$) of the entire region, which is $-$45.0 km s$^{-1}$ \citep{2022A&A...662A...8M}, while the $V_{\text{LSR}}$ of our observed region is $-$46.6 km s$^{-1}$, as detailed in Appendix \ref{appendix_0}. In this work, we investigated the relative roles of magnetic fields, turbulence, and gravity on the clump and core scales, aiming to understand how their interplay influences the fragmentation process.

{\color{black}
This paper is organized as follows. Section \ref{section_2} describes the observational data. In Sect. \ref{section_3} we estimate the magnetic field strength using the Davis–Chandrasekhar–Fermi (DCF) angular dispersion function (ADF) method and the Skalidis$-$Tassis (ST) method, utilizing non-thermal velocity dispersion and density, and examine its importance relative to turbulence and gravity. In Sect. \ref{section_4} we discuss the relative roles of turbulence, magnetic fields, and gravity at the core scale. The main results of our study are discussed in Sect. \ref{section_5}, and our conclusions are summarized in Sect. \ref{section_6}. 

}

\vspace{-2mm}

\section{Observation and data analysis}\label{section_2}

{\color{black}
We analysed the 1.2 mm continuum and two spectral lines HN$^{13}$C and H$^{13}$CO$^{+}$ and used the dust temperature map from the ALMA IMF large programme study \citep{2024A&A...687A.217D}. We discuss each separately below.
}

{\color{black}
\subsection{1.2 mm continuum and HN$^{13}$C ($J$ = 3$-$2) and H$^{13}$CO$^{+}$ ($J$ = 3$-$2) spectral lines }\label{subsection_2.1}
}

\begin{figure}[!ht]
        \centering 
        \includegraphics[width=3.5in,height=3.2in,angle=0]{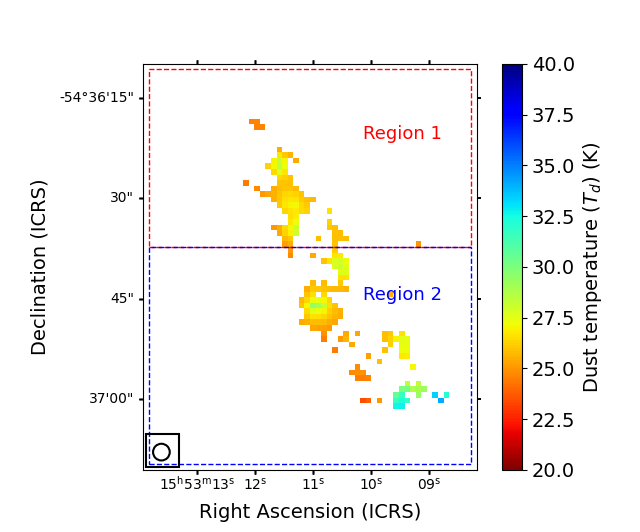}
        \caption{Dust temperature ($T_{\text{d}}$) map in the G327.29 region obtained from \cite{2024A&A...687A.217D}. We masked the regions where no 1.2 mm continuum emission was observed in our analysis. \texttt{Region 1} and \texttt{Region 2} are two distinct areas where we estimated the magnetic field separately.}
        \label{fig:fig11}
\end{figure}

ALMA 1.2 mm (Band 6) polarimetric observations towards the G327.29 protocluster were conducted between March 2019 and April 2021 in three sessions (project ID: 2019.1.01714.S; PI: Benjamin Wu) using the ALMA 12m array configuration {\color{black}(C43-2, C43-3)}. The observations achieved an angular resolution of $\sim$0.35$''$ (1000 AU) and a maximum recoverable scale (MRS) of $\sim$4.5$''$. Two partially overlapping fields, G327.29-MM15 (RA = 15$^{\text{h}}$53$^{\text{m}}$11.20$^{\text{s}}$, Dec = $-$54$^{\circ}$36$^{'}$32.81$^{''}$) and G327.29-MM19 (RA = 15$^{\text{h}}$53$^{\text{m}}$10.63$^{\text{s}}$, Dec = $-$54$^{\circ}$36$^{'}$48.95$^{''}$), were mosaicked together. Quasars J1617–5848, J1517–2422, and J1514–4748 served as the phase, flux and bandpass, and polarization calibrators, respectively. Data calibration and imaging were performed using \texttt{CASA 6.1.1.15}. Self-calibration was applied only to the Stokes I image, as the low signal-to-noise ratio (S/N) in Stokes Q and U prevented its application. Line-contaminated channels were masked in Stokes I, but not in Stokes Q and U due to their negligible effect \citep{2021ApJ...909..199O}. For each pointing, self-calibration was iteratively refined by decreasing the solution interval from 30 to 10 seconds, producing the final calibrated measurement sets. Imaging for all Stokes parameters was performed separately with the \texttt{tclean} task using Briggs weighting and a robust parameter of 1.0, resulting in an angular resolution of $\sim$0.35$''$. The RMS noise levels are $\sim$0.23 mJy beam$^{-1}$ for Stokes I and $\sim$0.023 mJy beam$^{-1}$ for Stokes Q and U, measured in the central region of the mosaic; the noise increases towards the edges according to the primary beam response. In addition to the continuum, we imaged the HN$^{13}$C ($J$ = 3$-$2) and H$^{13}$CO$^{+}$ ($J$ = 3$-$2) spectral lines at a resolution of $\sim$61 kHz (0.07 km s$^{-1}$). {\color{black} The rest frequencies of these lines are 261.2633101 and 260.255339 GHz, respectively. We used the \texttt{deconvolver = multiscale} with four scale parameters in geometric progression. The first scale was set to 0, the second to the beam size, and two additional larger scales were included to recover extended structures. We adopted a cleaning threshold of 3$\sigma$ and a \texttt{cycle factor} of 4 to minimize deconvolution artefacts. Similar to the continuum imaging, we also used Briggs weighting with a \texttt{robust} parameter of 1.0, resulting in an angular resolution of $\sim$ 0.35$''$. } 
{\color{black}

\subsection{Dust temperature ($T_{\text{d}}$) map }\label{subsection_2.2}
The temperatures of the region and most of the cores are derived from the dust temperature map of \cite{2024A&A...687A.217D}. This is illustrated in Fig. \ref{fig:fig11}. These maps were produced using the point process mapping technique, a Bayesian method that incorporates prior information on the opacity index ($\kappa_{\nu}$), dust temperature ($T_{\mathrm{d}}$), and the resulting spectral energy distribution. This approach provides line-of-sight–averaged physical properties. The analysis of \cite{2024A&A...687A.217D} combines 1.3 mm continuum data with Stratospheric Observatory for Infrared Astronomy (SOFIA)/High-resolution Airborne Wideband Camera-Plus (HAWC+) observations at 53, 89, and 214 $\mu$m, along with Atacama Pathfinder Experiment / Sub-millimeter APEX Bolometer Camera (APEX/SABOCA) (350 $\mu$m) and APEX/LABOCA (870 $\mu$m) data. The final dust temperature cubes have an angular resolution of 2.5$''$. Additional details are given in \cite{2024A&A...687A.217D}.

}

{\color{black}

\section{Magnetic field estimation and dynamical importance}\label{section_3}

}

We estimated the magnetic field strength ($B_{\text{tot}}$) using the DCF--ADF method and determined its relative importance with respect to turbulence and gravity. {\color{black}To estimate the magnetic field,} in addition to the ratio of turbulent to ordered magnetic field, the calculation requires the turbulent velocity dispersion ($\sigma_{\text{nth}}$) and the average mass density ($\rho_{\text{avg}}$). In the following subsections, we estimate each of these quantities individually and use them to calculate $B_{\text{tot}}$.

\subsection{Spectral line decomposition of HN$^{13}$C (J = 3$-$2) and  H$^{13}$CO$^{+}$ (J = 3$-$2) lines}\label{subsection_3.1}
To derive the turbulent velocity dispersion ($\sigma_{\text{nth}}$), we first performed a pixel-wise spectral line decomposition of the HN$^{13}$C (J = 3$-$2) and H$^{13}$CO$^{+}$ (J = 3$-$2) transitions using the \texttt{Gausspy+} module \citep{2019A&A...628A..78R}, which is based on the automated Gaussian decomposition (AGD) technique. The algorithm employs two smoothing parameters, \texttt{decompose.alpha1} and \texttt{decompose.alpha2}, which are automatically determined from representative spectra and optimized for different spectral features such as weak or narrow components. During the fitting process, the parameter $F_{1}$ quantifies the accuracy of the decomposition. The module also includes several additional parameters -- \texttt{significance}, \texttt{snr\_noise\_spike}, \texttt{refit\_rchi2}, and \texttt{min\_fwhm}. The last defines the minimum number of channels required for a Gaussian component, which we set to 2 to ensure adequate sampling of each spectral feature. The pixel-wise fitting yields the peak intensity, centroid velocity, and full width at half maximum (FWHM) for each Gaussian component. This procedure was applied to both the HN$^{13}$C and H$^{13}$CO$^{+}$ lines, producing pixel-wise decomposed component maps. From each of these decompositions, we computed the total velocity dispersion ($\sigma_{\text{tot}}$) for each pixel using the following expression \citep{2025A&A...702A.133K}:
\begin{equation}\label{eqn1}
    \sigma_{\text{tot}} = \sqrt{\frac{w_1}{w} \sigma_{1}^{2} + \frac{w_{2}}{w} \sigma_2^{2} + ..............}~~.
\end{equation}

\noindent Here $w_{1}$, $w_{2}$, ... represent the integrated intensities of the individual components, and $w$ denotes the total integrated intensity of the spectrum. Likewise, $\sigma_{1}$, $\sigma_{2}$, ... correspond to the velocity dispersions of each component. We avoided using the traditional velocity dispersion (moment 2) map because, in cases where multiple disjoint velocity components are present within a single pixel, the moment analysis tends to overestimate the velocity dispersion compared to the actual value. This is because the measured velocity spread can include contributions from cloud-to-cloud velocity differences within a pixel. Such velocity differences can arise from gravitational infall, rotation, or other large-scale motions, rather than from turbulence alone. This suggests that $\sigma_{\text{tot}}$ should be calculated using Eq.~\ref{eqn1}. The spatial distribution of the number of components for the spectral lines is shown in  Fig. \ref{fig:fig2}.

The observed region is divided into two subregions, \texttt{Region 1} and \texttt{Region 2}, which we discuss in Appendix \ref{appendix_1}. This division is made because the magnetic field analysis is carried out separately for these two regions, as discussed in a later section. Figure~\ref{fig:fig3} displays the spatial distribution of $\sigma_{\text{tot}}$ for both HN$^{13}$C and H$^{13}$CO$^{+}$ lines. For the HN$^{13}$C line, $\sigma_{\text{tot}}$ = 0.47 $\pm$ 0.26 km s$^{-1}$ in \texttt{Region 1} and 0.36 $\pm$ 0.21 km s$^{-1}$ in \texttt{Region 2}. Similarly, for the H$^{13}$CO$^{+}$ line, $\sigma_{\text{tot}}$ = 0.51 $\pm$ 0.35 km s$^{-1}$ in \texttt{Region 1} and 0.68 $\pm$ 0.35 km s$^{-1}$ in \texttt{Region 2}. Consequently, the average value of $\sigma_{\text{tot}}$ $\pm$ $\Delta \sigma_{\text{tot}}$ for \texttt{Region 1} is 0.49 $\pm$ 0.22 km s$^{-1}$. From the $T_{\text{d}}$ map, we obtained a dust temperature $T_{\text{d}}$ $\pm$ $\Delta T_{\text{d}}$ = 26.2 $\pm$ 0.8 K in \texttt{Region 1}. Using these values, we derived the non-thermal velocity dispersion $\sigma_{\text{nth}}$ $\pm$ $\Delta \sigma_{\text{nth}}$ = 0.48 $\pm$ 0.22 km s$^{-1}$. The following equations were used to calculate $\sigma_{\text{nth}}$ and $\Delta\sigma_{\text{nth}}$ \citep{2023PASA...40...46K}:

\begin{equation}
    \sigma_{\text{nth}} = \sqrt{\biggl(\sigma_{\text{tot}}^{2} - \frac{T_{\text{d}}}{a}\biggl)}~~,
\end{equation}

\begin{equation}
    \Delta\sigma_{\text{nth}} = \frac{\sqrt{\biggl(4~\sigma_{\text{tot}}^{2}~\Delta\sigma_{\text{tot}}^{2}  + \frac{\Delta T_{\text{d}}^{2}}{a^{2}} } \biggl)} {2\sqrt{\biggl(\sigma_{\text{tot}}^{2}-\frac{T_{\text{d}}}{a}}\biggl)}.
\end{equation}

\noindent Here, $a = \frac{m_{\text{HN}^{13}\text{C}} ~~\text{or}~~m_{\text{H}^{13}\text{CO$^{+}$}}    }{k_{\text{B}}}$, where $m_{\text{HN}^{13}\text{C}}$ and $m_{\text{H}^{13}\text{CO$^{+}$}}$ are the mass of the HN$^{13}$C and H$^{13}$CO$^{+}$ molecules, which are equal to 28$m_{\text{H}}$ and 30$m_{\text{H}}$, respectively. These values are almost similar. Here $m_{\text{H}}$ is the mass of a hydrogen atom ( = 1.67 $\times$ 10$^{-24}$ g). The Boltzmann constant $k_{\text{B}}$ is 1.38 $\times$ 10$^{-16}$ erg K$^{-1}$. For \texttt{Region 2}, the values of $\sigma_{\text{tot}}$ $\pm$ $\Delta \sigma_{\text{tot}}$ are 0.36 $\pm$ 0.21 km s$^{-1}$ and 0.68 $\pm$ 0.35 km s$^{-1}$ for the HN$^{13}$C and H$^{13}$CO$^{+}$ lines, respectively. Consequently, the average value of $\sigma_{\text{tot}}$ $\pm$ $\Delta \sigma_{\text{tot}}$ for this region is 0.52 $\pm$ 0.20 km s$^{-1}$. From the $T_{\text{d}}$ map, we obtain a dust temperature $T_{\text{d}}$ $\pm$ $\Delta T_{\text{d}}$ = 26.7 $\pm$ 2.0 K in \texttt{Region 2}. Using these parameters, we derived the non-thermal velocity dispersion $\sigma_{\text{nth}}$ $\pm$ $\Delta \sigma_{\text{nth}}$ = 0.51 $\pm$ 0.20 km s$^{-1}$.

\begin{figure}[!ht]
    \centering 
        \includegraphics[width=3.61in,height=4.91in,angle=0]{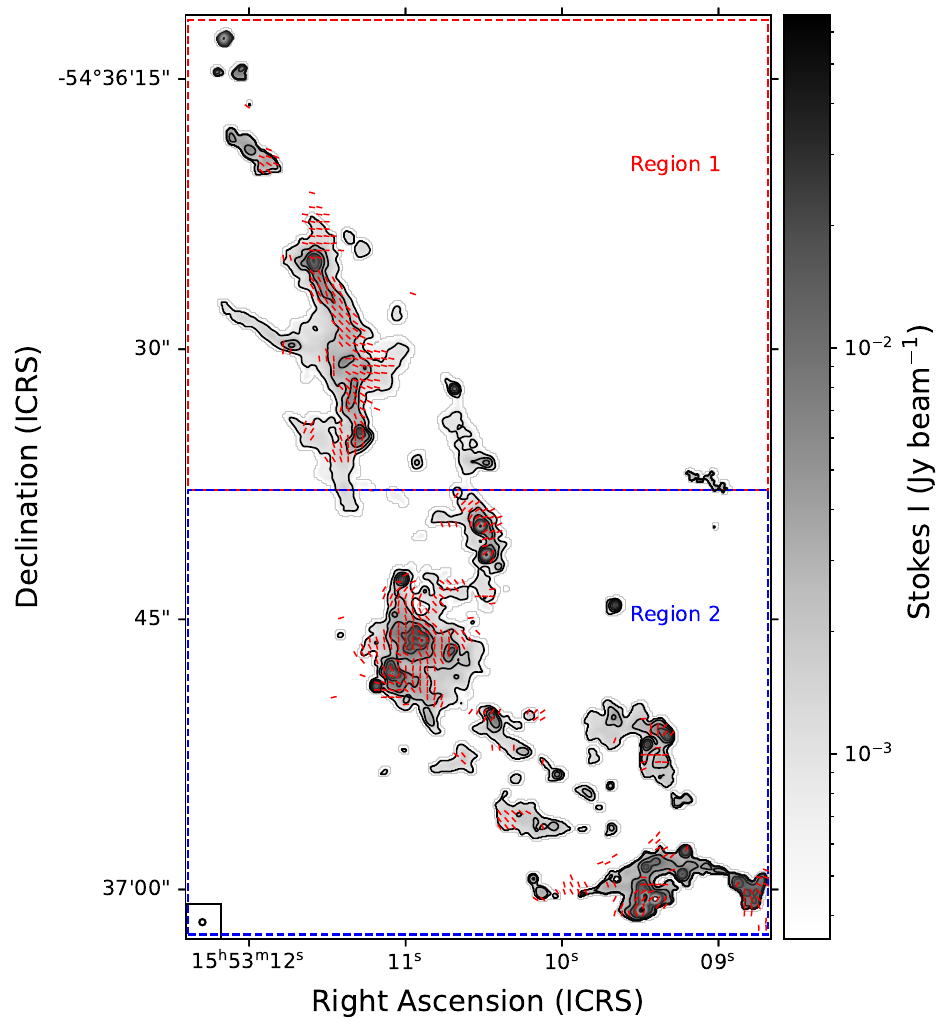}
        \caption{1.2 mm dust continuum map. Red line segments (arbitrary length) representing the magnetic field orientation are plotted for regions above the 3$\sigma$ level, where $\sigma$(rms) = 0.023 mJy beam$^{-1}$ (polarization vector rotated by 90$^{\circ}$). Contours represent the dust continuum emission at levels of 4, 10, 20, 40, 66, 100, 130, 160, and 320 times the rms noise, $\sigma$(rms) = $0.23~\text{mJy beam}^{-1}$. \texttt{Region 1} and \texttt{Region 2} denote the subregions where we calculated the magnetic fields separately. }
        \label{fig:fig5}
\end{figure}

\subsection{Average mass density ($\rho_{\text{avg}}$) of Region 1 and Region 2}\label{subsection_3.2}

In addition to the velocity dispersion ($\sigma_{\text{tot}}$), the average mass density ($\rho_{\text{avg}}$) of the regions is required to estimate the magnetic field strength. We derived $\rho_{\text{avg}}$ from the 1.2 mm dust continuum emission. For simplicity, we assumed that the dust emission is optically thin. Under the optically thin condition ($\tau_{\nu} \ll 1$), the mass of the region and its associated uncertainty are calculated using the following expressions \citep{2022A&A...664A..26P}:

\begin{equation}\label{Eqn4}
M_{\text{region}} = \frac{S^{\text{1.2mm}}_{\text{int}}~d^{2}}{\kappa_{\text{1.2mm}}~B_{\text{1.2mm}} (T_{\text{d}},\nu)},
\end{equation}

\begin{equation}\label{Eqn5}
    \Delta M_{\text{region}} = M_{\text{region}}\sqrt{\biggl(\frac{\Delta S^{\text{1.2mm}}_{\text{int}}}{S^{\text{1.2mm}}_{\text{int}}}\biggl)^{2} ~+~ \biggl(\frac{2\Delta d}{d}\biggl)^{2} ~+~  \biggl(\frac{\Delta B_{\text{1.2mm}} \Huge(T_{\text{d}},\nu)}{B_{\text{1.2mm}} (T_{\text{d}},\nu\Huge)}\biggl)^{2} }~~.
\end{equation}

\noindent Here, $S^{\text{1.2mm}}_{\text{int}}$ denotes the integrated intensity of the 1.2 mm continuum emission, $d$ is the distance to the source, and $T_{\text{d}}$ is the dust temperature in the region. The dust opacity at 1.2 mm, $\kappa_{\text{1.2 mm}}$, is taken as 0.01 $(\nu / 228.9~\text{GHz})^{\beta}$ cm$^{2}$ g$^{-1}$. In our analysis, the continuum was observed in three spectral windows in ALMA Band 6, centred approximately at 244 GHz, 246 GHz, and 258 GHz, giving a mean frequency $\nu$ $\approx$ 250 GHz. For \texttt{Region 1} and \texttt{Region 2}, the values of $S^{\text{1.2mm}}_{\text{int}}$ are 844.71 mJy and 2480.5 mJy, and the average dust temperatures ($T_{\text{d}}$) are 26.2 and 26.7 K, respectively.

To calculate the integrated flux from a FITS image plane, we used the following formula:

\begin{equation}\label{Eqn5_0}
   S_{\rm int} \text{(Jy)} = \sum_\text{i} S_\text{i} \text{(Jy beam$^{-1}$)} \left( \frac{A_{\rm pixel}}{\Omega_{\rm beam}} \right).
\end{equation}

\noindent Here, $S_\text{i}$ is the flux value in Jy beam$^{-1}$ in the $i^\text{th}$ pixel, $A_{\rm pixel}$ is the area of the pixel in arcseconds, and $\Omega_{\rm beam}$ is the beam area in arcseconds, which is 1.1331$\theta_\text{maj}$$\theta_\text{min}$. $\theta_\text{maj}$ and $\theta_\text{min}$ are the major and minor axes of the synthesized beam.

Substituting these values into Eqs.~\ref{Eqn4} and \ref{Eqn5}, we obtain $M_{\text{Region 1}}$ $\pm$ $\Delta M_{\text{Region 1}}$ = 60.1 $\pm$ 24.2 \(\textup{M}_\odot\) and $M_{\text{Region 2}}$ $\pm$ $\Delta$$M_{\text{Region 2}}$ = 170.4 $\pm$ 70.0 \(\textup{M}_\odot\). \texttt{Region 1} is mostly dominated by a filamentary structure with $S^{\text{1.2mm}}_{\text{int}}$ = 748 mJy and $M_{\text{filament}}$ = 52.6 $\pm$ 21.1 \(\textup{M}_\odot\).  Assuming a cylindrical geometry with a radius ($R$) of 0.20 $\pm$ 0.04 pc and a length ($L$) of 0.03 $\pm$ 0.006 pc, the average mass density ($\rho_{\text{avg}}$) and its associated uncertainty ($\Delta \rho_{\text{avg}}$) were computed using the following expressions:

\begin{equation}\label{Eqn6}
   \rho_{\text{avg}}  =  \frac{M_{\text{filament}}}{\pi R^{2} L},
\end{equation}


\begin{equation}\label{Eqn7}
   \Delta\rho_{\text{avg}} = \rho_{\text{avg}} \sqrt{\biggl(\frac{\Delta S^{\text{1.2mm}}_{\text{int}}}{S^{\text{1.2mm}}_{\text{int}}}\biggl)^{2} ~~+~~ \biggl(\frac{\Delta d}{d}\biggl)^{2} ~~+~~\biggl(\frac{\Delta B_{\text{1.2mm}} \Huge(T_{\text{d}},\nu)}{B_{\text{1.2mm}} (T_{\text{d}},\nu\Huge)}\biggl)^{2}  }.
\end{equation}

\noindent After inserting all the values into Eqs. \ref{Eqn6} and \ref{Eqn7}, we obtain $\rho_{\text{avg}}$ $\pm$ $\Delta \rho_{\text{avg}}$ = 5.8 $\pm$ 1.3 $\times$ 10$^{-18}$ g cm$^{-3}$. The derivation of $L$ and $R$ for the filament is discussed in Appendix \ref{appendix_2_0}.

To estimate the average mass density ($\rho_{\text{avg}}$) for \texttt{Region 2}, we first determined the effective radius ($r_{\text{eff}}$), defined as (\text{enclosed area}/$\pi$)$^{1/2}$. 
From this, we obtain $r_{\text{eff}}$ $\pm$ $\Delta r_{\text{eff}}$ = 0.06 $\pm$ 0.01~pc. An uncertainty of 20\% is adopted for $r_{\text{eff}}$, consistent with the assumed uncertainty in $d$. Using these values, we then computed the average mass density ($\rho_{\text{avg}}$) from the following expression:

\begin{equation}\label{Eqn8}
    \rho_{\text{avg}} = \frac{M_{\text{region}}}{(4/3)~\pi~r_{\text{eff}}^{3}}.
\end{equation}

For the uncertainty ($\Delta\rho_{\text{avg}}$), we used the same formula of Eq. \ref{Eqn7}. From the Eqs. \ref{Eqn8} and \ref{Eqn7}, we finally obtain $\rho_{\text{avg}}$ $\pm$ $\Delta \rho_{\text{avg}}$ = 9.2 $\pm$ 2.6 $\times$ 10$^{-18}$ g cm$^{-3}$. From the results of $\rho_{\text{avg}}$ for these two regions \texttt{Region 1} and \texttt{Region 2}, it appears that the average $\rho_{\text{avg}}$ is almost 1.6 times higher in \texttt{Region 2} than in \texttt{Region 1}.

\begin{figure*}[!ht]
        \centering 
    \includegraphics[width=3.5in,height=2.8in,angle=0]{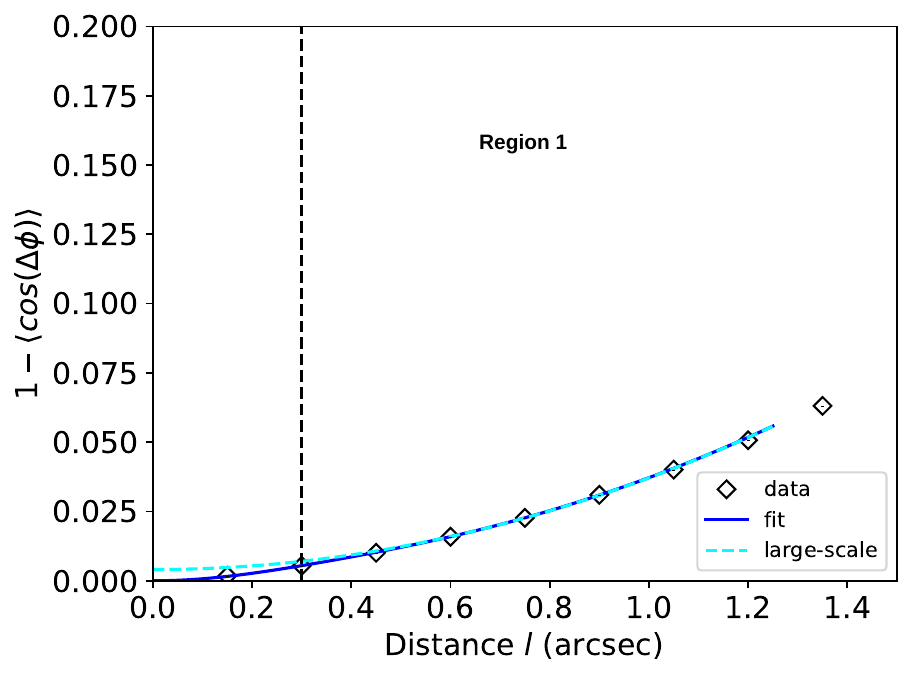}\includegraphics[width=3.5in,height=2.8in,angle=0]{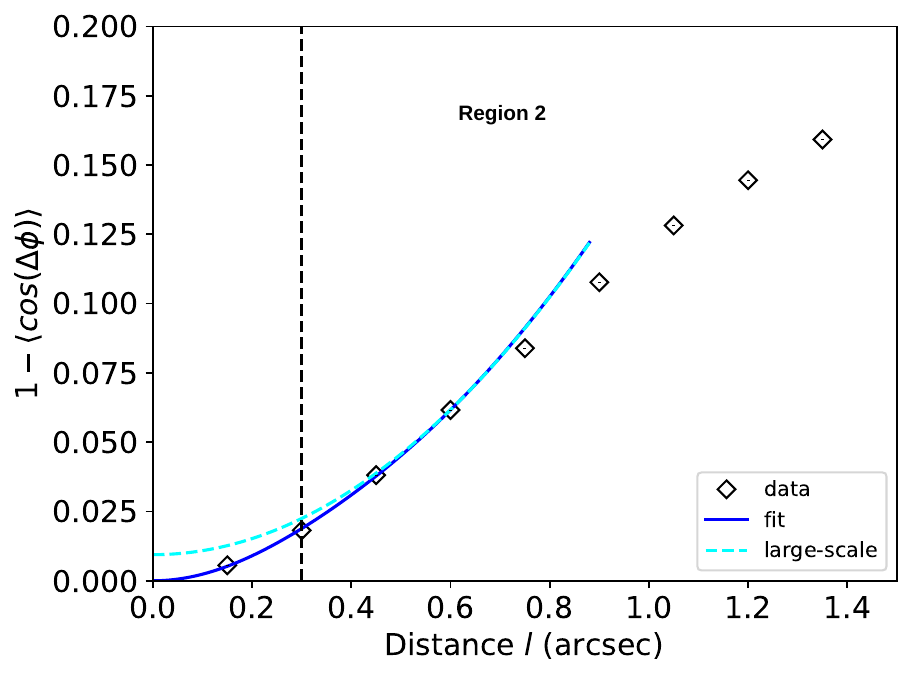}
    \caption{ADFs for \texttt{Region 1} (\textit{left}) and \texttt{Region 2} (\textit{right}). The diamond symbols denote the observed data points. The solid blue line represents the best-fit model. The dashed cyan line denotes the large-scale component of the fit. The vertical dashed~black line denotes the observed beam.}
        \label{fig:fig6}
\end{figure*}

\subsection{Estimation of magnetic field using the ADF method}\label{subsection_3.3}

After deriving $\sigma_{\text{tot}}$ and $\rho_{\text{avg}}$, we calculated the magnetic field using the primary beam–corrected Stokes I, Q, and U images. To do this, we first computed the pixel-wise polarization intensity ($I_\text{p}$), polarization fraction ($f_\text{p}$), and polarization angle ($\phi_\text{p}$), along with their associated uncertainties, using the following expressions:

\begin{equation}
    I_\text{p} = \sqrt{(Q^{2} + U^{2})} ~;~  \Delta I_\text{p} =  \sqrt{ \frac{( Q^{2} \Delta Q^{2} +  U^{2} \Delta U^{2} )}{(Q^{2} +  U^{2})}  }, 
\end{equation}

\begin{equation}
    f_\text{p} = \frac{\sqrt{(Q^{2} + U^{2})} } {I}~ ;~ \Delta f_\text{p}  =  \sqrt {\frac{(Q^2 \Delta Q^2 + U^2 \Delta U^2)}{(Q^2+U^2) I^2}  + \frac{(Q^2+U^2) \Delta I^2}{I^4} } ,
\end{equation}

\begin{equation}
    \phi_\text{p} = \frac{1}{2} \text{tan}^{-1} \biggl(\frac{U}{Q} \biggl)~;~\Delta \phi_\text{p} = \frac{1}{2} \biggl [\sqrt{\frac{(Q^2\Delta U^2 + U^2 \Delta Q^2)}{(Q^2+U^2)^2}}\biggl ]~(180/ \pi)~ \text{deg.}
\end{equation}

We note that the polarization intensity ($I_\text{p}$) and polarization fraction ($f_\text{p}$) were first de-biased following \cite{2006PASP..118.1340V} and \cite{2015JAI.....450005H}. We then derived the magnetic field from the observed polarization vectors using the ADF \citep{2009ApJ...696..567H,2009ApJ...706.1504H,2016ApJ...820...38H} and ST methods \citep{2021A&A...647A.186S}. The ADF method was adapted for interferometric observations by \cite{2016ApJ...820...38H}, accounting for variations in the large-scale magnetic field, signal integration along the line of sight (LOS) and within the beam, and, importantly, the large-scale filtering effects inherent to interferometric measurements. The ADF for interferometric observations is expressed as follows \citep{2016ApJ...820...38H}:

\begin{equation}\label{eq.1}
\begin{split}
 1- < \text{cos}[\Delta \Phi (l)] > ~=~ \sum_{j=1}^{\infty} a_{2j}~l^{2j} ~ + ~  \left[\frac{N}{1+N <B_\text{pos}^{2}>/<B_\text{t}^{2}>} \right]\\
 \times  \biggl \{  \frac{1}{N_{1}} \left [1 -e^{-l^{2}/2(\delta^{2} + 2W_{1}^{2})} \right ] + \frac{1}{N_{2}} \left [1 -e^{-l^{2}/2(\delta^{2} + 2W_{2}^{2})} \right ] \\
 - \frac{2}{N_{12}} \left [1 -e^{-l^{2}/2(\delta^{2} + W_{1}^{2} + W_{2}^{2})} \right ]  \biggl \}.
\end{split}
\end{equation}

\noindent Here, $\Delta \Phi (l)$ represents the angular difference between two points separated by a distance $l$, $\delta$ is the turbulent correlation length, and the summation in the Taylor expansion accounts for variations in the ordered component of the magnetic field. $B_\text{pos}$ denotes the ordered component of the magnetic field, while $B_\text{t}$ represents the turbulent component. For the interferometric observations with ALMA, large-scale structures accessible via total power measurements are not captured in the polarization data, and the corresponding spatial profile is given by

\begin{equation}
    H(r) =  \biggl( \frac{1}{2\pi W_{1}^{2}} \biggl )~e^{-r^{2}/2W_{1}^{2}} -   \biggl ( \frac{1}{2\pi W_{2}^{2}} \biggl )~e^{-r^{2}/2W_{2}^{2}}.
\end{equation}

\noindent Here, $W_{1}$ denotes the standard deviation of the observed beam, and $W_{2}$ represents the standard deviation corresponding to the MRS of the observation.

$N_{1}$ is the number of turbulent eddies sampled within the telescope beam, given by

\begin{equation}
    N_{1} = \frac{\biggl(\delta^{2} + 2W_{1}^{2}\biggl) \Delta'}{\sqrt{2\pi}\delta^{3}}.
\end{equation}

\noindent Here, $\Delta'$ represents the FWHM of the observed normalized autocorrelation function of the integrated normalized polarized flux. Similarly, $N_{2}$, $N_{12}$, and $N$ are defined as follows:

\begin{equation}
    N_{2} = \frac{\biggl(\delta^{2} + 2W_{2}^{2}\biggl) \Delta'}{\sqrt{2\pi}\delta^{3}}~,
\end{equation}

\begin{equation}
    N_{12} = \frac{\biggl(\delta^{2} + W_{1}^{2} + W_{2}^{2}\biggl)~ \Delta'}{\sqrt{2\pi}\delta^{3}}~,
\end{equation}

\begin{equation}
    N = \biggl ( \frac{1}{N_{1}} +   \frac{1}{N_{2}} -  \frac{2}{N_{12}}  \biggl ) ^{-1}.
\end{equation}

From Eq.~\ref{eq.1}, by fitting the observed polarization map shown in Fig. \ref{fig:fig5}, we obtained the parameters $\delta$, $\langle B_\text{t}^{2} \rangle / \langle B_\text{pos}^{2} \rangle$, and $a_{2j}$. The strength of the magnetic field on the plane of the sky ($B_\text{pos}$) and its associated uncertainty could then be calculated using the following expressions \citep{1951PhRv...81..890D,1953ApJ...118..113C}:

\begin{equation}\label{B_field_0}
    B_\text{pos,ADF} = \sqrt{4\pi\rho_{\text{avg}}}~\sigma_{\text{nth}}~\biggl [ \frac{<B_\text{t}^{2}>}{<B_\text{pos}^2>}  \biggl ]^{-1/2}~,
\end{equation}

\begin{equation}\label{error_B_field_0}
    \Delta B_\text{pos,ADF} = B_{\text{pos,ADF}} \sqrt{~ \biggl (\frac{\Delta \rho_{\text{avg}}}{2\rho_{\text{avg}}} \biggl )^{2}  ~ + ~  \biggl (\frac{\Delta \sigma_{\text{nth}}}{\sigma_{\text{nth}}} \biggl )^{2} ~+~\Biggl(\frac{\Delta(\frac{<B_\text{t}^{2}>}{<B_\text{pos}^{2}>})}{2(\frac{<B_\text{t}^{2}>}{<B_\text{pos}^{2}>})}                                \Biggl)^{2}}~.
\end{equation}

To convert the plane-of-sky magnetic field ($B_{\text{pos}}$) into the total magnetic field ($B_{\text{tot}}$), we applied a statistical correction factor ($f$) of $1.25$, following \cite{2023ApJ...949...30L} and \cite{2025ApJ...980...87S}. In \texttt{Region 1}, the ADF method underestimates the effect of LOS signal integration in high-density areas; to account for this, we adopted an additional numerical correction factor of 0.21 \citep{2023ApJ...949...30L,2025ApJ...980...87S}. From the ADF analysis, we derive $\left[\langle B_\text{t}^{2} \rangle / \langle B_\text{pos}^{2} \rangle \right]^{1/2} = $ 0.064 $\pm$ 0.0004 for \texttt{Region 1} and 0.19 $\pm$ 0.003 for \texttt{Region 2}. Figure \ref{fig:fig6} presents the ADFs for \texttt{Region 1} and \texttt{Region 2}. Each ADF is fitted using a reduced $\chi^{2}$ minimization approach. The fitting is performed over various maximum distance ($l$) ranges, and the optimal fit corresponds to the minimum reduced $\chi^{2}$ value. Using these values along with $\rho_{\text{avg}}$ and $\sigma_{\text{nth}}$ in Eqs.~\ref{B_field_0} and \ref{error_B_field_0}, we obtain $B_{\text{tot}}$ =  $1.7 \pm 0.8$ mG for \texttt{Region 1} and $3.6 \pm 1.5$ mG for \texttt{Region 2}. The average total magnetic field ($B_{\text{tot}}$) for the entire region is $2.7 \pm 0.9$ mG. 

The magnetic field strengths derived using the DCF--ADF method is subject to several underlying assumptions and may therefore be affected by systematic uncertainties. \cite{2021A&A...647A.186S} proposed a modified formulation (the ST method) to account for the limitations of the equipartition assumption between turbulent kinetic energy and magnetic fluctuation energy, which is derived under the framework of incompressible magnetohydrodynamic turbulence. Since molecular clouds are compressible environments and support both Alfvénic and compressible (slow and fast) magnetohydrodynamic modes, the equipartition relation may be altered. \cite{2021A&A...647A.186S}, and later \cite{2021A&A...656A.118S} and \cite{2025A&A...700A.256P}, argued that the coupling term of the magnetic-field component can possess energy comparable to that of the turbulent motions. Accordingly, the plane-of-sky magnetic field strength and its associated error are calculated using the following expressions:

\begin{equation}\label{B_field_1}
    B_\text{pos,ST} = \sqrt{2\pi\rho_{\text{avg}}}~\sigma_{\text{nth}}~\biggl [ \frac{<B_\text{t}^{2}>}{<B_\text{pos}^2>}  \biggl ]^{-1/4}~,
\end{equation}

\begin{equation}\label{error_B_field_1}
    \Delta B_\text{pos,ST} = B_{\text{pos,ST}} \sqrt{~ \biggl (\frac{\Delta \rho_{\text{avg}}}{2\rho_{\text{avg}}} \biggl )^{2}  ~ + ~  \biggl (\frac{\Delta \sigma_{\text{nth}}}{\sigma_{\text{nth}}} \biggl )^{2} ~+~\Biggl(\frac{\Delta(\frac{<B_\text{t}^{2}>}{<B_\text{pos}^{2}>})}{4(\frac{<B_\text{t}^{2}>}{<B_\text{pos}^{2}>})}                                \Biggl)^{2}}~.
\end{equation}

\cite{2025A&A...700A.256P} discussed the ST method and there the gas density ($\rho_\text{avg}$) was estimated using three different approaches: (a) inverse radiative transfer analysis, (b) assuming a cylindrical geometry, and (c) assuming a spherical geometry. Similarly, the non-thermal velocity dispersion ($\sigma_\text{nth}$) was derived using (a) the average spectrum, (b) the second-order moment map, and (c) a velocity-dispersion map obtained after subtracting the large-scale velocity gradient from the first order moment map. For the polarization angle dispersion ($\Delta\Phi$), three methods were considered: (a) direct calculation of the dispersion angle, (b) Gaussian fitting of the polarization-angle distribution, and (c) the method of \cite{2016ApJ...820...38H}. By comparing the resulting magnetic field estimates with the true magnetic field strengths in simulations viewed at different inclination angles ($\gamma$), they found that the most accurate results were obtained when the $\rho_\text{avg}$ was derived through inverse radiative transfer analysis, the $\sigma_\text{nth}$ was measured from the second-order moment map, and the $\Delta\Phi$ was determined using Gaussian fitting of the polarization-angle distribution (with the \cite{2016ApJ...820...38H} method yielding similarly consistent results). This combination provided magnetic field strengths that most closely matched the actual values in the simulations.

We also estimated the magnetic field strength using the ST method. However, we were unable to follow the exact procedure adopted in the study of \cite{2025A&A...700A.256P}. In particular, owing to the difficulty of deriving the gas density through inverse radiative transfer analysis, we retained our density estimates obtained from the 1.2 mm dust continuum emission. Furthermore, since our target region is dense and strongly influenced by self-gravity, where ordered motions such as infall and rotation can contribute significantly to the observed kinematics, we did not adopt the second-order moment map to estimate the non-thermal velocity dispersion. Instead, we used the mean intensity-weighted velocity dispersion, consistent with our DCF--ADF analysis.

For the polarization angle dispersion, we employed the method of \cite{2016ApJ...820...38H}. In the study of \cite{2025A&A...700A.256P}, both Gaussian fitting of the polarization-angle distribution and the \cite{2016ApJ...820...38H} method were found to provide reliable estimates of the angular dispersion, although the ST formulation introduces a slight modification to the \cite{2016ApJ...820...38H} approach. Nevertheless, for consistency with our previous analysis, we retained the \cite{2016ApJ...820...38H} method. Under these assumptions, from Eqs.~\ref{B_field_1} and \ref{error_B_field_1}. the plane of sky magnetic field strength ($B_\text{pos}$) are 0.5 $\pm$ 0.2 mG for \texttt{Region 1} and 0.9 $\pm$ 0.4 mG for \texttt{Region 2}. Using the DCF--ADF method, we derived $B_{\rm pos} = 1.4 \pm 0.6\,\mathrm{mG}$ for \texttt{Region 1} and $B_{\rm pos}=2.9 \pm 1.2\,\mathrm{mG}$ for \texttt{Region 2}. Consequently, the  DCF--ADF method appears to overestimate the magnetic field strength by approximately a factor of three relative to the ST method.

However, it should be noted that both the DCF--ADF method and the ST method are based on assumptions that may break down in regions where self-gravity becomes dynamically important. This is discussed in Sect. 6 and Appendix A in \cite{2025A&A...700A.256P}. In self-gravitating environments, ordered motions such as gravitational infall or in the region where rotation or outflow can contribute significantly to the observed velocity field, violating the assumption that the velocity dispersion is solely due to turbulence. Consequently, additional terms, including kinetic-energy cross terms associated with these coherent motions, may need to be considered for a more accurate determination of the magnetic field strength. At present, the treatment of such effects remains an open problem and requires further theoretical investigation.

As both methods are known to have limitations in the dense, self-gravitating regions of molecular clouds, a condition that is likely applicable to our target region. Therefore, for the calculation of the virial parameters of the regions and the cores, we adopted the magnetic field strength derived using the \cite{2016ApJ...820...38H} method. Furthermore, the virial parameters of the regions and the cores themselves are associated with substantial uncertainties, and the difference between the magnetic field strengths obtained from the two methods does not significantly affect our conclusions.

\subsection{Energy balance between magnetic fields, turbulence, and gravity}\label{subsection_3.4}
{\color{black}We calculated the energy balance by comparing the contributions from turbulence, magnetic fields, and gravity in each region. The detailed calculations for \texttt{Region 1} and \texttt{Region 2} are presented below.}
\subsubsection{\texttt{Region 1}}

\texttt{Region 1} is mainly dominated by a filament. For a filament of mass $M$, radius $R$, and length $L$, the gravitational potential energy ($E_{\text{G}}$) is \citep{2000MNRAS.311...85F, 2024MNRAS.528.1460R}

\begin{equation}
   E_{G} = - \frac{GM^{2}}{L},~~~~~~~~~ \text{when R $<<$ L. }
\end{equation}

\noindent The kinetic energy of the filament is \citep{2018A&A...609A..43X}

\begin{equation}
   E_{\text{k}}  = \frac{3}{2} M \sigma_{\text{tot}}^{2}.
\end{equation}

\noindent The magnetic energy of the filament is

\begin{equation}
   E_{\text{B}}  = \frac{B_{\text{tot}}^{2}R^{2}L}{8}.
\end{equation}

According to the earlier analysis, the filament has $M =$ $52.6 \pm 21.1$~\(\textup{M}_\odot\), $\sigma_{\text{nth}} = 0.48 \pm 0.22$ km s$^{-1}$, $L = 0.20 \pm 0.04$ pc, and $B_\text{tot}$ = 1.7 $\pm$ 0.8 mG. To find the $\sigma_{\text{tot}}$ of the region, we needed to calculate the thermal contribution ($\sigma_{\text{th}}$), which is equal to $\sqrt{\frac{k_{\text{B}}T_{\text{d}}}{\mu m_{\text{H}}}}$. Here $\mu$ is mean molecular weight (2.37), $m_{\text{H}}$ is the hydrogen mass (1.67 $\times$ 10$^{-24}$ g), and $T_{\text{d}}$ is the dust temperature in the region. This gives $\sigma_{\text{tot}} = 0.57 \pm 0.22$ km s$^{-1}$. From these values we obtain a virial parameter $\alpha_{\text{vir}} \pm \Delta\alpha_{\text{vir}}$ = $2.8 \pm 1.9$. The detailed derivations of $\alpha_\text{vir}$ and $\Delta\alpha_{\text{vir}}$ are presented in Eqs. \ref{eqn25} and \ref{eqn26}. Given the uncertainties of these values, both scenarios remain plausible. We then examined the relative importance of turbulence and magnetic fields in \texttt{Region 1} by calculating the ratio of turbulent energy to magnetic energy ($\beta_{1}$):

\begin{eqnarray}
    \beta_{1} & = & \frac{2E_{\text{nth}}}{E_{\text{B}}} \nonumber \\
          & = & \frac{24M\sigma_{\text{nth}}^{2}}{B_\text{{tot}}^{2}R^{2}H}. 
\end{eqnarray}

\noindent From this, we obtain $\beta_{1}$ $\sim$ 0.34. From the value of $\beta_{1}$, it indicates that in \texttt{Region 1}, the magnetic field is more dynamically important than the turbulence.

\subsubsection{\texttt{Region 2}}

The shape of \texttt{Region 2} is arbitrary. We considered this region to be a spherical system of mass $M$, radius $R$, and density $\rho$ that varies with the radial distance ($r$) as $r^{-\alpha}$. The gravitational potential energy of the system is obtained as \citep{2023PASA...40...53K}

\begin{equation}\label{eqn27}
    E_{\text{G}} =- \biggl(\frac{3-\alpha}{5-2\alpha}\biggl)~ \biggl(\frac{G M^{2}}{R} \biggl).
\end{equation}

\hspace{-5mm}For $\alpha$ = 0; $E_{\text{G}}$ = - $\frac{3}{5}$$\biggl(\frac{GM^{2}}{R}\biggl)$ and $\alpha$ = 2; $E_{\text{G}}$ = - $\biggl(\frac{GM^{2}}{R}\biggl)$.

\noindent Here we consider the centrally peaked density profile where $\alpha$ = 2 rather than the uniform density profile where $\alpha$ = 0. This is because towards the cores, which are there in the region, the density is much higher than in the surrounding environment. The kinetic and magnetic energy of the region are

\begin{equation}\label{eqn28}
   E_{\text{k}}  = \frac{3}{2} M \sigma_{\text{tot}}^{2}
\end{equation}

\hspace{-3mm}and

\begin{equation}\label{eqn29}
   E_{\text{B}}  = \frac{B_{\text{tot}}^{2}R^{3}}{6}.
\end{equation}

\begin{figure}[!ht]
        \centering 
        \includegraphics[width=3.4in,height=2.7in,angle=0]{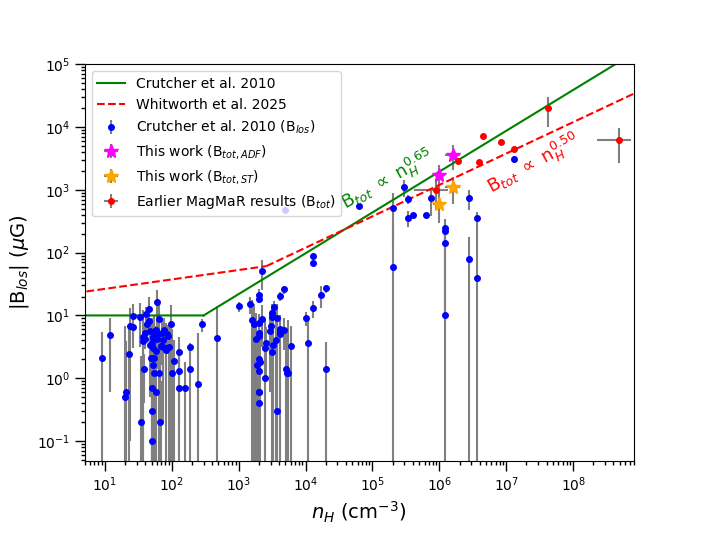}
        \caption{Correlation between the magnetic field ($B_{\text{tot}}$) and number density ($n_{\text{H}}$) obtained from \cite{2010ApJ...725..466C}. The blue dots and grey lines are measured LOS magnetic fields ($B_{\text{los}}$) and the corresponding errors (1$\sigma$) obtained from Zeeman splitting measurements. A power law index ($\alpha$) of 0.65 is obtained from the observed data using Bayesian statistics beyond a $n_{\text{H}}$ of 300 cm$^{-3}$ from \cite{2010ApJ...725..466C} and 0.50 beyond a $n_{\text{H}}$ of 2600 cm$^{-3}$ from \cite{2025MNRAS.540.2762W}. The magenta asterisks are $B_{\text{tot}}$ values (after statistical correction) obtained from our work using the ADF method, and the orange asterisks are $B_{\text{tot}}$ determined using the ST method. The red dots are $B_{\text{tot}}$ values from the earlier \texttt{MagMar} survey \citep{2021ApJ...923..204C,2021ApJ...915L..10S,2024ApJ...974..257Z,2024ApJ...972L...6S, 2025ApJ...980...87S,2026AJ....171...50H}.}
        \label{fig:fig7}
\end{figure}

In \texttt{Region 2}, $M_{\text{region,2}}$ = 170.4 $\pm$ 70.0 \(\textup{M}_\odot\), $\sigma_{\text{nth}}$ = 0.51 $\pm$ 0.30 km s$^{-1}$, $d$ = 2.5 $\pm$ 0.5 pc, and $B_\text{tot}$ = 3.6 $\pm$ 1.5 mG. Similar to Region 1, we calculated the contribution of $\sigma_{\text{th}}$ for the $T_{\text{d}}$ = 26.7 $\pm$ 2.0 K and obtain $\sigma_{\text{tot}}$ = 0.59 $\pm$ 0.20 km s$^{-1}$. From the above equations we thus obtain $\alpha_{\text{vir}}$ $\pm$ $\Delta \alpha_{\text{vir}}$ = 0.5 $\pm$ 0.4. The values of $\alpha_{\text{vir}}$, along with their associated uncertainties, indicate that the region could be either gravitationally bound or unbound. Under these conditions, the two possibilities remain equally plausible. The detailed derivations of $\alpha_\text{vir}$ and $\Delta\alpha_{\text{vir}}$ are presented in Appendix \ref{appendix_4}.

We calculated the ratio of turbulent energy to magnetic energy. This parameter is defined by

\begin{equation}
    \beta_{2} = \frac{18M\sigma_{\text{nth}}^{2}}{B_{\text{tot}}^{2}R^{3}}.
\end{equation}

\noindent We determined $\beta_{2}$ to be $\sim$ 0.17. This indicates that, like in \texttt{Region 1}, magnetic field is more dynamically important than turbulence.

\begin{figure*}[!ht]
        \includegraphics[width=2.5in,height=2.0in,angle=0]{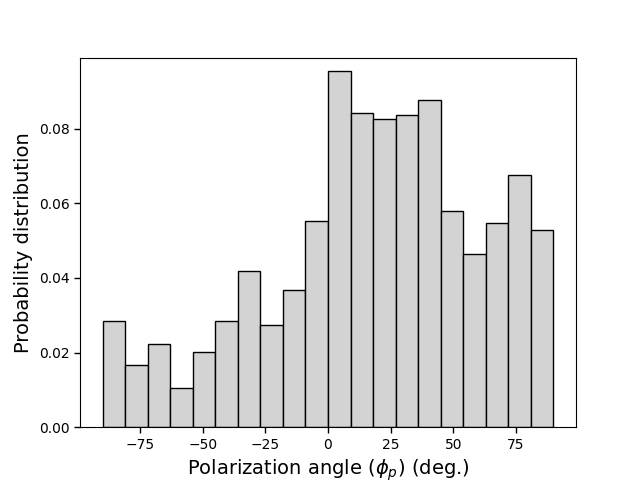}
   \includegraphics[width=2.5in,height=2.0in,angle=0]{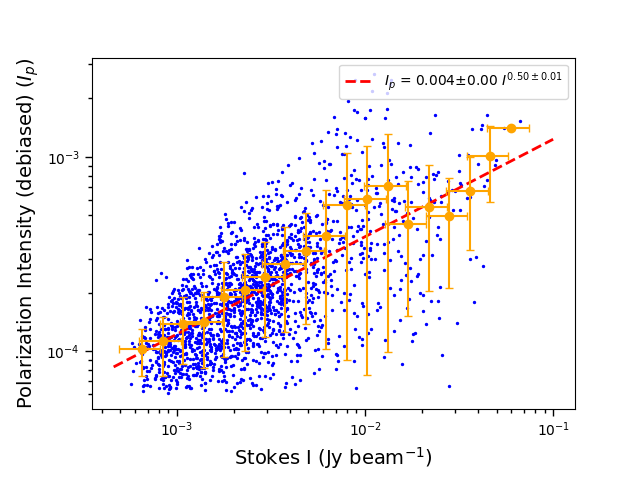}\includegraphics[width=2.5in,height=2.0in,angle=0]{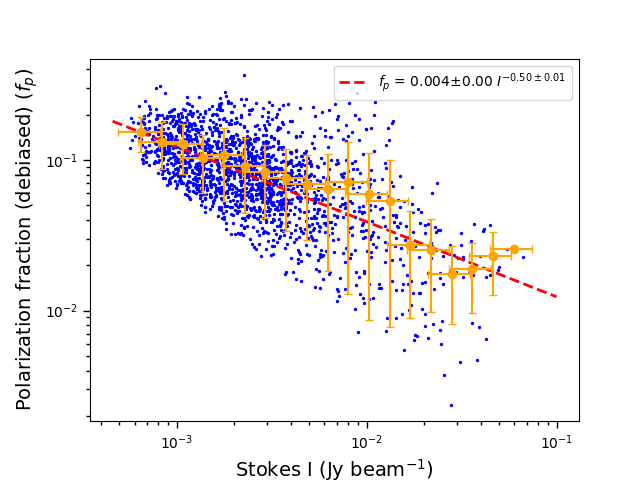}\\
    \caption{\textit{Left}: Histogram plot of polarization angle ($\phi_\text{p}$) for the entire observed region (\texttt{Region 1} plus \texttt{Region 2}). \textit{Middle}: Correlation between  Stokes I and polarization intensity ($I_\text{p}$). \textit{Right}: Correlation between Stokes I and the polarization fraction ($f_\text{p}$) in the same region.}
        \label{fig:fig8}
\end{figure*}

\begin{figure}[hbt!]
        \centering 
        \includegraphics[width=3.6in,height=3.4in,angle=0]{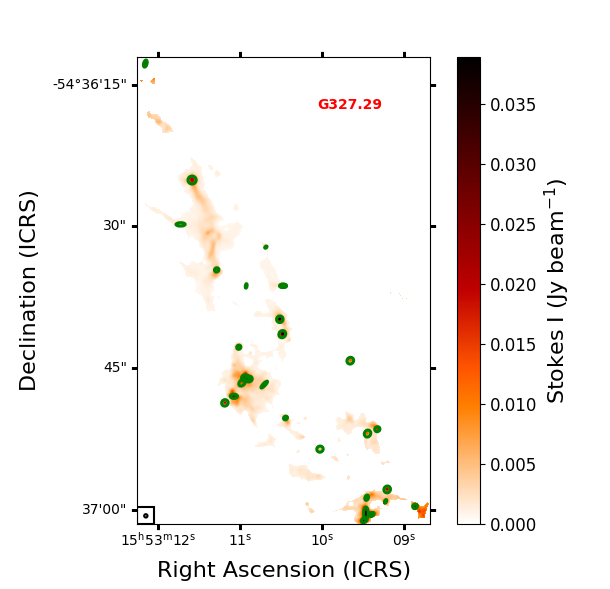}\\
        \caption{Spatial distribution of 1.2 mm cores extracted using {\color{black}\texttt{Astrodendro} \citep{2008ApJ...679.1338R} and \texttt{TGIF} \citep[][]{tgif} modules}. Here the green circles and ellipses are the 1.2 mm continuum cores, and the background emission is the 1.2 mm continuum.}
        \label{fig:fig10}
\end{figure}

\begin{figure*}[ht!]
        \centering 
        \includegraphics[width=3.6in,height=2.8in,angle=0]{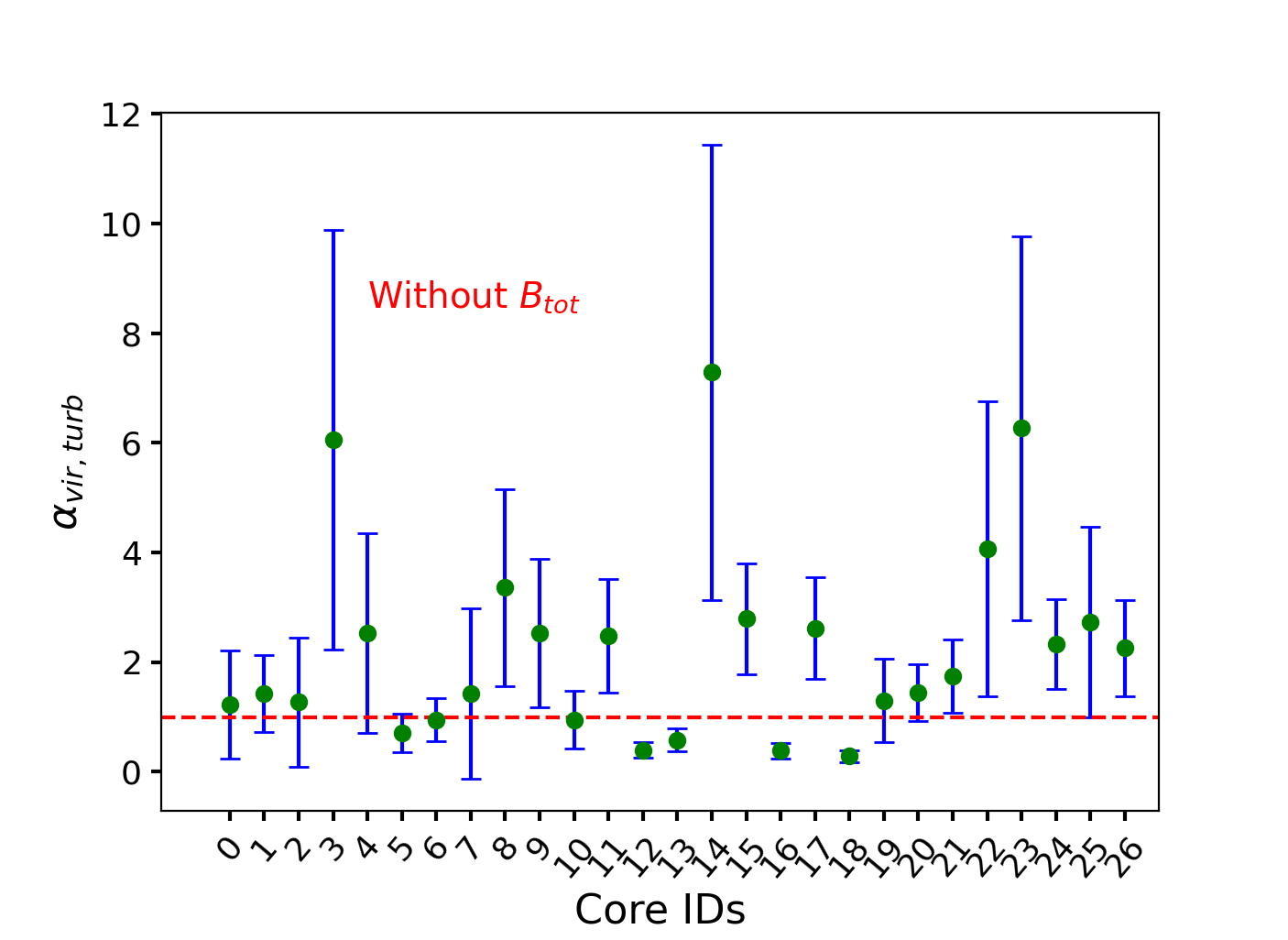}\includegraphics[width=3.6in,height=2.8in,angle=0]{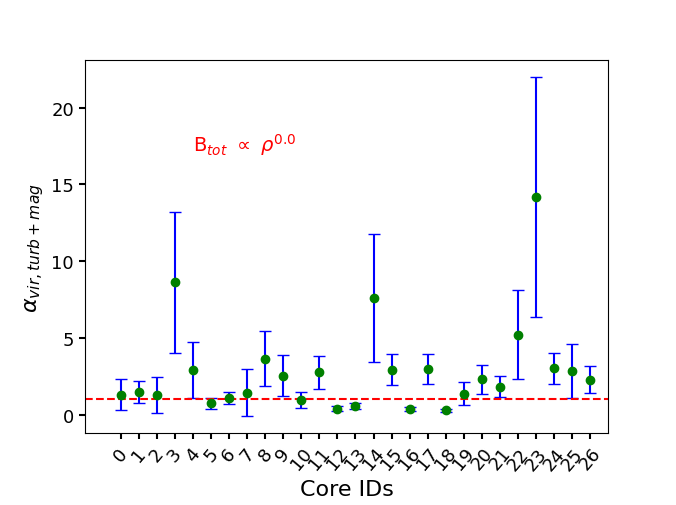}\\
    \includegraphics[width=3.6in,height=2.8in,angle=0]{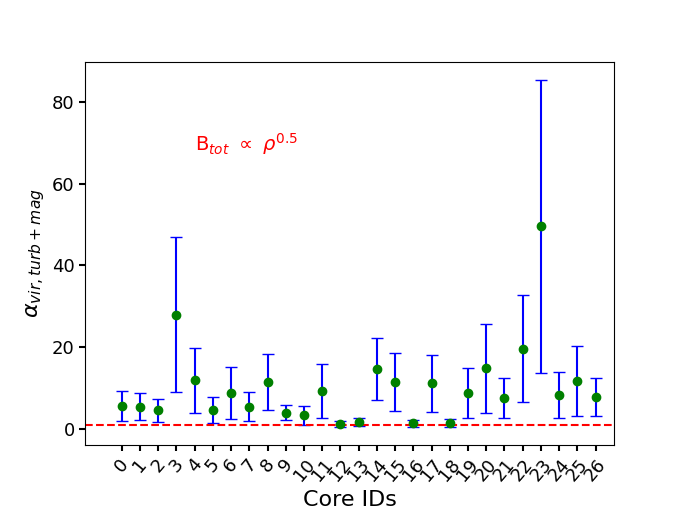}\includegraphics[width=3.6in,height=2.8in,angle=0]{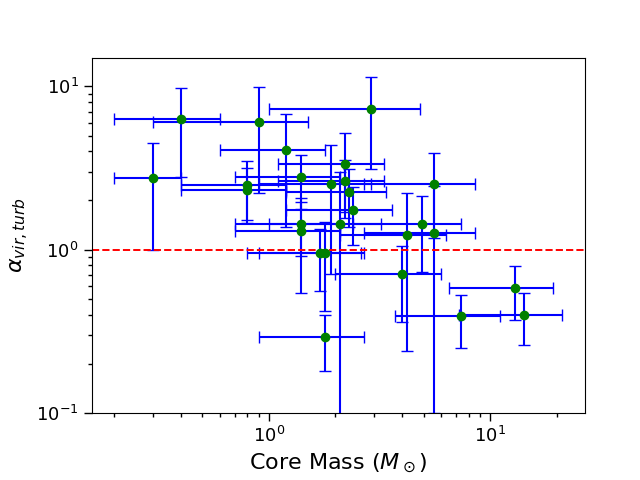}\\
        \caption{\textit{Top left}: Virial parameter ($\alpha_{\text{vir, turb}}$) for all the cores without taking the magnetic field ($B_{\text{tot}}$) into account. \textit{Top right}: Virial parameter ($\alpha_{\text{vir, turb + mag}}$) after considering the effect of the magnetic field ($B_{\text{tot}}$), where $B_{\text{tot}}$ $\propto$ $\rho^{0.0}$. \textit{Bottom left}: Same as the upper right but for $B_{\text{tot}}$ $\propto$ $\rho^{0.5}$. \textit{Bottom right}: Virial parameter ($\alpha_{\text{vir, turb}}$) as a function of core mass.}
        \label{fig:fig12}
\end{figure*}

\begin{figure}[ht!]
        \centering 
        \includegraphics[width=3.6in,height=3.0in,angle=0]{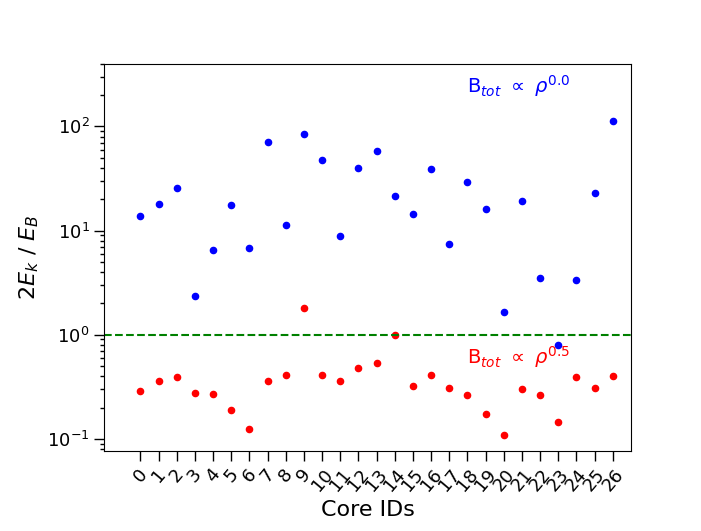}\\
        \caption{Ratio of the kinetic energy (2$E_\mathrm{k}$) to magnetic energy ($E_\mathrm{B}$) towards the cores for two cases: $B_\mathrm{tot} \propto \rho^{0.0}$ (shown in blue) and $B_\mathrm{tot} \propto \rho^{0.5}$ (shown in red).}
        \label{fig:fig13}
\end{figure}

\subsection{Magnetic field versus density correlation}\label{subsection_3.5}

 Magnetic field ($B_{\text{tot}}$) and density ($\rho$) are correlated in the relatively dense regime of the interstellar medium. \cite{2010ApJ...725..466C} studied the correlation between them in the neutral interstellar medium and the relatively high-density regime of the interstellar medium using the Zeeman splitting measurements of H{\sc i}, OH, and CN radicals. From their analysis, they observed that up to the number density ($n$) of $\sim$ 300 cm$^{-3}$ no correlation exists. {\color{black}Beyond this density,} a correlation is evident and the power law index ($\alpha$) is $\sim$ 0.65. This is portrayed in Fig. \ref{fig:fig7}. On the other hand, \cite{2025MNRAS.540.2762W} obtained the value of $\alpha$ $\sim$ 0.50 from the Zeeman splitting technique and $\sim$ 0.78 from dust polarization measurements. The value of $\alpha$ varies based on how the gas assembles from the low density to the high density medium \citep{2015MNRAS.451.4384T}. For example, if a uniform magnetic field is threaded in the plane of the disc and the gas accumulation occurs perpendicular to the plane of the disc, then $B_{\text{tot}}$ $\propto$ $\rho$. In the same system, if the cloud is allowed to contract along the field lines then $B_{\text{tot}}$ $\propto$ $\rho^{0}$. Likewise, if the magnetic field is orientated perpendicular to the plane of the disc, matter accumulation occurs along the magnetic field and due to the pinching effect of the gravity, an hourglass type morphology of the magnetic field is observed. In this scenario, $B_{\text{tot}}$ varies with $\rho^{1/2}$, which comes under the assumption of isothermal contraction and equipartition between magnetic and thermal pressures \citep{1976ApJ...206..753M}. Furthermore, if the gas cloud collapses almost isotopically under the effect of gravity, then, based on the conservation of mass and magnetic flux, $B_{\text{tot}}$ scales with $\rho^{2/3}$ \citep{1965QJRAS...6..265M}.
In Fig. \ref{fig:fig7} we also show the magnetic fields ($B_\text{{tot}}$) for \texttt{Region 1} and \texttt{Region 2} and $B_\text{{tot}}$ from the earlier \texttt{MagMar} survey \citep{2021ApJ...915L..10S,2021ApJ...923..204C, 2024ApJ...974..257Z, 2024ApJ...972L...6S, 2025ApJ...980...87S}. Our analysis yields $B_{\text{tot}}$ of 1.7 $\pm$ 0.8 and 3.6 $\pm$ 1.5 mG at densities 1.5 $\pm$ 0.3 $\times$ 10$^{6}$ and 2.3 $\pm$ 0.7 $\times$ 10$^{6}$ cm$^{-3}$, respectively. This was obtained from DCF--ADF method. From the ST method, we obtain the magnetic fields of 0.6 $\pm$ 0.3 and 1.1 $\pm$ 0.5 mG. Likewise, in the \texttt{MagMar} survey, \cite{2021ApJ...915L..10S} studied the magnetic field towards the IRAS 18089-1732 star-forming region and observed a complex spiral pattern of magnetic field towards the envelope region. From their analysis, they obtained a $B_{\text{tot}}$ of 4.4 mG at $n$ = 1.3 $\times$ 10$^{7}$ cm$^{-3}$. \cite{2021ApJ...923..204C} show that the total magnetic field strength in the NGC 6334I (N) region is 20 $\pm$ 10 mG, which corresponds to a number density of 4.24 $\times$ 10$^{7}$ cm$^{-3}$. In addition, \cite{2024ApJ...974..257Z} show that the magnetic field strength in the O-type protostellar system IRAS 16547–4247 ranges from $\sim$ 2 to 6 mG, corresponding to a number density between $\sim$ 2 $\times$ 10$^{6}$ and 5 $\times$ 10$^{6}$ cm$^{-3}$. Likewise, \cite{2024ApJ...972L...6S} observed an hourglass-type morphology of the magnetic field towards this region. From the modelling of this pattern, they measured a $B_{\text{tot}}$ of 5.7 mG at $n$ = 8.4 $\times$ 10$^{7}$ cm$^{-3}$. Furthermore, \cite{2025ApJ...980...87S} and \cite{ 2026AJ....171...50H} studied the young high-mass core G11.92 MM2 and the G35.20–0.74N high-mass star-forming region and obtained a $B_{\text{tot}}$ of 6.2 $\pm$ 3.5 mG at $n$ = 4.8 $\pm$ 2.5 $\times$ 10$^{8}$ cm$^{-3}$ and 1.0 $\pm$ 0.8 mG at $n$ = 8.8 $\pm$ 4.6 $\times$ 10$^{5}$ cm$^{-3}$, respectively. Unlike in \cite{2025ApJ...980...87S}, other \texttt{MagMar} results as well as our results follow a power law that is closely related to the observed result of \cite{2010ApJ...725..466C}.  For the case of G11.92 MM2, this effect may be due to the different physical environment and the evolutionary stage of the core.

\subsection{Properties of the linear polarized emission}\label{subsection_3.6}
In Fig. \ref{fig:fig8} we show the histogram plot of the polarization angle ($\phi_{\text{p}}$). From the distribution, we observe that $\phi_{\text{p}}$ is almost uniformly distributed from $-$90 to $+$90 deg. This indicates that in the observed region the magnetic field has complex morphology. We also show the correlations between de biased polarization intensity ($I_{\text{p}}$) and Stokes I and between the de-biased polarization fraction ($f_{\text{p}}$) and Stokes I. We notice that $I_{\text{p}}$ increases with  Stokes I with a power law index ($\alpha$) of 0.50 and $f_{\text{p}}$ decreases with Stokes I with a $\alpha$ value of $-$0.50. One correlation is simply obtained from the other. We note that, prior to fitting the data points, the pixel size was adjusted such that each beam contained only two pixels. This reduces the effect of oversampling in the analysis. The same procedure was also adopted in the calculation of the magnetic field using the ADF and ST methods.

Earlier studies have also obtained a similar decrease in $f_{\text{p}}$ with Stokes I \citep{2003ApJ...592..233W, 2009ApJ...695.1399T, 2013ApJ...763..135T,2018ApJ...861...65S,2021ApJ...913...29F, 2024A&A...682A..81B}. For example, \cite{2024A&A...682A..81B} studied the correlation in 20 massive star-forming regions and noticed that the value of $\alpha$ is $-$0.62. Likewise, \cite{2021ApJ...913...29F} studied the linear polarization in the ultracompact H {\sc ii} region G5.89–0.39 and noticed that the value of $\alpha$ is $-$0.72. This anti-correlation comes from the depolarization effect of the polarized emission \citep{2003ApJ...592..233W, 2007MNRAS.378..910L,2009ApJ...695.1399T, 2013ApJ...763..135T, 2024A&A...682A..81B}.

\section{Turbulence and magnetic fields at core scale}\label{section_4}

After examining the effect of magnetic fields, turbulence, and gravity on \texttt{Region 1} and \texttt{Region 2}, we studied the effect of these on core dynamics. For that we extracted the cores and calculated their mass, size, temperature, and the total velocity dispersion. In addition to that, we also examined the average spectra of the two spectral lines HN$^{13}$C and H$^{13}$CO$^{+}$ towards the dense cores. This enables us to determine the effect of turbulence on core dynamics. Furthermore, to examine the effect of magnetic fields, we considered two scenarios: one in which the magnetic field is the same as the surrounding environment, and the other in which we consider some scaling relation between magnetic fields and density.

\subsection{Core extraction using the 1.2 mm continuum}\label{subsection_4.1}
Firstly, for extracting the cores from the 1.2 mm continuum in this observed region, we used the \texttt{Astrodendro} and \texttt{TGIF} modules \citep[][]{2008ApJ...679.1338R,tgif}. The combination of these two modules allows us to accurately determine the positions of the cores and subtract the background emissions while measuring the integrated intensities of the cores. First, we used the \texttt{Astrodendro} module to identify the cores in the 1.2 mm continuum emission. We set the parameters \texttt{min\_value} = 3$\sigma$, \texttt{min\_delta} = 1.5$\sigma$ and \texttt{min\_pix} equal to 2 times the beam FWHM. This configuration allows us to extract all hierarchical intensity structures from the position-position image plane. However, after a visual inspection, we notice that some structures (five) identified as cores in the \texttt{Astrodendro} module are not real cores. We consider them false detections and did not take them into account in our analysis. We note that due to the uniform level of noise, we first used the primary beam uncorrected image plane to find the core positions (leaves in the \texttt{Astrodendro} module). We then inserted those positions into the TGIF module and obtained the physical parameters of the cores from the primary beam-corrected image plane. The TGIF module first fits a 2D Gaussian at the specified position. This then subtracts the background of the surrounding diffuse emissions. After that, this module again fits a 2D Gaussian at the specified position in the background-subtracted image plane to obtain the various parameters of the cores, for example the integrated intensity ($S_{\text{int,core}}^\text{1.2mm}$), peak flux ($S_{\text{peak,core}}^\text{1.2mm}$), and major ($\theta_{\text{maj}}$) and minor ($\theta_{\text{min}}$) axes of the cores (before and after beam correction).  We note that we did not consider the fixed values of the fitting parameters \texttt{bkg\_inner\_width}, \texttt{bkg\_annulas\_width}, \texttt{fitting\_size\_default}, rather we varied {\color{black}them} slightly in a few cases. For example, we increased the \texttt{bkg\_inner\_width} to 6 in cases where multiple cores are located adjacently. In this case, the default value of 4 of the parameter \texttt{bkg\_inner\_width} will take the nearby cores as the background. Consequently, the background-subtracted value of the target core will be less than the original. Likewise, we changed the parameter \texttt{fitting\_size\_default} from 0.9 to 0.6 depending on whether the core is isolated or adjacent cores are present nearby. We extracted a total of 28 cores from the region. An example of 2D Gaussian modelling for the ALMA 1.2 mm flux measurements of the cores in the G327.29 protocluster is shown in Fig. \ref{fig:fig9}, while the remaining examples are presented in Koley\_zenodo\_2026\footnote{\url{https://zenodo.org/records/22541324}}. In Fig. \ref{fig:fig10} we display these cores on top of the 1.2 mm continuum emission. The properties of the cores, for example their spatial position (RA and Dec), deconvolved size ($\theta_\text{{maj}}$ $\times$ $\theta_\text{{min}}$), size ($S$), peak flux ($S_{\text{peak,core}}^\text{1.2mm}$), and integrated intensity ($S_\text{{int,core}}^\text{1.2mm}$), are listed in Table \ref{tab:table1}.

{\color{black}

\subsection{Core temperature estimation}\label{subsection_4.2.1}
We adopted the point process mapping dust temperatures for all cores, except for the hot core candidates, for which we instead applied the method proposed by \cite{2024A&A...687A.163B} from the ALMA-IMF work. We cross-checked the 1.2 mm continuum with the peak position of Methyl formate (CH$_{3}$OCHO), which is formed on dust grain surfaces in lukewarm environments (30$-$40 K) and is released into the gas phase once the temperature reaches about $\sim$100 K. We assigned a temperature of 100$\pm$50 K to the 2 sources whose positions coincide with peaks of extended methyl formate emission peaks, and classified these sources as hot core candidates. For the other cores, the mean temperatures were taken from the dust map and the uncertainties associated with the mean temperature were calculated using the same procedure as described in \cite{2024A&A...690A..33L}. The temperature values for each core ($T_{\text{core}}$) are listed in Table \ref{tab:table1}.

}

\subsection{Core mass estimation using the 1.2 mm continuum}\label{subsection_4.2}

After extracting the cores and obtaining the temperature, we calculated the mass of the cores from the measured properties of the cores. In general, 1.2 mm continuum emission is optically thin. However, previous studies have shown that in the extremely high density regime, where the number density is 10$^{7}$ to 10$^{8}$ cm$^{-3}$ \citep{2017MNRAS.468.3694C,2018ARA&A..56...41M,2025A&A...694A..24M}, the emission becomes optically thick. In this case, an optically thin assumption will lead to an underestimation of the core mass. Consequently, in our analysis, we used the optical depth ($\tau_{\nu}$) corrected core masses and its associated errors, which are obtained using the formulae \citep{2022A&A...664A..26P}

\begin{equation}
\begin{split}
M_{\text{core},\tau \ge 1} =- \Omega^{1.2\text{mm}}_{\text{beam}} ~ \frac{d^{2}}{\kappa_{1.2 \text{mm}}}~\frac{S^{{1.2 \text{mm}}}_{\text{int,core}}}{S^{\text{1.2mm}}_{\text{peak,core}}}~\text{ln}~\biggl(1-\\
\frac{S^{1.2 \text{mm}}_{\text{peak,core}}} {\Omega^{\text{1.2 mm}}_{\text{beam}}~B_\text{1.2mm}(T_{\text{core}},\nu)_\text{}}  \biggl )~, 
\end{split}
\end{equation}

\begin{equation}
\Delta M_{\text{core},\tau\ge 1}\approx M_{\text{core},\tau\ge 1}
\sqrt{
\begin{aligned} \bigg(\frac{2\Delta d}{d}\bigg)^{2}+\bigg(\frac{\Delta S^{{1.2 \text{mm}}}_{\text{int, core}}}{S^{{1.2 \text{mm}}}_{\text{int, core}}}\bigg)^{2}+\\
\bigg(\frac{\Delta B_\text{1.2mm}(T_{\text{core}},\nu)_\text{}}{B_\text{1.2mm}(T_{\text{core}},\nu)_\text{}}\bigg)^{2}.
\end{aligned} 
}  
\end{equation}

\noindent Here $\Omega^{1.2\text{mm}}_{\text{beam}} $ is the solid angle of the beam at 1.2 mm, $d$ is the distance of the source, $\kappa_{1.2 \text{mm}}$ is the opacity at 1.2 mm (which is equal to 0.01 ($\nu /228.9~\text{GHz}$)$^{\beta}$ cm$^{-2}$g$^{-1}$), and $\nu$ is the central frequency at 1.2 mm continuum and $\beta$ = 1.5, which is the opacity index for dense cold gas on the core scale \citep{2022A&A...664A..26P,2024A&A...690A..33L,2025A&A...696A..11V}. $S^{{1.2 \text{mm}}}_{\text{int,core}}$ and $S^{{1.2 \text{mm}}}_{\text{peak,core}}$ are the peak and integrated intensities of the core. $T_{\text{core}}$ is the core temperature and $B_\text{1.2mm}(T_{\text{core}},\nu)_\text{}$ is the Planck function at frequency $\nu$ and temperature $T_{\text{core}}$. After obtaining all the necessary parameters, we calculated the core masses. The derived mass values and their associated errors are presented in Table \ref{tab:table1}.

{\color{black}
\subsection{Core velocity dispersion measurement}\label{subsection_4.2.2}
To calculate the total velocity dispersion ($\sigma_{\text{tot,core}}$) of the cores, we used the mean spectra of HN$^{13}$C and H$^{13}$CO$^{+}$ towards each core. The spectra are extracted over the area corresponding to twice the FWHM of the core size, similar to the procedure used in \cite{2023A&A...678A.194C}. The velocity dispersion for each spectral line is then calculated using Eq. \ref{eqn1}, and the mean value ($\sigma_{\text{tot,avg}}$) and its associated errors ($\Delta\sigma_{\text{tot,avg}}$) towards each core are obtained by averaging the results of the two lines. Spectral line fitting is performed using the \texttt{GaussPy+} module \citep{2019A&A...628A..78R}. The average spectra for HN$^{13}$C and H$^{13}$CO$^{+}$  towards core 9 is shown in  Appendix \ref{appendix_3_0_0} and the rest are shown in Koley\_zenodo\_2026\footnote{\url{https://zenodo.org/records/22541324}}. After obtaining $\sigma_{\text{tot,avg}}$ and its associated error $\Delta\sigma_{\text{tot,avg}}$, we subtracted the thermal part of the species using the core temperature. As the non-thermal part is independent of species \citep{2019MNRAS.483..593K,2022MNRAS.516..185K}, we then added the thermal part of the mean molecular weight (2.37) of the gas and obtain the total velocity dispersion of the cores ($\sigma_{\text{tot,core}}$) and its associated errors ($\Delta\sigma_{\text{tot,core}}$), which we list in Table \ref{tab:table2}.
}

\subsection{Relative importance of turbulence and magnetic fields at core scale}\label{subsection_4.3}
After determining the masses, sizes, temperatures, and total velocity dispersions of the cores, we calculated the virial parameter ($\alpha_{\text{vir}}$) both without and with the inclusion of magnetic field effects. We obtained the magnetic field in \texttt{Region 1} and \texttt{Region 2} at densities of $\sim$ 1.5 $\times 10^6$ cm$^{-3}$ and $\sim$ $2.3 \times 10^6$ cm$^{-3}$, respectively, while the typical core density is $\sim10^{7-8}$ cm$^{-3}$. Since direct measurements of the magnetic field at core densities are not available, we considered two possible scenarios when including the effects of magnetic fields. In the first scenario, the magnetic field strength ($B_{\text{tot}}$) varies with density ($\rho$) following a power law index ($p$) of 0.0, extending from the relatively low-density region where the magnetic field is estimated using the ADF method to the relatively high-density core regime. In the second scenario, the $p$ is taken to be 0.5. \cite{2010ApJ...725..466C} obtained a value of 0.65, while \cite{2025MNRAS.540.2762W} reported 0.50 from an ensemble study. Since the power law index varies with how the gas accumulates from low to high density, we adopted $p$ = 0.50, which corresponds to a magnetically dominated system \citep{1976ApJ...206..753M,2015MNRAS.451.4384T}.

As the cores are not resolved due to their large distance, this contrasts with low-mass cores in the Taurus molecular cloud \citep{2022MNRAS.516L..48K,2022MNRAS.516..185K,2023PASA...40...53K}, where the cores are well resolved and their internal density profiles can be determined. Here we therefore assumed that the density of the core is approximately uniform. We used Eqs. \ref{eqn33} and  \ref{eqn34} to calculate the virial parameters and their associated errors of the cores when we did not take into account the effect of the magnetic field. Equations \ref{eqn35} and  \ref{eqn36} were used when the magnetic fields of the cores are the same as those of the surrounding environment or the value of $p$ = 0.0 and  Eqs. \ref{eqn37}--\ref{eqn39} were used when we assume $p$ =  0.5 from the surrounding environment to the core scales.

When the effects of magnetic fields are not considered, we find that about four cores have a virial parameter, $\alpha_{\text{vir,turb}} < 1$, even after accounting for the uncertainties, twelve cores have $\alpha_{\text{vir,turb}} > 1$ when the errors are considered, and the remaining cores have values that can be below or above unity. The minimum and maximum values of $\alpha_{\text{vir,turb}}$ are 0.29 and 7.29, respectively, with a mean value of 2.27. After including the effects of magnetic fields with $p = 0.0$, we find that about four cores have a virial parameter, $\alpha_{\text{vir, turb+mag}} < 1$, even after accounting for uncertainties, thirteen cores have $\alpha_{\text{vir, turb+mag}} > 1$ when the errors are considered, and the remaining cores have values that are consistent with unity within the uncertainties. In this case, the distribution of $\alpha_{\text{vir, turb+mag}}$ shows values between 0.30 and 14.18, giving an average of 2.86. In the case of $p = 0.5$, 21 cores have $\alpha_{\text{vir, turb+mag}} > 1$ including uncertainty; the rest of the cores could be gravitationally bound or unbound given the large uncertainties. In this case, $\alpha_{\text{vir, turb+mag}}$ spans a broad range from 1.24 to 49.60, resulting in an average value of 10.09. All of these results are shown in Fig. \ref{fig:fig12} and  all these values are listed in Table \ref{tab:table3}. In the bottom-right panel of Fig. \ref{fig:fig12}, we present the virial parameter ($\alpha_{\text{vir,turb}}$) of the cores as a function of the mass of the core. We find that the high-mass cores ($> 6M_{\odot}$) are gravitationally bound, expecting one low mass core (including uncertainties). This trend is consistent with findings from previous studies \citep{2023ApJ...949..109L,2024ApJ...967..157L,2025A&A...696A..11V}. 

In Fig. \ref{fig:fig13} we show the ratio of kinetic energy (two times) to magnetic energy for two power-law indices, $p = 0.0$ and $p = 0.5$. When $p = 0.0$, the turbulent energy is more dynamically important than the magnetic energy, with a ratio ranging from $\sim$ 0.79 to $\sim$ 113.0 and a median value of $\sim$ 26.06. However, when $p = 0.5$, the magnetic energy becomes more dynamically important than the turbulent energy, with the ratio ranging from $\sim$ 0.11 to $\sim$ 1.79 and a mean value of $\sim$ 0.39. Since we do not have an accurate measurement of the magnetic field strength at core scales, we cannot directly determine the relative importance of the magnetic field compared to turbulence. Instead, we can only conclude that it depends on the assumed power-law scaling of the magnetic field from the low- to high-density regime. These values are presented in Table \ref{tab:table3}.

\section{Discussion}\label{section_5}

Our study of the G327.29 protocluster provides valuable insights into the interplay between magnetic fields, turbulence, and gravity in shaping the dynamics of star-forming regions. The total magnetic field strengths in \texttt{Region 1} and \texttt{Region 2} are measured as $1.7 \pm 0.8$ mG and $3.6 \pm 1.5$ mG, respectively, with corresponding number densities ($n$) of $1.5 \pm 0.3 \times 10^6$ cm$^{-3}$ and $2.3 \pm 0.7 \times 10^6$ cm$^{-3}$. This is obtained from the ADF method. At the same densities the derived magnetic fields are $0.6 \pm 0.3$ mG and $1.1 \pm 0.5$ mG obtained from the ST method. These range of these values are consistent with previous magnetic field surveys of high-mass star-forming regions \citep{2021ApJ...923..204C,2021ApJ...915L..10S,2024ApJ...974..257Z,2024ApJ...972L...6S} and are in agreement with the findings of the ensemble study of \cite{2010ApJ...725..466C}, \cite{2025MNRAS.540.2762W}, and \cite{2026AJ....171...50H}.

The virial parameters ($\alpha_{\text{vir}}$) for these regions, $2.8 \pm 1.9$ and $0.5 \pm 0.4$, do not allow us to conclude on whether \texttt{Region 1} and \texttt{Region 2} are gravitationally bound or unbound. In addition, the ratios of kinetic energy ($E_{\text{k}}$) to magnetic energy ($E_{\text{B}}$) in \texttt{Region 1} and \texttt{Region 2} are approximately 0.34 and 0.17, respectively, suggesting that the magnetic field dominates turbulent motions in these regions. This finding aligns with the results of \cite{2015ApJ...799...74P}, who reported that magnetic fields play a crucial role in regulating star cluster formation, emphasizing the need to account for magnetic fields in theoretical models. In a separate study, \cite{2013ApJ...779..182Q} demonstrated that the magnetic field in the high-mass protocluster G35.2–0.74 N is well ordered and concluded that it plays a dynamically significant role in the formation of the dense clump. Likewise, \cite{2019ApJ...883...95S} investigated the infrared-dark cloud G34.43+0.24 and demonstrated that the magnetic field plays a crucial role in guiding the gravitational contraction of the cloud. Furthermore, \citet{2024ApJ...966..120L} investigated the infrared-dark cloud G28.24 and found that the magnetic field is dynamically important in the relatively diffuse gas. However, within the cloud, the field morphology appears distorted by gravity and influenced by star formation activity in the high-density regions. In conjunction with magnetic field strength diagnostics, the morphological evidence of an hourglass-shaped structure also supports the idea that the magnetic field plays a  crucial role in the evolution of a molecular cloud \citep{2021ApJ...923..204C,2024ApJ...972L...6S}.

At the core scale, since direct measurements of the magnetic field are unavailable, the relative importance of magnetic versus turbulent energy depends on the assumed power law scaling from low- to high-density regions. Furthermore, because of the significant uncertainties in the virial parameters, without a magnetic field  almost eleven, with a power law index of 0.0 almost ten, and with a power law index of 0.0 almost six cores could be gravitationally bound or unbound. Several previous studies have also reported large uncertainties in the magnetic field and consequently in the virial parameters at the core scale, which are consistent with our results \citep{2021ApJ...912..159P,2024ApJ...967..157L}.

In addition to the virial analysis at the clump and core scales, we find that the magnetic field exhibits a complex morphology in the region, with polarization angles uniformly distributed throughout the range from $-$90$^\circ$ to $+$90$^\circ$. This suggests that although the magnetic field is overall dynamically more important than turbulence, the ordered magnetic field exhibits a complex pattern governed by the intricate gas dynamics of the system. Such a complex magnetic field morphology has also been revealed in numerous earlier investigations \citep{2014ApJ...797...99K,2019ApJ...883...95S}. Moreover, the polarization intensity shows an anti-correlation with the Stokes I emission, following a power law index of $-$0.50. The depolarization effect has been observed in molecular clouds in earlier studies \citep{2003ApJ...592..233W, 2009ApJ...695.1399T, 2013ApJ...763..135T,2018ApJ...861...65S,2021ApJ...913...57K, 2021ApJ...913...29F, 2024A&A...682A..81B}. Although the power law index ($\alpha$) varies from study to study, the anti-correlation between the polarization fraction ($f_{\text{p}}$) and the total intensity ($I$) is common in all these studies. There are several plausible reasons for the depolarization effect. For example, due to the finite beam size effect of the telescope, the magnetic field can twist along the LOS and also on the plane of the sky, which reduces the $f_{\text{p}}$. This effect is severe in single-dish studies. Similarly, the decrease in the asymmetry of the dust grains during the growth process of the grains in the dense region may also cause the inefficiency of the grain alignment due to the radiative torque. Furthermore, photon attenuation towards high-density regions can also lead to a decrease in the radiative torque of the dust grains. These effects are discussed in detail in \cite{2007MNRAS.378..910L}, \cite{2014ApJS..213...13H}, and \cite{2018ApJ...861...65S}. 
\section{Conclusions}\label{section_6}
We have investigated the magnetic fields and turbulence in the subregion of the G327.29 protocluster. The principal findings of this study are summarized as follows:\begin{itemize}

\item The magnetic field in the region exhibits a complex morphology, with polarization angles uniformly distributed between $-90^\circ$ and $+90^\circ$. This suggests that the ordered magnetic field exhibits a complex structure across the observed region.

\item The magnetic field strengths ($B_{\text{tot}}$) in the two subregions, \texttt{Region 1} and \texttt{Region 2}, were measured to be $\sim$ 1.7 $\pm$ 0.8 mG and $\sim$ 3.6 $\pm$ 1.5 mG,  respectively, using the ADF method and $\sim$ 0.6 $\pm$ 0.3 mG and $\sim$ 1.1 $\pm$ 0.5 mG  using the ST method.

\item The virial parameters ($\alpha_{\text{vir}}$) for these regions, $2.8 \pm 1.9$ and $0.5 \pm 0.4$, do not allow us to conclude on whether \texttt{Region 1} and \texttt{Region 2} are gravitationally bound or unbound.

\item The ratios of kinetic (2$E_{\text{k}}$) to magnetic energy ($E_{\text{B}}$) in \texttt{Region 1} and \texttt{Region 2} are $\sim$ 0.34 and 0.17, respectively, indicating that the magnetic field dominates the turbulence.

\item Dense cores exhibit a mix of gravitationally bound and unbound states. The number of cores in each state depends on whether magnetic fields are included in the virial analysis and on how the magnetic field strength scales from low- to high-density regions.\end{itemize}

\noindent Overall, this study investigated the relative roles of magnetic fields, turbulence, and gravity on both the clump and core scales within a subregion of the G327.29 protocluster.

 \begin{acknowledgements}
We thank the anonymous Reviewer for the valuable and constructive comments, which have helped us significantly improve the manuscript. A.K. would like to acknowledge support from  Fondecyt postdoctoral project (project id: 3250070, 2025). {\color{black}A.S. gratefully acknowledges support from Fondecyt Regular (project code 1220610), ANID BASAL project FB210003, and the China-Chile Joint Research Fund (CCJRF No. 2312). P.S. was partially supported by a Grant-in-Aid for Scientific Research (KAKENHI Number JP23H01221) of JSPS. P.S. was partially supported by a Grant-in-Aid for Scientific Research (KAKENHI No JP24K17100) of the Japan Society for the Promotion of Science (JSPS).}
\end{acknowledgements}

\bibliographystyle{aa} 
\bibliography{a.bib}

@ARTICLE{2015JAI.....450005H,
       author = {{Hull}, Charles L.~H. and {Plambeck}, Richard L.},
        title = "{The 1.3mm Full-Stokes Polarization System at CARMA}",
      journal = {Journal of Astronomical Instrumentation},
         year = 2015,
        month = jun,
       volume = {4},
          eid = {1550005},
        pages = {1550005},
          doi = {10.1142/S2251171715500051},
archivePrefix = {arXiv},
       eprint = {1506.04771},
 primaryClass = {astro-ph.IM},
       adsurl = {https://ui.adsabs.harvard.edu/abs/2015JAI.....450005H}
}

@misc{Koley_zenodo,
  author = {Koley, Atanu},
  year = {2026},
  title = {Your Zenodo Dataset Title},
  publisher = {Zenodo},
  note = {doi:10.5281/zenodo.22541324}
}

@ARTICLE{2006PASP..118.1340V,
       author = {{Vaillancourt}, John E.},
        title = "{Placing Confidence Limits on Polarization Measurements}",
      journal = {\pasp},
         year = 2006,
        month = sep,
       volume = {118},
       number = {847},
        pages = {1340-1343},
          doi = {10.1086/507472},
archivePrefix = {arXiv},
       eprint = {astro-ph/0603110},
 primaryClass = {astro-ph},
       adsurl = {https://ui.adsabs.harvard.edu/abs/2006PASP..118.1340V}
}

@ARTICLE{2024A&A...687A.163B,
       author = {{Bonfand}, M. and {Csengeri}, T. and {Bontemps}, S. and {Brouillet}, N. and {Motte}, F. and {Louvet}, F. and {Ginsburg}, A. and {Cunningham}, N. and {Galv{\'a}n-Madrid}, R. and {Herpin}, F. and {Wyrowski}, F. and {Valeille-Manet}, M. and {Stutz}, A.~M. and {Di Francesco}, J. and {Gusdorf}, A. and {Fern{\'a}ndez-L{\'o}pez}, M. and {Lefloch}, B. and {Liu}, H.-L. and {Sanhueza}, P. and {{\'A}lvarez-Guti{\'e}rrez}, R.~H. and {Olguin}, F. and {Nony}, T. and {Lopez-Sepulcre}, A. and {Dell'Ova}, P. and {Pouteau}, Y. and {Jeff}, D. and {Chen}, H.-R.~V. and {Armante}, M. and {Towner}, A. and {Bronfman}, L. and {Kessler}, N.},
        title = "{ALMA-IMF. XI. The sample of hot core candidates: A rich population of young high-mass protostars unveiled by the emission of methyl formate}",
      journal = {\aap},
         year = 2024,
        month = jul,
       volume = {687},
          eid = {A163},
        pages = {A163},
          doi = {10.1051/0004-6361/202347856},
archivePrefix = {arXiv},
       eprint = {2402.15023},
 primaryClass = {astro-ph.GA},
       adsurl = {https://ui.adsabs.harvard.edu/abs/2024A&A...687A.163B}
}

@ARTICLE{2023A&A...678A.194C,
       author = {{Cunningham}, N. and {Ginsburg}, A. and {Galv{\'a}n-Madrid}, R. and {Motte}, F. and {Csengeri}, T. and {Stutz}, A.~M. and {Fern{\'a}ndez-L{\'o}pez}, M. and {{\'A}lvarez-Guti{\'e}rrez}, R.~H. and {Armante}, M. and {Baug}, T. and {Bonfand}, M. and {Bontemps}, S. and {Braine}, J. and {Brouillet}, N. and {Busquet}, G. and {D{\'\i}az-Gonz{\'a}lez}, D.~J. and {Di Francesco}, J. and {Gusdorf}, A. and {Herpin}, F. and {Liu}, H. and {L{\'o}pez-Sepulcre}, A. and {Louvet}, F. and {Lu}, X. and {Maud}, L. and {Nony}, T. and {Olguin}, F.~A. and {Pouteau}, Y. and {Rivera-Soto}, R. and {Sandoval-Garrido}, N.~A. and {Sanhueza}, P. and {Tatematsu}, K. and {Towner}, A.~P.~M. and {Valeille-Manet}, M.},
        title = "{ALMA-IMF. VII. First release of the full spectral line cubes: Core kinematics traced by DCN J = (3‒2)}",
      journal = {\aap},
         year = 2023,
        month = oct,
       volume = {678},
          eid = {A194},
        pages = {A194},
          doi = {10.1051/0004-6361/202245429},
archivePrefix = {arXiv},
       eprint = {2306.14710},
 primaryClass = {astro-ph.GA},
       adsurl = {https://ui.adsabs.harvard.edu/abs/2023A&A...678A.194C}
}

@ARTICLE{2025A&A...700A.256P,
       author = {{Polychronakis}, A. and {Tritsis}, A. and {Skalidis}, R. and {Tassis}, K.},
        title = "{A three-step approach to reliably estimate magnetic field strengths in star-forming regions}",
      journal = {\aap},
         year = 2025,
        month = aug,
       volume = {700},
          eid = {A256},
        pages = {A256},
          doi = {10.1051/0004-6361/202553774},
archivePrefix = {arXiv},
       eprint = {2507.06297},
 primaryClass = {astro-ph.GA},
       adsurl = {https://ui.adsabs.harvard.edu/abs/2025A&A...700A.256P}
}

@ARTICLE{2009ApJ...696..567H,
       author = {{Hildebrand}, Roger H. and {Kirby}, Larry and {Dotson}, Jessie L. and {Houde}, Martin and {Vaillancourt}, John E.},
        title = "{Dispersion of Magnetic Fields in Molecular Clouds. I}",
      journal = {\apj},
         year = 2009,
        month = may,
       volume = {696},
       number = {1},
        pages = {567-573},
          doi = {10.1088/0004-637X/696/1/567},
archivePrefix = {arXiv},
       eprint = {0811.0813},
 primaryClass = {astro-ph},
       adsurl = {https://ui.adsabs.harvard.edu/abs/2009ApJ...696..567H}
}

@ARTICLE{2009ApJ...706.1504H,
       author = {{Houde}, Martin and {Vaillancourt}, John E. and {Hildebrand}, Roger H. and {Chitsazzadeh}, Shadi and {Kirby}, Larry},
        title = "{Dispersion of Magnetic Fields in Molecular Clouds. II.}",
      journal = {\apj},
         year = 2009,
        month = dec,
       volume = {706},
       number = {2},
        pages = {1504-1516},
          doi = {10.1088/0004-637X/706/2/1504},
archivePrefix = {arXiv},
       eprint = {0909.5227},
 primaryClass = {astro-ph.GA},
       adsurl = {https://ui.adsabs.harvard.edu/abs/2009ApJ...706.1504H}
}

@ARTICLE{2021A&A...647A.186S,
       author = {{Skalidis}, Raphael and {Tassis}, Konstantinos},
        title = "{High-accuracy estimation of magnetic field strength in the interstellar medium from dust polarization}",
      journal = {\aap},
         year = 2021,
        month = mar,
       volume = {647},
          eid = {A186},
        pages = {A186},
          doi = {10.1051/0004-6361/202039779},
archivePrefix = {arXiv},
       eprint = {2010.15141},
 primaryClass = {astro-ph.GA},
       adsurl = {https://ui.adsabs.harvard.edu/abs/2021A&A...647A.186S}
}

@ARTICLE{2021A&A...656A.118S,
       author = {{Skalidis}, R. and {Sternberg}, J. and {Beattie}, J.~R. and {Pavlidou}, V. and {Tassis}, K.},
        title = "{Why take the square root? An assessment of interstellar magnetic field strength estimation methods}",
      journal = {\aap},
         year = 2021,
        month = dec,
       volume = {656},
          eid = {A118},
        pages = {A118},
          doi = {10.1051/0004-6361/202142045},
archivePrefix = {arXiv},
       eprint = {2109.10925},
 primaryClass = {astro-ph.GA},
       adsurl = {https://ui.adsabs.harvard.edu/abs/2021A&A...656A.118S}
}

@ARTICLE{2016ApJ...820...38H,
       author = {{Houde}, Martin and {Hull}, Charles L.~H. and {Plambeck}, Richard L. and {Vaillancourt}, John E. and {Hildebrand}, Roger H.},
        title = "{Dispersion of Magnetic Fields in Molecular Clouds. IV. Analysis of Interferometry Data}",
      journal = {\apj},
         year = 2016,
        month = mar,
       volume = {820},
       number = {1},
          eid = {38},
        pages = {38},
          doi = {10.3847/0004-637X/820/1/38},
archivePrefix = {arXiv},
       eprint = {1602.01873},
 primaryClass = {astro-ph.GA},
       adsurl = {https://ui.adsabs.harvard.edu/abs/2016ApJ...820...38H}
}

@ARTICLE{2022A&A...662A...8M,
       author = {{Motte}, F. and {Bontemps}, S. and {Csengeri}, T. and {Pouteau}, Y. and {Louvet}, F. and {Stutz}, A.~M. and {Cunningham}, N. and {L{\'o}pez-Sepulcre}, A. and {Brouillet}, N. and {Galv{\'a}n-Madrid}, R. and {Ginsburg}, A. and {Maud}, L. and {Men'shchikov}, A. and {Nakamura}, F. and {Nony}, T. and {Sanhueza}, P. and {{\'A}lvarez-Guti{\'e}rrez}, R.~H. and {Armante}, M. and {Baug}, T. and {Bonfand}, M. and {Busquet}, G. and {Chapillon}, E. and {D{\'\i}az-Gonz{\'a}lez}, D. and {Fern{\'a}ndez-L{\'o}pez}, M. and {Guzm{\'a}n}, A.~E. and {Herpin}, F. and {Liu}, H. -L. and {Olguin}, F. and {Towner}, A.~P.~M. and {Bally}, J. and {Battersby}, C. and {Braine}, J. and {Bronfman}, L. and {Chen}, H. -R.~V. and {Dell'Ova}, P. and {Di Francesco}, J. and {Gonz{\'a}lez}, M. and {Gusdorf}, A. and {Hennebelle}, P. and {Izumi}, N. and {Joncour}, I. and {Lee}, Y. -N. and {Lefloch}, B. and {Lesaffre}, P. and {Lu}, X. and {Menten}, K.~M. and {Mignon-Risse}, R. and {Molet}, J. and {Moraux}, E. and {Mundy}, L. and {Nguyen Luong}, Q. and {Reyes}, N. and {Reyes Reyes}, S.~D. and {Robitaille}, J. -F. and {Rosolowsky}, E. and {Sandoval-Garrido}, N.~A. and {Schuller}, F. and {Svoboda}, B. and {Tatematsu}, K. and {Thomasson}, B. and {Walker}, D. and {Wu}, B. and {Whitworth}, A.~P. and {Wyrowski}, F.},
        title = "{ALMA-IMF. I. Investigating the origin of stellar masses: Introduction to the Large Program and first results}",
      journal = {\aap},
         year = 2022,
        month = jun,
       volume = {662},
          eid = {A8},
        pages = {A8},
          doi = {10.1051/0004-6361/202141677},
archivePrefix = {arXiv},
       eprint = {2112.08182},
 primaryClass = {astro-ph.GA},
       adsurl = {https://ui.adsabs.harvard.edu/abs/2022A&A...662A...8M}
}

@ARTICLE{2024ApJS..274...15G,
       author = {{Galv{\'a}n-Madrid}, Roberto and {D{\'\i}az-Gonz{\'a}lez}, Daniel J. and {Motte}, Fr{\'e}d{\'e}rique and {Ginsburg}, Adam and {Cunningham}, Nichol and {Menten}, Karl M. and {Armante}, M{\'e}lanie and {Bonfand}, M{\'e}lisse and {Braine}, Jonathan and {Csengeri}, Timea and {Dell'Ova}, Pierre and {Louvet}, Fabien and {Nony}, Thomas and {Rivera-Soto}, Rudy and {Sanhueza}, Patricio and {Stutz}, Amelia M. and {Wyrowski}, Friedrich and {{\'A}lvarez-Guti{\'e}rrez}, Rodrigo H. and {Baug}, Tapas and {Bontemps}, Sylvain and {Bronfman}, Leonardo and {Fern{\'a}ndez-L{\'o}pez}, Manuel and {Gusdorf}, Antoine and {Koley}, Atanu and {Liu}, Hong-Li and {Salinas}, Javiera and {Towner}, Allison P.~M. and {Whitworth}, Anthony P.},
        title = "{ALMA-IMF. XIV. Free{\textendash}Free Templates Derived from H41{\ensuremath{\alpha}} and Ionized Gas Content in 15 Massive Protoclusters}",
      journal = {\apjs},
         year = 2024,
        month = sep,
       volume = {274},
       number = {1},
          eid = {15},
        pages = {15},
          doi = {10.3847/1538-4365/ad61e6},
archivePrefix = {arXiv},
       eprint = {2407.07359},
 primaryClass = {astro-ph.GA},
       adsurl = {https://ui.adsabs.harvard.edu/abs/2024ApJS..274...15G}
}

@ARTICLE{1951PhRv...81..890D,
  author = {Davis, Leverett, Jr.},
  title = {The Strength of Interstellar Magnetic Fields},
  journal = {Physical Review},
  year = {1951},
  volume = {81},
  pages = {890},
  doi = {10.1103/PhysRev.81.890.2}
}

@ARTICLE{1953ApJ...118..113C,
  author = {{Chandrasekhar}, S. and {Fermi}, E.},
  title = {Magnetic Fields in Spiral Arms},
  journal = {The Astrophysical Journal},
  year = 1953,
  volume = 118,
  pages = {113--115},
  doi = {10.1086/145732}
}

@ARTICLE{1965QJRAS...6..265M,
  author = {{Mestel}, L.},
  title = {...},
  journal = {Quarterly Journal of the Royal Astronomical Society},
  year = 1965,
  volume = 6,
  pages = {265},
  adsurl = {https://ui.adsabs.harvard.edu/abs/1965QJRAS...6..265M}
}

@ARTICLE{1976ApJ...206..753M,
  author = {{Mouschovias}, T. Ch.},
  title = {Nonhomologous Contraction and Equilibria of Self-Gravitating, Magnetic Interstellar Clouds Embedded in an Intercloud Medium: Star Formation. I. Formulation of the Problem and Method of Solution},
  journal = {The Astrophysical Journal},
  year = 1976,
  volume = {206},
  pages = {753--767},
  doi = {10.1086/154436},
  adsurl = {https://ui.adsabs.harvard.edu/abs/1976ApJ...206..753M}
}

@ARTICLE{2019A&A...628A..78R,
       author = {{Riener}, M. and {Kainulainen}, J. and {Henshaw}, J.~D. and {Orkisz}, J.~H. and {Murray}, C.~E. and {Beuther}, H.},
        title = "{GAUSSPY+: A fully automated Gaussian decomposition package for emission line spectra}",
      journal = {\aap},
         year = 2019,
        month = aug,
       volume = {628},
          eid = {A78},
        pages = {A78},
          doi = {10.1051/0004-6361/201935519},
archivePrefix = {arXiv},
       eprint = {1906.10506},
 primaryClass = {astro-ph.IM},
       adsurl = {https://ui.adsabs.harvard.edu/abs/2019A&A...628A..78R}
}

@ARTICLE{2009A&A...504..415S,
       author = {{Schuller}, F. and {Menten}, K.~M. and {Contreras}, Y. and {Wyrowski}, F. and {Schilke}, P. and {Bronfman}, L. and {Henning}, T. and {Walmsley}, C.~M. and {Beuther}, H. and {Bontemps}, S. and {Cesaroni}, R. and {Deharveng}, L. and {Garay}, G. and {Herpin}, F. and {Lefloch}, B. and {Linz}, H. and {Mardones}, D. and {Minier}, V. and {Molinari}, S. and {Motte}, F. and {Nyman}, L. -{\r{A}}. and {Reveret}, V. and {Risacher}, C. and {Russeil}, D. and {Schneider}, N. and {Testi}, L. and {Troost}, T. and {Vasyunina}, T. and {Wienen}, M. and {Zavagno}, A. and {Kovacs}, A. and {Kreysa}, E. and {Siringo}, G. and {Wei{\ss}}, A.},
        title = "{ATLASGAL - The APEX telescope large area survey of the galaxy at 870 {\ensuremath{\mu}}m}",
      journal = {\aap},
         year = 2009,
        month = sep,
       volume = {504},
       number = {2},
        pages = {415-427},
          doi = {10.1051/0004-6361/200811568},
archivePrefix = {arXiv},
       eprint = {0903.1369},
 primaryClass = {astro-ph.GA},
       adsurl = {https://ui.adsabs.harvard.edu/abs/2009A&A...504..415S}
}

@ARTICLE{2003PASP..115..953B,
       author = {{Benjamin}, Robert A. and {Churchwell}, E. and {Babler}, Brian L. and {Bania}, T.~M. and {Clemens}, Dan P. and {Cohen}, Martin and {Dickey}, John M. and {Indebetouw}, R{\'e}my and {Jackson}, James M. and {Kobulnicky}, Henry A. and {Lazarian}, Alex and {Marston}, A.~P. and {Mathis}, John S. and {Meade}, Marilyn R. and {Seager}, Sara and {Stolovy}, S.~R. and {Watson}, C. and {Whitney}, Barbara A. and {Wolff}, Michael J. and {Wolfire}, Mark G.},
        title = "{GLIMPSE. I. An SIRTF Legacy Project to Map the Inner Galaxy}",
      journal = {\pasp},
         year = 2003,
        month = aug,
       volume = {115},
       number = {810},
        pages = {953-964},
          doi = {10.1086/376696},
archivePrefix = {arXiv},
       eprint = {astro-ph/0306274},
 primaryClass = {astro-ph},
       adsurl = {https://ui.adsabs.harvard.edu/abs/2003PASP..115..953B}
}

@software{tgif,
  author       = {Yoo, Taehwa and
                  Ginsburg, Adam},
  title        = {TGIF: Two d Gaussian In Fitting},
  month        = oct,
  year         = 2024,
  publisher    = {Zenodo},
  doi          = {10.5281/zenodo.13973837},
  url          = {https://doi.org/10.5281/zenodo.13973837}
}

@ARTICLE{2008ApJ...679.1338R,
       author = {{Rosolowsky}, E.~W. and {Pineda}, J.~E. and {Kauffmann}, J. and {Goodman}, A.~A.},
        title = "{Structural Analysis of Molecular Clouds: Dendrograms}",
      journal = {\apj},
         year = 2008,
        month = jun,
       volume = {679},
       number = {2},
        pages = {1338-1351},
          doi = {10.1086/587685},
archivePrefix = {arXiv},
       eprint = {0802.2944},
 primaryClass = {astro-ph},
       adsurl = {https://ui.adsabs.harvard.edu/abs/2008ApJ...679.1338R}
}

@ARTICLE{2022A&A...664A..26P,
       author = {{Pouteau}, Y. and {Motte}, F. and {Nony}, T. and {Galv{\'a}n-Madrid}, R. and {Men'shchikov}, A. and {Bontemps}, S. and {Robitaille}, J. -F. and {Louvet}, F. and {Ginsburg}, A. and {Herpin}, F. and {L{\'o}pez-Sepulcre}, A. and {Dell'Ova}, P. and {Gusdorf}, A. and {Sanhueza}, P. and {Stutz}, A.~M. and {Brouillet}, N. and {Thomasson}, B. and {Armante}, M. and {Baug}, T. and {Bonfand}, M. and {Busquet}, G. and {Csengeri}, T. and {Cunningham}, N. and {Fern{\'a}ndez-L{\'o}pez}, M. and {Liu}, H. -L. and {Olguin}, F. and {Towner}, A.~P.~M. and {Bally}, J. and {Braine}, J. and {Bronfman}, L. and {Joncour}, I. and {Gonz{\'a}lez}, M. and {Hennebelle}, P. and {Lu}, X. and {Menten}, K.~M. and {Moraux}, E. and {Tatematsu}, K. and {Walker}, D. and {Whitworth}, A.~P.},
        title = "{ALMA-IMF. III. Investigating the origin of stellar masses: top-heavy core mass function in the W43-MM2\&MM3 mini-starburst}",
      journal = {\aap},
         year = 2022,
        month = aug,
       volume = {664},
          eid = {A26},
        pages = {A26},
          doi = {10.1051/0004-6361/202142951},
archivePrefix = {arXiv},
       eprint = {2203.03276},
 primaryClass = {astro-ph.GA},
       adsurl = {https://ui.adsabs.harvard.edu/abs/2022A&A...664A..26P}
}

@ARTICLE{2025A&A...696A..11V,
       author = {{Valeille-Manet}, M. and {Bontemps}, S. and {Csengeri}, T. and {Nony}, T. and {Motte}, F. and {Stutz}, A.~M. and {Gusdorf}, A. and {Ginsburg}, A. and {Galv{\'a}n-Madrid}, R. and {Sanhueza}, P. and {Bonfand}, M. and {Brouillet}, N. and {Dell'Ova}, P. and {Louvet}, F. and {Cunningham}, N. and {Fern{\'a}ndez-L{\'o}pez}, M. and {Herpin}, F. and {Wyrowski}, F. and {{\'A}lvarez-Guti{\'e}rrez}, R.~H. and {Armante}, M. and {Guzm{\'a}n}, A.~E. and {Kessler}, N. and {Koley}, A. and {Salinas}, J. and {Yoo}, T. and {Bronfman}, L. and {Le Nestour}, N.},
        title = "{ALMA-IMF: XVII. Census and lifetime of high-mass prestellar cores in 14 massive protoclusters}",
      journal = {\aap},
         year = 2025,
        month = apr,
       volume = {696},
          eid = {A11},
        pages = {A11},
          doi = {10.1051/0004-6361/202451291},
archivePrefix = {arXiv},
       eprint = {2502.09426},
 primaryClass = {astro-ph.GA},
       adsurl = {https://ui.adsabs.harvard.edu/abs/2025A&A...696A..11V}
}

@ARTICLE{2023PASA...40...46K,
       author = {{Koley}, Atanu},
        title = "{Turbulence measurements in the neutral ISM from HI-21 cm emission-absorption spectra}",
      journal = {\pasa},
         year = 2023,
        month = sep,
       volume = {40},
          eid = {e046},
        pages = {e046},
          doi = {10.1017/pasa.2023.43},
archivePrefix = {arXiv},
       eprint = {2308.01808},
 primaryClass = {astro-ph.GA},
       adsurl = {https://ui.adsabs.harvard.edu/abs/2023PASA...40...46K}
}

@ARTICLE{2025ApJ...980...87S,
       author = {{Sanhueza}, Patricio and {Liu}, Junhao and {Morii}, Kaho and {Girart}, Josep Miquel and {Zhang}, Qizhou and {Stephens}, Ian W. and {Jackson}, James M. and {Cort{\'e}s}, Paulo C. and {Koch}, Patrick M. and {Cyganowski}, Claudia J. and {Saha}, Piyali and {Beuther}, Henrik and {Zhang}, Suinan and {Beltr{\'a}n}, Maria T. and {Cheng}, Yu and {Olguin}, Fernando A. and {Lu}, Xing and {Choudhury}, Spandan and {Pattle}, Kate and {Fern{\'a}ndez-L{\'o}pez}, Manuel and {Hwang}, Jihye and {Kang}, Ji-hyun and {Karoly}, Janik and {Ginsburg}, Adam and {Lyo}, A. -Ran and {Taniguchi}, Kotomi and {Jiao}, Wenyu and {Eswaraiah}, Chakali and {Luo}, Qiu-yi and {Wang}, Jia-Wei and {Commer{\c{c}}on}, Beno{\^\i}t and {Li}, Shanghuo and {Xu}, Fengwei and {Chen}, Huei-Ru Vivien and {Zapata}, Luis A. and {Chung}, Eun Jung and {Nakamura}, Fumitaka and {Panigrahy}, Sandhyarani and {Sakai}, Takeshi},
        title = "{Magnetic Fields in Massive Star-forming Regions (MagMaR). V. The Magnetic Field at the Onset of High-mass Star Formation}",
      journal = {\apj},
         year = 2025,
        month = feb,
       volume = {980},
       number = {1},
          eid = {87},
        pages = {87},
          doi = {10.3847/1538-4357/ad9d40},
archivePrefix = {arXiv},
       eprint = {2412.08790},
 primaryClass = {astro-ph.GA},
       adsurl = {https://ui.adsabs.harvard.edu/abs/2025ApJ...980...87S}
}

@ARTICLE{2023ApJ...949...30L,
       author = {{Liu}, Junhao and {Zhang}, Qizhou and {Liu}, Hauyu Baobab and {Qiu}, Keping and {Li}, Shanghuo and {Li}, Zhi-Yun and {Ho}, Paul T.~P. and {Girart}, Josep Miquel and {Ching}, Tao-Chung and {Chen}, Huei-Ru Vivien and {Lai}, Shih-Ping and {Rao}, Ramprasad and {Tang}, Ya-wen},
        title = "{Deviation from a Continuous and Universal Turbulence Cascade in NGC 6334 due to Massive Star Formation Activity}",
      journal = {\apj},
         year = 2023,
        month = may,
       volume = {949},
       number = {1},
          eid = {30},
        pages = {30},
          doi = {10.3847/1538-4357/acc4c0},
archivePrefix = {arXiv},
       eprint = {2303.08170},
 primaryClass = {astro-ph.GA},
       adsurl = {https://ui.adsabs.harvard.edu/abs/2023ApJ...949...30L}
}

@ARTICLE{2024A&A...690A..33L,
       author = {{Louvet}, F. and {Sanhueza}, P. and {Stutz}, A. and {Men'shchikov}, A. and {Motte}, F. and {Galv{\'a}n-Madrid}, R. and {Bontemps}, S. and {Pouteau}, Y. and {Ginsburg}, A. and {Csengeri}, T. and {Di Francesco}, J. and {Dell'Ova}, P. and {Gonz{\'a}lez}, M. and {Didelon}, P. and {Braine}, J. and {Cunningham}, N. and {Thomasson}, B. and {Lesaffre}, P. and {Hennebelle}, P. and {Bonfand}, M. and {Gusdorf}, A. and {{\'A}lvarez-Guti{\'e}rrez}, R.~H. and {Nony}, T. and {Busquet}, G. and {Olguin}, F. and {Bronfman}, L. and {Salinas}, J. and {Fernandez-Lopez}, M. and {Moraux}, E. and {Liu}, H.~L. and {Lu}, X. and {Huei-Ru}, V. and {Towner}, A. and {Valeille-Manet}, M. and {Brouillet}, N. and {Herpin}, F. and {Lefloch}, B. and {Baug}, T. and {Maud}, L. and {L{\'o}pez-Sepulcre}, A. and {Svoboda}, B.},
        title = "{ALMA-IMF: XV. Core mass function in the high-mass star formation regime}",
      journal = {\aap},
         year = 2024,
        month = oct,
       volume = {690},
          eid = {A33},
        pages = {A33},
          doi = {10.1051/0004-6361/202345986},
archivePrefix = {arXiv},
       eprint = {2407.18719},
 primaryClass = {astro-ph.GA},
       adsurl = {https://ui.adsabs.harvard.edu/abs/2024A&A...690A..33L}
}

@ARTICLE{2017MNRAS.468.3694C,
       author = {{Cyganowski}, C.~J. and {Brogan}, C.~L. and {Hunter}, T.~R. and {Smith}, R. and {Kruijssen}, J.~M.~D. and {Bonnell}, I.~A. and {Zhang}, Q.},
        title = "{Simultaneous low- and high-mass star formation in a massive protocluster: ALMA observations of G11.92-0.61$^{★}$}",
      journal = {\mnras},
         year = 2017,
        month = jul,
       volume = {468},
       number = {3},
        pages = {3694-3708},
          doi = {10.1093/mnras/stx043},
archivePrefix = {arXiv},
       eprint = {1701.02802},
 primaryClass = {astro-ph.GA},
       adsurl = {https://ui.adsabs.harvard.edu/abs/2017MNRAS.468.3694C}
}

@ARTICLE{2018ARA&A..56...41M,
       author = {{Motte}, Fr{\'e}d{\'e}rique and {Bontemps}, Sylvain and {Louvet}, Fabien},
        title = "{High-Mass Star and Massive Cluster Formation in the Milky Way}",
      journal = {\araa},
         year = 2018,
        month = sep,
       volume = {56},
        pages = {41-82},
          doi = {10.1146/annurev-astro-091916-055235},
archivePrefix = {arXiv},
       eprint = {1706.00118},
 primaryClass = {astro-ph.GA},
       adsurl = {https://ui.adsabs.harvard.edu/abs/2018ARA&A..56...41M}
}

@ARTICLE{2024A&A...687A.217D,
       author = {{Dell'Ova}, P. and {Motte}, F. and {Gusdorf}, A. and {Pouteau}, Y. and {Men'shchikov}, A. and {D{\'\i}az-Gonz{\'a}lez}, D. and {Galv{\'a}n-Madrid}, R. and {Lesaffre}, P. and {Didelon}, P. and {Stutz}, A.~M. and {Towner}, A.~P.~M. and {Marsh}, K. and {Whitworth}, A. and {Armante}, M. and {Bonfand}, M. and {Nony}, T. and {Valeille-Manet}, M. and {Bontemps}, S. and {Csengeri}, T. and {Cunningham}, N. and {Ginsburg}, A. and {Louvet}, F. and {{\'A}lvarez-Guti{\'e}rrez}, R.~H. and {Brouillet}, N. and {Salinas}, J. and {Sanhueza}, P. and {Nakamura}, F. and {Nguyen Luong}, Q. and {Baug}, T. and {Fern{\'a}ndez-L{\'o}pez}, M. and {Liu}, H. -L. and {Olguin}, F.},
        title = "{ALMA-IMF. XII. Point-process mapping of 15 massive protoclusters}",
      journal = {\aap},
         year = 2024,
        month = jul,
       volume = {687},
          eid = {A217},
        pages = {A217},
          doi = {10.1051/0004-6361/202348984},
archivePrefix = {arXiv},
       eprint = {2407.07610},
 primaryClass = {astro-ph.GA},
       adsurl = {https://ui.adsabs.harvard.edu/abs/2024A&A...687A.217D}
}

@ARTICLE{2021ApJ...915L..10S,
       author = {{Sanhueza}, Patricio and {Girart}, Josep Miquel and {Padovani}, Marco and {Galli}, Daniele and {Hull}, Charles L.~H. and {Zhang}, Qizhou and {Cortes}, Paulo and {Stephens}, Ian W. and {Fern{\'a}ndez-L{\'o}pez}, Manuel and {Jackson}, James M. and {Frau}, Pau and {Kock}, Patrick M. and {Wu}, Benjamin and {Zapata}, Luis A. and {Olguin}, Fernando and {Lu}, Xing and {Silva}, Andrea and {Tang}, Ya-Wen and {Sakai}, Takeshi and {Guzm{\'a}n}, Andr{\'e}s E. and {Tatematsu}, Ken'ichi and {Nakamura}, Fumitaka and {Chen}, Huei-Ru Vivien},
        title = "{Gravity-driven Magnetic Field at  1000 au Scales in High-mass Star Formation}",
      journal = {\apjl},
         year = 2021,
        month = jul,
       volume = {915},
       number = {1},
          eid = {L10},
        pages = {L10},
          doi = {10.3847/2041-8213/ac081c},
archivePrefix = {arXiv},
       eprint = {2106.03866},
 primaryClass = {astro-ph.GA},
       adsurl = {https://ui.adsabs.harvard.edu/abs/2021ApJ...915L..10S}
}

@ARTICLE{2024ApJ...972L...6S,
       author = {{Saha}, Piyali and {Sanhueza}, Patricio and {Padovani}, Marco and {Girart}, Josep M. and {Cort{\'e}s}, Paulo C. and {Morii}, Kaho and {Liu}, Junhao and {S{\'a}nchez-Monge}, {\'A}. and {Galli}, Daniele and {Basu}, Shantanu and {Koch}, Patrick M. and {Beltr{\'a}n}, Maria T. and {Li}, Shanghuo and {Beuther}, Henrik and {Stephens}, Ian W. and {Nakamura}, Fumitaka and {Zhang}, Qizhou and {Jiao}, Wenyu and {Fern{\'a}ndez-L{\'o}pez}, M. and {Hwang}, Jihye and {Chung}, Eun Jung and {Pattle}, Kate and {Zapata}, Luis A. and {Xu}, Fengwei and {Olguin}, Fernando A. and {Kang}, Ji-hyun and {Karoly}, Janik and {Law}, Chi-Yan and {Wang}, Jia-Wei and {Csengeri}, Timea and {Lu}, Xing and {Cheng}, Yu and {Kim}, Jongsoo and {Choudhury}, Spandan and {Chen}, Huei-Ru Vivien and {Hull}, Charles L.~H.},
        title = "{Magnetic Fields in Massive Star-forming Regions (MagMaR): Unveiling an Hourglass Magnetic Field in G333.46{\textendash}0.16 Using ALMA}",
      journal = {\apjl},
         year = 2024,
        month = sep,
       volume = {972},
       number = {1},
          eid = {L6},
        pages = {L6},
          doi = {10.3847/2041-8213/ad660c},
archivePrefix = {arXiv},
       eprint = {2407.16654},
 primaryClass = {astro-ph.GA},
       adsurl = {https://ui.adsabs.harvard.edu/abs/2024ApJ...972L...6S}
}

@ARTICLE{2023PASA...40...53K,
       author = {{Koley}, Atanu},
        title = "{Studying the internal structures of the central region of prestellar core L1517B in Taurus molecular cloud using ammonia (NH$_{3}$) (1,1) and (2,2) lines}",
      journal = {\pasa},
         year = 2023,
        month = nov,
       volume = {40},
          eid = {e053},
        pages = {e053},
          doi = {10.1017/pasa.2023.53},
archivePrefix = {arXiv},
       eprint = {2210.00524},
 primaryClass = {astro-ph.GA},
       adsurl = {https://ui.adsabs.harvard.edu/abs/2023PASA...40...53K}
}

@ARTICLE{2024ApJ...974..257Z,
       author = {{Zapata}, Luis A. and {Fern{\'a}ndez-L{\'o}pez}, Manuel and {Sanhueza}, Patricio and {Girart}, Josep M. and {Rodr{\'\i}guez}, Luis F. and {Cort{\'e}s}, Paulo and {Koch}, Patrick and {Beltr{\'a}n}, Maria T. and {Pattle}, Kate and {Beuther}, Henrik and {Saha}, Piyali and {Jiao}, Wenyu and {Xu}, Fengwei and {Lu}, Xing Walker and {Olguin}, Fernando and {Li}, Shanghuo and {Stephens}, Ian W. and {Kang}, Ji-hyun and {Cheng}, Yu and {Choudhury}, Spandan and {Morii}, Kaho and {Chung}, Eun Jung and {Wang}, Jia-Wei and {Hwang}, Jihye and {Lyo}, A. -Ran and {Zhang}, Q. and {Chen}, Huei-Ru Vivien},
        title = "{Magnetic Fields in Massive Star-forming Regions (MagMaR). IV. Tracing the Magnetic Fields in the O-type Protostellar System IRAS 16547{\textendash}4247}",
      journal = {\apj},
         year = 2024,
        month = oct,
       volume = {974},
       number = {2},
          eid = {257},
        pages = {257},
          doi = {10.3847/1538-4357/ad701d},
archivePrefix = {arXiv},
       eprint = {2408.10199},
 primaryClass = {astro-ph.SR},
       adsurl = {https://ui.adsabs.harvard.edu/abs/2024ApJ...974..257Z}
}

@ARTICLE{2021ApJ...923..204C,
       author = {{Cort{\'e}s}, Paulo C. and {Sanhueza}, Patricio and {Houde}, Martin and {Mart{\'\i}n}, Sergio and {Hull}, Charles L.~H. and {Girart}, Josep M. and {Zhang}, Qizhou and {Fernandez-Lopez}, Manuel and {Zapata}, Luis A. and {Stephens}, Ian W. and {Li}, Hua-bai and {Wu}, Benjamin and {Olguin}, Fernando and {Lu}, Xing and {Guzm{\'a}n}, Andres E. and {Nakamura}, Fumitaka},
        title = "{Magnetic Fields in Massive Star-forming Regions (MagMaR). II. Tomography through Dust and Molecular Line Polarization in NGC 6334I(N)}",
      journal = {\apj},
         year = 2021,
        month = dec,
       volume = {923},
       number = {2},
          eid = {204},
        pages = {204},
          doi = {10.3847/1538-4357/ac28a1},
archivePrefix = {arXiv},
       eprint = {2109.09270},
 primaryClass = {astro-ph.GA},
       adsurl = {https://ui.adsabs.harvard.edu/abs/2021ApJ...923..204C}
}

@ARTICLE{2025A&A...702A.133K,
       author = {{Koley}, A. and {Stutz}, A.~M. and {Louvet}, F. and {Motte}, F. and {Ginsburg}, A. and {Galv{\'a}n-Madrid}, R. and {{\'A}lvarez-Guti{\'e}rrez}, R.~H. and {Sanhueza}, P. and {Baug}, T. and {Sandoval-Garrido}, N. and {Salinas}, J. and {Busquet}, G. and {Braine}, J. and {Liu}, H.-L. and {Csengeri}, T. and {Gusdorf}, A. and {Fern{\'a}ndez-L{\'o}pez}, M. and {Cunningham}, N. and {Bronfman}, L. and {Bonfand}, M.},
        title = "{ALMA-IMF: XIX. C$^{18}$O (J = 2─1): Measurements of turbulence in 15 massive protoclusters}",
      journal = {\aap},
         year = 2025,
        month = oct,
       volume = {702},
          eid = {A133},
        pages = {A133},
          doi = {10.1051/0004-6361/202553830},
archivePrefix = {arXiv},
       eprint = {2507.14502},
 primaryClass = {astro-ph.GA},
       adsurl = {https://ui.adsabs.harvard.edu/abs/2025A&A...702A.133K}
}

@ARTICLE{2010ApJ...725..466C,
       author = {{Crutcher}, Richard M. and {Wandelt}, Benjamin and {Heiles}, Carl and {Falgarone}, Edith and {Troland}, Thomas H.},
        title = "{Magnetic Fields in Interstellar Clouds from Zeeman Observations: Inference of Total Field Strengths by Bayesian Analysis}",
      journal = {\apj},
         year = 2010,
        month = dec,
       volume = {725},
       number = {1},
        pages = {466-479},
          doi = {10.1088/0004-637X/725/1/466},
       adsurl = {https://ui.adsabs.harvard.edu/abs/2010ApJ...725..466C}
}

@ARTICLE{2014ApJS..213...13H,
       author = {{Hull}, Charles L.~H. and {Plambeck}, Richard L. and {Kwon}, Woojin and {Bower}, Geoffrey C. and {Carpenter}, John M. and {Crutcher}, Richard M. and {Fiege}, Jason D. and {Franzmann}, Erica and {Hakobian}, Nicholas S. and {Heiles}, Carl and {Houde}, Martin and {Hughes}, A. Meredith and {Lamb}, James W. and {Looney}, Leslie W. and {Marrone}, Daniel P. and {Matthews}, Brenda C. and {Pillai}, Thushara and {Pound}, Marc W. and {Rahman}, Nurur and {Sandell}, G{\"o}ran and {Stephens}, Ian W. and {Tobin}, John J. and {Vaillancourt}, John E. and {Volgenau}, N.~H. and {Wright}, Melvyn C.~H.},
        title = "{TADPOL: A 1.3 mm Survey of Dust Polarization in Star-forming Cores and Regions}",
      journal = {\apjs},
         year = 2014,
        month = jul,
       volume = {213},
       number = {1},
          eid = {13},
        pages = {13},
          doi = {10.1088/0067-0049/213/1/13},
archivePrefix = {arXiv},
       eprint = {1310.6653},
 primaryClass = {astro-ph.SR},
       adsurl = {https://ui.adsabs.harvard.edu/abs/2014ApJS..213...13H}
}

@ARTICLE{2007MNRAS.378..910L,
       author = {{Lazarian}, A. and {Hoang}, Thiem},
        title = "{Radiative torques: analytical model and basic properties}",
      journal = {\mnras},
         year = 2007,
        month = jul,
       volume = {378},
       number = {3},
        pages = {910-946},
          doi = {10.1111/j.1365-2966.2007.11817.x},
archivePrefix = {arXiv},
       eprint = {0707.0886},
 primaryClass = {astro-ph},
       adsurl = {https://ui.adsabs.harvard.edu/abs/2007MNRAS.378..910L}
}

@ARTICLE{2021ApJ...912..159P,
       author = {{Palau}, Aina and {Zhang}, Qizhou and {Girart}, Josep M. and {Liu}, Junhao and {Rao}, Ramprasad and {Koch}, Patrick M. and {Estalella}, Robert and {Chen}, Huei-Ru Vivien and {Liu}, Hauyu Baobab and {Qiu}, Keping and {Li}, Zhi-Yun and {Zapata}, Luis A. and {Bontemps}, Sylvain and {Ho}, Paul T.~P. and {Beuther}, Henrik and {Ching}, Tao-Chung and {Shinnaga}, Hiroko and {Ahmadi}, Aida},
        title = "{Does the Magnetic Field Suppress Fragmentation in Massive Dense Cores?}",
      journal = {\apj},
         year = 2021,
        month = may,
       volume = {912},
       number = {2},
          eid = {159},
        pages = {159},
          doi = {10.3847/1538-4357/abee1e},
archivePrefix = {arXiv},
       eprint = {2010.12099},
 primaryClass = {astro-ph.GA},
       adsurl = {https://ui.adsabs.harvard.edu/abs/2021ApJ...912..159P}
}

@ARTICLE{2024ApJ...967..157L,
       author = {{Law}, Chi-Yan and {Tan}, Jonathan C. and {Skalidis}, Raphael and {Morgan}, Larry and {Xu}, Duo and {de Oliveira Alves}, Felipe and {Barnes}, Ashley T. and {Butterfield}, Natalie and {Caselli}, Paola and {Cosentino}, Giuliana and {Fontani}, Francesco and {Henshaw}, Jonathan D. and {Jimenez-Serra}, Izaskun and {Lim}, Wanggi},
        title = "{Polarized Light from Massive Protoclusters (POLIMAP). I. Dissecting the Role of Magnetic Fields in the Massive Infrared Dark Cloud G28.37+0.07}",
      journal = {\apj},
         year = 2024,
        month = jun,
       volume = {967},
       number = {2},
          eid = {157},
        pages = {157},
          doi = {10.3847/1538-4357/ad39e0},
archivePrefix = {arXiv},
       eprint = {2401.11560},
 primaryClass = {astro-ph.GA},
       adsurl = {https://ui.adsabs.harvard.edu/abs/2024ApJ...967..157L}
}

@ARTICLE{2014ApJ...797...99K,
       author = {{Koch}, Patrick M. and {Tang}, Ya-Wen and {Ho}, Paul T.~P. and {Zhang}, Qizhou and {Girart}, Josep M. and {Chen}, Huei-Ru Vivien and {Frau}, Pau and {Li}, Hua-Bai and {Li}, Zhi-Yun and {Liu}, Hau-Yu Baobab and {Padovani}, Marco and {Qiu}, Keping and {Yen}, Hsi-Wei and {Chen}, How-Huan and {Ching}, Tao-Chung and {Lai}, Shih-Ping and {Rao}, Ramprasad},
        title = "{The Importance of the Magnetic Field from an SMA-CSO-combined Sample of Star-forming Regions}",
      journal = {\apj},
         year = 2014,
        month = dec,
       volume = {797},
       number = {2},
          eid = {99},
        pages = {99},
          doi = {10.1088/0004-637X/797/2/99},
archivePrefix = {arXiv},
       eprint = {1411.3830},
 primaryClass = {astro-ph.GA},
       adsurl = {https://ui.adsabs.harvard.edu/abs/2014ApJ...797...99K}
}

@ARTICLE{2019ApJ...883...95S,
       author = {{Soam}, Archana and {Liu}, Tie and {Andersson}, B.-G. and {Lee}, Chang Won and {Liu}, Junhao and {Juvela}, Mika and {Li}, Pak Shing and {Goldsmith}, Paul F. and {Zhang}, Qizhou and {Koch}, Patrick M. and {Kim}, Kee-Tae and {Qiu}, Keping and {Evans}, II, Neal J. and {Johnstone}, Doug and {Thompson}, Mark and {Ward-Thompson}, Derek and {Di Francesco}, James and {Tang}, Ya-Wen and {Montillaud}, Julien and {Kim}, Gwanjeong and {Mairs}, Steve and {Sanhueza}, Patricio and {Kim}, Shinyoung and {Berry}, David and {Gordon}, Michael S. and {Tatematsu}, Ken'ichi and {Liu}, Sheng-Yuan and {Pattle}, Kate and {Eden}, David and {McGehee}, Peregrine M. and {Wang}, Ke and {Ristorcelli}, I. and {Graves}, Sarah F. and {Alina}, Dana and {Lacaille}, Kevin M. and {Montier}, Ludovic and {Park}, Geumsook and {Kwon}, Woojin and {Chung}, Eun Jung and {Pelkonen}, Veli-Matti and {Micelotta}, Elisabetta R. and {Saajasto}, Mika and {Fuller}, Gary},
        title = "{Magnetic Fields in the Infrared Dark Cloud G34.43+0.24}",
      journal = {\apj},
         year = 2019,
        month = sep,
       volume = {883},
       number = {1},
          eid = {95},
        pages = {95},
          doi = {10.3847/1538-4357/ab39dd},
archivePrefix = {arXiv},
       eprint = {1908.03624},
 primaryClass = {astro-ph.GA},
       adsurl = {https://ui.adsabs.harvard.edu/abs/2019ApJ...883...95S}
}

@ARTICLE{2023ApJ...949..109L,
       author = {{Li}, Shanghuo and {Sanhueza}, Patricio and {Zhang}, Qizhou and {Guido}, Garay and {Sabatini}, Giovanni and {Morii}, Kaho and {Lu}, Xing and {Tafoya}, Daniel and {Nakamura}, Fumitaka and {Izumi}, Natsuko and {Tatematsu}, Ken'ichi and {Li}, Fei},
        title = "{The ALMA Survey of 70 {\ensuremath{\mu}}m Dark High-mass Clumps in Early Stages (ASHES). VIII. Dynamics of Embedded Dense Cores}",
      journal = {\apj},
         year = 2023,
        month = jun,
       volume = {949},
       number = {2},
          eid = {109},
        pages = {109},
          doi = {10.3847/1538-4357/acc58f},
archivePrefix = {arXiv},
       eprint = {2304.01718},
 primaryClass = {astro-ph.GA},
       adsurl = {https://ui.adsabs.harvard.edu/abs/2023ApJ...949..109L}
}

@ARTICLE{2024A&A...689A..74A,
       author = {{{\'A}lvarez-Guti{\'e}rrez}, R.~H. and {Stutz}, A.~M. and {Sandoval-Garrido}, N. and {Louvet}, F. and {Motte}, F. and {Galv{\'a}n-Madrid}, R. and {Cunningham}, N. and {Sanhueza}, P. and {Bonfand}, M. and {Bontemps}, S. and {Gusdorf}, A. and {Ginsburg}, A. and {Csengeri}, T. and {Reyes}, S.~D. and {Salinas}, J. and {Baug}, T. and {Bronfman}, L. and {Busquet}, G. and {D{\'\i}az-Gonz{\'a}lez}, D.~J. and {Fernandez-Lopez}, M. and {Guzm{\'a}n}, A. and {Koley}, A. and {Liu}, H.-L. and {Olguin}, F.~A. and {Valeille-Manet}, M. and {Wyrowski}, F.},
        title = "{ALMA-IMF: XIII. N$_{2}$H$^{+}$ kinematic analysis of the intermediate protocluster G353.41}",
      journal = {\aap},
         year = 2024,
        month = sep,
       volume = {689},
          eid = {A74},
        pages = {A74},
          doi = {10.1051/0004-6361/202450321},
archivePrefix = {arXiv},
       eprint = {2404.07363},
 primaryClass = {astro-ph.GA},
       adsurl = {https://ui.adsabs.harvard.edu/abs/2024A&A...689A..74A}
}

@ARTICLE{2025A&A...696A.202S,
       author = {{Sandoval-Garrido}, N.~A. and {Stutz}, A.~M. and {{\'A}lvarez-Guti{\'e}rrez}, R.~H. and {Galv{\'a}n-Madrid}, R. and {Motte}, F. and {Ginsburg}, A. and {Cunningham}, N. and {Reyes-Reyes}, S. and {Redaelli}, E. and {Bonfand}, M. and {Salinas}, J. and {Koley}, A. and {Bernal-Mesina}, G. and {Braine}, J. and {Bronfman}, L. and {Busquet}, G. and {Csengeri}, T. and {Di Francesco}, J. and {Fern{\'a}ndez-L{\'o}pez}, M. and {Garcia}, P. and {Gusdorf}, A. and {Liu}, H.-L. and {Sanhueza}, P.},
        title = "{ALMA-IMF: XVIII. The assembly of a star cluster: Dense N$_{2}$H$^{+}$ (1─0) kinematics in the massive G351.77 protocluster}",
      journal = {\aap},
         year = 2025,
        month = apr,
       volume = {696},
          eid = {A202},
        pages = {A202},
          doi = {10.1051/0004-6361/202452589},
archivePrefix = {arXiv},
       eprint = {2410.09843},
 primaryClass = {astro-ph.GA},
       adsurl = {https://ui.adsabs.harvard.edu/abs/2025A&A...696A.202S}
}

@ARTICLE{2026arXiv260113473K,
       author = {{Koley}, A. and {Stutz}, A.~M. and {Lazarian}, A. and {Hu}, Y. and {Sanhueza}, P. and {Saha}, P. and {Alvarez-Gutierrez}, R.~H. and {Sandoval-Garrido}, N.~S. and {Castro-Toledo}, N. and {Bernal Mesina}, G.},
        title = "{Magnetic field morphological diagnostics with ALMA in the G327.29 protocluster: VGT versus dust polarization}",
      journal = {arXiv e-prints},
         year = 2026,
        month = jan,
          eid = {arXiv:2601.13473},
        pages = {arXiv:2601.13473},
          doi = {10.48550/arXiv.2601.13473},
archivePrefix = {arXiv},
       eprint = {2601.13473},
 primaryClass = {astro-ph.GA},
       adsurl = {https://ui.adsabs.harvard.edu/abs/2026arXiv260113473K}
}

@ARTICLE{2025arXiv251003447S,
       author = {{Salinas}, J. and {Stutz}, A.~M. and {{\'A}lvarez-Guti{\'e}rrez}, R.~H. and {Sandoval-Garrido}, N.~A. and {Louvet}, F. and {Galv{\'a}n-Madrid}, R. and {Motte}, F. and {Armante}, M. and {Csengeri}, T. and {Braine}, J. and {Ginsburg}, A. and {Valeille-Manet}, M. and {Bronfman}, L. and {Sanhueza}, P. and {D{\'\i}az}, D. and {Busquet}, G. and {Koley}, A. and {Bonfand}, M. and {Fern{\'a}ndez-L{\'o}pez}, M. and {Castro-Toledo}, N. and {Veyry}, R. and {Bernal-Mesina}, G.},
        title = "{ALMA-IMF. XXI.: N$_2$H$^+$ kinematics in the G012.80 protocluster: Evidence for filament rotation and evolution}",
      journal = {arXiv e-prints},
         year = 2025,
        month = oct,
          eid = {arXiv:2510.03447},
        pages = {arXiv:2510.03447},
          doi = {10.48550/arXiv.2510.03447},
archivePrefix = {arXiv},
       eprint = {2510.03447},
 primaryClass = {astro-ph.GA},
       adsurl = {https://ui.adsabs.harvard.edu/abs/2025arXiv251003447S}
}

@ARTICLE{2018ApJ...861...65S,
       author = {{Soam}, Archana and {Pattle}, Kate and {Ward-Thompson}, Derek and {Lee}, Chang Won and {Sadavoy}, Sarah and {Koch}, Patrick M. and {Kim}, Gwanjeong and {Kwon}, Jungmi and {Kwon}, Woojin and {Arzoumanian}, Doris and {Berry}, David and {Hoang}, Thiem and {Tamura}, Motohide and {Lee}, Sang-Sung and {Liu}, Tie and {Kim}, Kee-Tae and {Johnstone}, Doug and {Nakamura}, Fumitaka and {Lyo}, A. -Ran and {Onaka}, Takashi and {Kim}, Jongsoo and {Furuya}, Ray S. and {Hasegawa}, Tetsuo and {Lai}, Shih-Ping and {Bastien}, Pierre and {Chung}, Eun Jung and {Kim}, Shinyoung and {Parsons}, Harriet and {Rawlings}, Mark G. and {Mairs}, Steve and {Graves}, Sarah F. and {Robitaille}, Jean-Franois and {Liu}, Hong-Li and {Whitworth}, Anthony P. and {Eswaraiah}, Chakali and {Rao}, Ramprasad and {Yoo}, Hyunju and {Houde}, Martin and {Kang}, Ji-hyun and {Doi}, Yasuo and {Choi}, Minho and {Kang}, Miju and {Coud{\'e}}, Simon and {Li}, Hua-bai and {Matsumura}, Masafumi and {Matthews}, Brenda C. and {Pon}, Andy and {Di Francesco}, James and {Hayashi}, Saeko S. and {Kawabata}, Koji S. and {Inutsuka}, Shu-ichiro and {Qiu}, Keping and {Franzmann}, Erica and {Friberg}, Per and {Greaves}, Jane S. and {Kirk}, Jason M. and {Li}, Di and {Shinnaga}, Hiroko and {van Loo}, Sven and {Aso}, Yusuke and {Byun}, Do-Young and {Chen}, Huei-Ru and {Chen}, Mike C. -Y. and {Chen}, Wen Ping and {Ching}, Tao-Chung and {Cho}, Jungyeon and {Chrysostomou}, Antonio and {Drabek-Maunder}, Emily and {Eyres}, Stewart P.~S. and {Fiege}, Jason and {Friesen}, Rachel K. and {Fuller}, Gary and {Gledhill}, Tim and {Griffin}, Matt J. and {Gu}, Qilao and {Hatchell}, Jennifer and {Holland}, Wayne and {Inoue}, Tsuyoshi and {Iwasaki}, Kazunari and {Jeong}, Il-Gyo and {Kang}, Sung-ju and {Kemper}, Francisca and {Kim}, Kyoung Hee and {Kim}, Mi-Ryang and {Lacaille}, Kevin M. and {Lee}, Jeong-Eun and {Li}, Dalei and {Liu}, Junhao and {Liu}, Sheng-Yuan and {Moriarty-Schieven}, Gerald H. and {Nakanishi}, Hiroyuki and {Ohashi}, Nagayoshi and {Peretto}, Nicolas and {Pyo}, Tae-Soo and {Qian}, Lei and {Retter}, Brendan and {Richer}, John and {Rigby}, Andrew and {Savini}, Giorgio and {Scaife}, Anna M.~M. and {Tang}, Ya-Wen and {Tomisaka}, Kohji and {Wang}, Hongchi and {Wang}, Jia-Wei and {Yen}, Hsi-Wei and {Yuan}, Jinghua and {Zhang}, Chuan-Peng and {Zhang}, Guoyin and {Zhou}, Jianjun and {Zhu}, Lei and {Andr{\'e}}, Philippe and {Dowell}, C. Darren and {Falle}, Sam and {Tsukamoto}, Yusuke and {Kanamori}, Yoshihiro and {Kataoka}, Akimasa and {Kobayashi}, Masato I.~N. and {Nagata}, Tetsuya and {Saito}, Hiro and {Seta}, Masumichi and {Hwang}, Jihye and {Han}, Ilseung and {Lee}, Hyeseung and {Zenko}, Tetsuya},
        title = "{Magnetic Fields toward Ophiuchus-B Derived from SCUBA-2 Polarization Measurements}",
      journal = {\apj},
         year = 2018,
        month = jul,
       volume = {861},
       number = {1},
          eid = {65},
        pages = {65},
          doi = {10.3847/1538-4357/aac4a6},
archivePrefix = {arXiv},
       eprint = {1805.06131},
 primaryClass = {astro-ph.GA},
       adsurl = {https://ui.adsabs.harvard.edu/abs/2018ApJ...861...65S}
}

@ARTICLE{2013ApJ...763..135T,
       author = {{Tang}, Ya-Wen and {Ho}, Paul T.~P. and {Koch}, Patrick M. and {Guilloteau}, Stephane and {Dutrey}, Anne},
        title = "{Dust Continuum and Polarization from Envelope to Cores in Star Formation: A Case Study in the W51 North Region}",
      journal = {\apj},
         year = 2013,
        month = feb,
       volume = {763},
       number = {2},
          eid = {135},
        pages = {135},
          doi = {10.1088/0004-637X/763/2/135},
archivePrefix = {arXiv},
       eprint = {1212.0656},
 primaryClass = {astro-ph.GA},
       adsurl = {https://ui.adsabs.harvard.edu/abs/2013ApJ...763..135T}
}

@ARTICLE{2009ApJ...695.1399T,
       author = {{Tang}, Ya-Wen and {Ho}, Paul T.~P. and {Girart}, Josep Miquel and {Rao}, Ramprasad and {Koch}, Patrick and {Lai}, Shih-Ping},
        title = "{Evolution of Magnetic Fields in High Mass Star Formation: Submillimeter Array Dust Polarization Image of the Ultracompact H II Region G5.89-0.39}",
      journal = {\apj},
         year = 2009,
        month = apr,
       volume = {695},
       number = {2},
        pages = {1399-1412},
          doi = {10.1088/0004-637X/695/2/1399},
archivePrefix = {arXiv},
       eprint = {0812.3444},
 primaryClass = {astro-ph},
       adsurl = {https://ui.adsabs.harvard.edu/abs/2009ApJ...695.1399T}
}

@ARTICLE{2003ApJ...592..233W,
       author = {{Wolf}, Sebastian and {Launhardt}, R. and {Henning}, T.},
        title = "{Magnetic Field Evolution in Bok Globules}",
      journal = {\apj},
         year = 2003,
        month = jul,
       volume = {592},
       number = {1},
        pages = {233-244},
          doi = {10.1086/375622},
archivePrefix = {arXiv},
       eprint = {astro-ph/0303652},
 primaryClass = {astro-ph},
       adsurl = {https://ui.adsabs.harvard.edu/abs/2003ApJ...592..233W}
}

@ARTICLE{2024A&A...682A..81B,
       author = {{Beuther}, H. and {Gieser}, C. and {Soler}, J.~D. and {Zhang}, Q. and {Rao}, R. and {Semenov}, D. and {Henning}, Th. and {Pudritz}, R. and {Peters}, T. and {Klaassen}, P. and {Beltr{\'a}n}, M.~T. and {Palau}, A. and {M{\"o}ller}, T. and {Johnston}, K.~G. and {Zinnecker}, H. and {Urquhart}, J. and {Kuiper}, R. and {Ahmadi}, A. and {S{\'a}nchez-Monge}, {\'A}. and {Feng}, S. and {Leurini}, S. and {Ragan}, S.~E.},
        title = "{Density distributions, magnetic field structures, and fragmentation in high-mass star formation}",
      journal = {\aap},
         year = 2024,
        month = feb,
       volume = {682},
          eid = {A81},
        pages = {A81},
          doi = {10.1051/0004-6361/202348117},
archivePrefix = {arXiv},
       eprint = {2311.11874},
 primaryClass = {astro-ph.GA},
       adsurl = {https://ui.adsabs.harvard.edu/abs/2024A&A...682A..81B}
}

@ARTICLE{2021ApJ...913...29F,
       author = {{Fern{\'a}ndez-L{\'o}pez}, M. and {Sanhueza}, P. and {Zapata}, L.~A. and {Stephens}, I. and {Hull}, C. and {Zhang}, Q. and {Girart}, J.~M. and {Koch}, P.~M. and {Cort{\'e}s}, P. and {Silva}, A. and {Tatematsu}, K. and {Nakamura}, F. and {Guzm{\'a}n}, A.~E. and {Nguyen Luong}, Q. and {Guzm{\'a}n Ccolque}, E. and {Tang}, Y. -W. and {Chen}, H. -R.~V.},
        title = "{Magnetic Fields in Massive Star-forming Regions (MagMaR). I. Linear Polarized Imaging of the Ultracompact H II Region G5.89-0.39}",
      journal = {\apj},
         year = 2021,
        month = may,
       volume = {913},
       number = {1},
          eid = {29},
        pages = {29},
          doi = {10.3847/1538-4357/abf2b6},
archivePrefix = {arXiv},
       eprint = {2104.03331},
 primaryClass = {astro-ph.GA},
       adsurl = {https://ui.adsabs.harvard.edu/abs/2021ApJ...913...29F}
}

@ARTICLE{2025A&A...696A.150S,
       author = {{S{\'a}nchez-Monge}, {\'A}. and {Brogan}, C.~L. and {Hunter}, T.~R. and {Ahmadi}, A. and {Avison}, A. and {Beltr{\'a}n}, M.~T. and {Beuther}, H. and {Coletta}, A. and {Fuller}, G.~A. and {Johnston}, K.~G. and {Jones}, B. and {Liu}, S. -Y. and {Mininni}, C. and {Molinari}, S. and {Schilke}, P. and {Schisano}, E. and {Su}, Y. -N. and {Traficante}, A. and {Zhang}, Q. and {Battersby}, C. and {Benedettini}, M. and {Elia}, D. and {Ho}, P.~T.~P. and {Klaassen}, P.~D. and {Klessen}, R.~S. and {Law}, C.~Y. and {Lis}, D.~C. and {Liu}, T. and {Maud}, L. and {M{\"o}ller}, T. and {Moscadelli}, L. and {Pezzuto}, S. and {Rygl}, K.~L.~J. and {Sanhueza}, P. and {Soler}, J.~D. and {Stroud}, G. and {Tang}, Y. and {van der Tak}, F.~F.~S. and {Walker}, D.~L. and {Wallace}, J. and {Walch}, S. and {Wells}, M.~R.~A. and {Wyrowski}, F. and {Zhang}, T. and {Allande}, J. and {Bronfman}, L. and {Dann}, E. and {De Angelis}, F. and {Fontani}, F. and {Henning}, Th. and {Kim}, W. -J. and {Kuiper}, R. and {Merello}, M. and {Nakamura}, F. and {Nucara}, A. and {Rigby}, A.~J.},
        title = "{ALMAGAL: II. The ALMA evolutionary study of high-mass protocluster formation in the Galaxy: ALMA data processing and pipeline}",
      journal = {\aap},
         year = 2025,
        month = apr,
       volume = {696},
          eid = {A150},
        pages = {A150},
          doi = {10.1051/0004-6361/202452703},
archivePrefix = {arXiv},
       eprint = {2503.05559},
 primaryClass = {astro-ph.GA},
       adsurl = {https://ui.adsabs.harvard.edu/abs/2025A&A...696A.150S}
}

@ARTICLE{2025ApJ...994..233Y,
       author = {{Yoo}, T. and {Ginsburg}, A. and {Braine}, J. and {Budaiev}, N. and {Louvet}, F. and {Motte}, F. and {Stutz}, A.~M. and {Thomasson}, B. and {Armante}, M. and {Bonfand}, M. and {Bontemps}, S. and {Bronfman}, L. and {Busquet}, G. and {Csengeri}, T. and {Cunningham}, N. and {Di Francesco}, J. and {D{\'\i}az-Gonz{\'a}lez}, D.~J. and {Fern{\'a}ndez-Lopez}, M. and {Galv{\'a}n-Madrid}, R. and {Goddi}, C. and {Gusdorf}, A. and {Kessler}, N. and {Koley}, A. and {Liu}, H.-L. and {Nony}, T. and {Olguin}, F. and {Sanhueza}, P. and {Valeille-Manet}, M. and {Zapata}, L.~A. and {Zhang}, Q.},
        title = "{ALMA-IMF. XX. Core Fragmentation in the W51 High-mass Star-forming Region}",
      journal = {\apj},
         year = 2025,
        month = dec,
       volume = {994},
       number = {2},
          eid = {233},
        pages = {233},
          doi = {10.3847/1538-4357/ae0619},
archivePrefix = {arXiv},
       eprint = {2509.06749},
 primaryClass = {astro-ph.GA},
       adsurl = {https://ui.adsabs.harvard.edu/abs/2025ApJ...994..233Y}
}

@ARTICLE{2025A&A...694A..24M,
       author = {{Motte}, F. and {Pouteau}, Y. and {Nony}, T. and {Dell'Ova}, P. and {Gusdorf}, A. and {Brouillet}, N. and {Stutz}, A.~M. and {Bontemps}, S. and {Ginsburg}, A. and {Csengeri}, T. and {Men'shchikov}, A. and {Valeille-Manet}, M. and {Louvet}, F. and {Bonfand}, M. and {Galv{\'a}n-Madrid}, R. and {{\'A}lvarez-Guti{\'e}rrez}, R.~H. and {Armante}, M. and {Bronfman}, L. and {Chen}, H.-R.~V. and {Cunningham}, N. and {D{\'\i}az-Gonz{\'a}lez}, D. and {Didelon}, P. and {Fern{\'a}ndez-L{\'o}pez}, M. and {Herpin}, F. and {Kessler}, N. and {Koley}, A. and {Lefloch}, B. and {Le Nestour}, N. and {Liu}, H.-L. and {Moraux}, E. and {Nguyen Luong}, Q. and {Olguin}, F. and {Salinas}, J. and {Sandoval-Garrido}, N.~A. and {Sanhueza}, P. and {Veyry}, R. and {Yoo}, T.},
        title = "{ALMA-IMF: XVI. Mass-averaged temperature of cores and protostellar luminosities in the ALMA-IMF protoclusters}",
      journal = {\aap},
         year = 2025,
        month = feb,
       volume = {694},
          eid = {A24},
        pages = {A24},
          doi = {10.1051/0004-6361/202451931},
archivePrefix = {arXiv},
       eprint = {2412.02011},
 primaryClass = {astro-ph.GA},
       adsurl = {https://ui.adsabs.harvard.edu/abs/2025A&A...694A..24M}
}

@ARTICLE{2022MNRAS.516..185K,
       author = {{Koley}, Atanu},
        title = "{Studying the chemical and kinematical structures of dense cores TMC-1C, L1544, and TMC-1 in the Taurus molecular cloud using CCS and NH$_{3}$ observations}",
      journal = {\mnras},
         year = 2022,
        month = oct,
       volume = {516},
       number = {1},
        pages = {185-196},
          doi = {10.1093/mnras/stac1935},
archivePrefix = {arXiv},
       eprint = {2208.00968},
 primaryClass = {astro-ph.GA},
       adsurl = {https://ui.adsabs.harvard.edu/abs/2022MNRAS.516..185K}
}

@ARTICLE{2022MNRAS.516L..48K,
       author = {{Koley}, Atanu and {Roy}, Nirupam and {Momjian}, Emmanuel and {Sarma}, Anuj P. and {Datta}, Abhirup},
        title = "{Magnetic field measurement in TMC-1C using 22.3 GHz CCS Zeeman splitting}",
      journal = {\mnras},
         year = 2022,
        month = oct,
       volume = {516},
       number = {1},
        pages = {L48-L52},
          doi = {10.1093/mnrasl/slac085},
archivePrefix = {arXiv},
       eprint = {2207.12604},
 primaryClass = {astro-ph.GA},
       adsurl = {https://ui.adsabs.harvard.edu/abs/2022MNRAS.516L..48K}
}

@ARTICLE{2019MNRAS.483..593K,
       author = {{Koley}, Atanu and {Roy}, Nirupam},
        title = "{Estimating the kinetic temperature from H I 21-cm absorption studies: correction for turbulence broadening}",
      journal = {\mnras},
         year = 2019,
        month = feb,
       volume = {483},
       number = {1},
        pages = {593-598},
          doi = {10.1093/mnras/sty3152},
archivePrefix = {arXiv},
       eprint = {1811.07352},
 primaryClass = {astro-ph.GA},
       adsurl = {https://ui.adsabs.harvard.edu/abs/2019MNRAS.483..593K}
}

@ARTICLE{2024MNRAS.529.2220R,
       author = {{Reyes-Reyes}, S.~D. and {Stutz}, A.~M. and {Megeath}, S.~T. and {Xu}, Fengwei and {{\'A}lvarez-Guti{\'e}rrez}, R.~H. and {Sandoval-Garrido}, N. and {Liu}, H. -L.},
        title = "{Benchmarking the IRDC G351.77-0.53: Gaia DR3 distance, mass distribution, and star formation content}",
      journal = {\mnras},
         year = 2024,
        month = apr,
       volume = {529},
       number = {3},
        pages = {2220-2233},
          doi = {10.1093/mnras/stae631},
archivePrefix = {arXiv},
       eprint = {2403.02456},
 primaryClass = {astro-ph.GA},
       adsurl = {https://ui.adsabs.harvard.edu/abs/2024MNRAS.529.2220R}
}

@ARTICLE{2017A&A...601A..60C,
       author = {{Csengeri}, T. and {Bontemps}, S. and {Wyrowski}, F. and {Megeath}, S.~T. and {Motte}, F. and {Sanna}, A. and {Wienen}, M. and {Menten}, K.~M.},
        title = "{The ATLASGAL survey: The sample of young massive cluster progenitors}",
      journal = {\aap},
         year = 2017,
        month = may,
       volume = {601},
          eid = {A60},
        pages = {A60},
          doi = {10.1051/0004-6361/201628254},
archivePrefix = {arXiv},
       eprint = {1701.01563},
 primaryClass = {astro-ph.GA},
       adsurl = {https://ui.adsabs.harvard.edu/abs/2017A&A...601A..60C}
}

@ARTICLE{2010ApJ...724L.113B,
       author = {{Beuther}, H. and {Vlemmings}, W.~H.~T. and {Rao}, R. and {van der Tak}, F.~F.~S.},
        title = "{Magnetic Field Structure in a High-mass Outflow/Disk System}",
      journal = {\apjl},
         year = 2010,
        month = nov,
       volume = {724},
       number = {1},
        pages = {L113-L117},
          doi = {10.1088/2041-8205/724/1/L113},
archivePrefix = {arXiv},
       eprint = {1010.3635},
 primaryClass = {astro-ph.SR},
       adsurl = {https://ui.adsabs.harvard.edu/abs/2010ApJ...724L.113B}
}

@ARTICLE{2016A&A...593L..14F,
       author = {{Fontani}, F. and {Commer{\c{c}}on}, B. and {Giannetti}, A. and {Beltr{\'a}n}, M.~T. and {S{\'a}nchez-Monge}, A. and {Testi}, L. and {Brand}, J. and {Caselli}, P. and {Cesaroni}, R. and {Dodson}, R. and {Longmore}, S. and {Rioja}, M. and {Tan}, J.~C. and {Walmsley}, C.~M.},
        title = "{Magnetically regulated fragmentation of a massive, dense, and turbulent clump}",
      journal = {\aap},
         year = 2016,
        month = sep,
       volume = {593},
          eid = {L14},
        pages = {L14},
          doi = {10.1051/0004-6361/201629442},
archivePrefix = {arXiv},
       eprint = {1608.08083},
 primaryClass = {astro-ph.GA},
       adsurl = {https://ui.adsabs.harvard.edu/abs/2016A&A...593L..14F}
}

@ARTICLE{2009Sci...324.1408G,
       author = {{Girart}, Josep M. and {Beltr{\'a}n}, Maria T. and {Zhang}, Qizhou and {Rao}, Ramprasad and {Estalella}, Robert},
        title = "{Magnetic Fields in the Formation of Massive Stars}",
      journal = {Science},
         year = 2009,
        month = jun,
       volume = {324},
       number = {5933},
        pages = {1408},
          doi = {10.1126/science.1171807},
       adsurl = {https://ui.adsabs.harvard.edu/abs/2009Sci...324.1408G}
}

@ARTICLE{2023ApJ...951...68C,
       author = {{Chung}, Eun Jung and {Lee}, Chang Won and {Kwon}, Woojin and {Tafalla}, Mario and {Kim}, Shinyoung and {Soam}, Archana and {Cho}, Jungyeon},
        title = "{Magnetic Fields and Fragmentation of Filaments in the Hub of California-X}",
      journal = {\apj},
         year = 2023,
        month = jul,
       volume = {951},
       number = {1},
          eid = {68},
        pages = {68},
          doi = {10.3847/1538-4357/acd540},
archivePrefix = {arXiv},
       eprint = {2305.09949},
 primaryClass = {astro-ph.GA},
       adsurl = {https://ui.adsabs.harvard.edu/abs/2023ApJ...951...68C}
}

@ARTICLE{2013ApJ...772...69G,
       author = {{Girart}, J.~M. and {Frau}, P. and {Zhang}, Q. and {Koch}, P.~M. and {Qiu}, K. and {Tang}, Y. -W. and {Lai}, S. -P. and {Ho}, P.~T.~P.},
        title = "{DR 21(OH): A Highly Fragmented, Magnetized, Turbulent Dense Core}",
      journal = {\apj},
         year = 2013,
        month = jul,
       volume = {772},
       number = {1},
          eid = {69},
        pages = {69},
          doi = {10.1088/0004-637X/772/1/69},
archivePrefix = {arXiv},
       eprint = {1305.6509},
 primaryClass = {astro-ph.GA},
       adsurl = {https://ui.adsabs.harvard.edu/abs/2013ApJ...772...69G}
}

@ARTICLE{2024ApJ...966..120L,
       author = {{Liu}, Junhao and {Zhang}, Qizhou and {Lin}, Yuxin and {Qiu}, Keping and {Koch}, Patrick M. and {Liu}, Hauyu Baobab and {Li}, Zhi-Yun and {Girart}, Josep Miquel and {Pillai}, Thushara G.~S. and {Li}, Shanghuo and {Chen}, Huei-Ru Vivien and {Ching}, Tao-Chung and {Ho}, Paul T.~P. and {Lai}, Shih-Ping and {Rao}, Ramprasad and {Tang}, Ya-Wen and {Wang}, Ke},
        title = "{Dark Dragon Breaks Magnetic Chain: Dynamical Substructures of IRDC G28.34 Form in Supported Environments}",
      journal = {\apj},
         year = 2024,
        month = may,
       volume = {966},
       number = {1},
          eid = {120},
        pages = {120},
          doi = {10.3847/1538-4357/ad3105},
archivePrefix = {arXiv},
       eprint = {2403.03437},
 primaryClass = {astro-ph.GA},
       adsurl = {https://ui.adsabs.harvard.edu/abs/2024ApJ...966..120L}
}

@ARTICLE{2020ApJ...905..158W,
       author = {{Wang}, Jia-Wei and {Koch}, Patrick M. and {Galv{\'a}n-Madrid}, Roberto and {Lai}, Shih-Ping and {Liu}, Hauyu Baobab and {Lin}, Sheng-Jun and {Pattle}, Kate},
        title = "{Formation of the Hub-Filament System G33.92+0.11: Local Interplay between Gravity, Velocity, and Magnetic Field}",
      journal = {\apj},
         year = 2020,
        month = dec,
       volume = {905},
       number = {2},
          eid = {158},
        pages = {158},
          doi = {10.3847/1538-4357/abc74e},
archivePrefix = {arXiv},
       eprint = {2011.01555},
 primaryClass = {astro-ph.GA},
       adsurl = {https://ui.adsabs.harvard.edu/abs/2020ApJ...905..158W}
}

@ARTICLE{2025arXiv250810128C,
       author = {{Chen}, Huei-Ru Vivien and {Zhang}, Qizhou and {Ching}, Tao-Chung and {Beuther}, H. and {Wang}, Kuo-Song},
        title = "{Pinched Magnetic Fields in the High-mass Protocluster W3 IRS5}",
      journal = {arXiv e-prints},
         year = 2025,
        month = aug,
          eid = {arXiv:2508.10128},
        pages = {arXiv:2508.10128},
          doi = {10.48550/arXiv.2508.10128},
archivePrefix = {arXiv},
       eprint = {2508.10128},
 primaryClass = {astro-ph.GA},
       adsurl = {https://ui.adsabs.harvard.edu/abs/2025arXiv250810128C}
}

@ARTICLE{2017ApJ...847...92H,
       author = {{Hull}, Charles L.~H. and {Girart}, Josep M. and {Tychoniec}, {\L}ukasz and {Rao}, Ramprasad and {Cort{\'e}s}, Paulo C. and {Pokhrel}, Riwaj and {Zhang}, Qizhou and {Houde}, Martin and {Dunham}, Michael M. and {Kristensen}, Lars E. and {Lai}, Shih-Ping and {Li}, Zhi-Yun and {Plambeck}, Richard L.},
        title = "{ALMA Observations of Dust Polarization and Molecular Line Emission from the Class 0 Protostellar Source Serpens SMM1}",
      journal = {\apj},
         year = 2017,
        month = oct,
       volume = {847},
       number = {2},
          eid = {92},
        pages = {92},
          doi = {10.3847/1538-4357/aa7fe9},
archivePrefix = {arXiv},
       eprint = {1707.03827},
 primaryClass = {astro-ph.GA},
       adsurl = {https://ui.adsabs.harvard.edu/abs/2017ApJ...847...92H}
}

@ARTICLE{2004Ap&SS.292..225C,
       author = {{Crutcher}, Richard M.},
        title = "{What Drives Star Formation?}",
      journal = {\apss},
         year = 2004,
        month = jul,
       volume = {292},
       number = {1},
        pages = {225-237},
          doi = {10.1023/B:ASTR.0000045021.42255.95},
       adsurl = {https://ui.adsabs.harvard.edu/abs/2004Ap&SS.292..225C}
}

@ARTICLE{2021MNRAS.501.4825K,
       author = {{Koley}, Atanu and {Roy}, Nirupam and {Menten}, Karl M. and {Jacob}, Arshia M. and {Pillai}, Thushara G.~S. and {Rugel}, Michael R.},
        title = "{The magnetic field in the dense photodissociation region of DR 21}",
      journal = {\mnras},
         year = 2021,
        month = mar,
       volume = {501},
       number = {4},
        pages = {4825-4836},
          doi = {10.1093/mnras/staa3898},
archivePrefix = {arXiv},
       eprint = {2012.08253},
 primaryClass = {astro-ph.GA},
       adsurl = {https://ui.adsabs.harvard.edu/abs/2021MNRAS.501.4825K}
}

@ARTICLE{2016A&A...590A...2S,
       author = {{Stutz}, Amelia M. and {Gould}, Andrew},
        title = "{Slingshot mechanism in Orion: Kinematic evidence for ejection of protostars by filaments}",
      journal = {\aap},
         year = 2016,
        month = may,
       volume = {590},
          eid = {A2},
        pages = {A2},
          doi = {10.1051/0004-6361/201527979},
archivePrefix = {arXiv},
       eprint = {1512.04944},
 primaryClass = {astro-ph.SR},
       adsurl = {https://ui.adsabs.harvard.edu/abs/2016A&A...590A...2S}
}

@ARTICLE{2018MNRAS.473.4890S,
       author = {{Stutz}, Amelia M.},
        title = "{Slingshot mechanism for clusters: Gas density regulates star density in the Orion Nebula Cluster (M42)}",
      journal = {\mnras},
         year = 2018,
        month = feb,
       volume = {473},
       number = {4},
        pages = {4890-4899},
          doi = {10.1093/mnras/stx2629},
archivePrefix = {arXiv},
       eprint = {1705.05838},
 primaryClass = {astro-ph.GA},
       adsurl = {https://ui.adsabs.harvard.edu/abs/2018MNRAS.473.4890S}
}

@ARTICLE{2018A&A...609A..43X,
       author = {{Xu}, Jin-Long and {Xu}, Ye and {Zhang}, Chuan-Peng and {Liu}, Xiao-Lan and {Yu}, Naiping and {Ning}, Chang-Chun and {Ju}, Bing-Gang},
        title = "{Gas kinematics and star formation in the filamentary molecular cloud G47.06+0.26}",
      journal = {\aap},
         year = 2018,
        month = jan,
       volume = {609},
          eid = {A43},
        pages = {A43},
          doi = {10.1051/0004-6361/201629189},
archivePrefix = {arXiv},
       eprint = {1708.09098},
 primaryClass = {astro-ph.GA},
       adsurl = {https://ui.adsabs.harvard.edu/abs/2018A&A...609A..43X}
}

@ARTICLE{2025MNRAS.540.2762W,
       author = {{Whitworth}, D.~J. and {Srinivasan}, S. and {Pudritz}, R.~E. and {Mac Low}, M. -M. and {Eadie}, G. and {Palau}, A. and {Soler}, J.~D. and {Smith}, R.~J. and {Pattle}, K. and {Robinson}, H. and {Pillsworth}, R. and {Wadsley}, J. and {Brucy}, N. and {Lebreuilly}, U. and {Hennebelle}, P. and {Girichidis}, P. and {Gent}, F.~A. and {Marin}, J. and {S{\'a}nchez Valido}, L. and {Camacho}, V. and {Klessen}, R.~S. and {V{\'a}zquez-Semadeni}, E.},
        title = "{On the relation between magnetic field strength and gas density in the interstellar medium: a multiscale analysis}",
      journal = {\mnras},
         year = 2025,
        month = jul,
       volume = {540},
       number = {3},
        pages = {2762-2786},
          doi = {10.1093/mnras/staf901},
archivePrefix = {arXiv},
       eprint = {2407.18293},
 primaryClass = {astro-ph.GA},
       adsurl = {https://ui.adsabs.harvard.edu/abs/2025MNRAS.540.2762W}
}

@ARTICLE{2013ApJ...779..182Q,
       author = {{Qiu}, Keping and {Zhang}, Qizhou and {Menten}, Karl M. and {Liu}, Hauyu B. and {Tang}, Ya-Wen},
        title = "{From Poloidal to Toroidal: Detection of a Well-ordered Magnetic Field in the High-mass Protocluster G35.2-0.74 N}",
      journal = {\apj},
         year = 2013,
        month = dec,
       volume = {779},
       number = {2},
          eid = {182},
        pages = {182},
          doi = {10.1088/0004-637X/779/2/182},
archivePrefix = {arXiv},
       eprint = {1311.0566},
 primaryClass = {astro-ph.GA},
       adsurl = {https://ui.adsabs.harvard.edu/abs/2013ApJ...779..182Q}
}

@ARTICLE{2015ApJ...799...74P,
       author = {{Pillai}, T. and {Kauffmann}, J. and {Tan}, J.~C. and {Goldsmith}, P.~F. and {Carey}, S.~J. and {Menten}, K.~M.},
        title = "{Magnetic Fields in High-mass Infrared Dark Clouds}",
      journal = {\apj},
         year = 2015,
        month = jan,
       volume = {799},
       number = {1},
          eid = {74},
        pages = {74},
          doi = {10.1088/0004-637X/799/1/74},
archivePrefix = {arXiv},
       eprint = {1410.7390},
 primaryClass = {astro-ph.GA},
       adsurl = {https://ui.adsabs.harvard.edu/abs/2015ApJ...799...74P}
}

@ARTICLE{2021ApJ...913...57K,
       author = {{K{\"o}nyves}, Vera and {Ward-Thompson}, Derek and {Pattle}, Kate and {Di Francesco}, James and {Arzoumanian}, Doris and {Chen}, Zhiwei and {Diep}, Pham Ngoc and {Eswaraiah}, Chakali and {Fanciullo}, Lapo and {Furuya}, Ray S. and {Hoang}, Thiem and {Hull}, Charles L.~H. and {Hwang}, Jihye and {Johnstone}, Doug and {Kang}, Ji-hyun and {Karoly}, Janik and {Kirchschlager}, Florian and {Kirk}, Jason M. and {Koch}, Patrick M. and {Kwon}, Jungmi and {Lee}, Chang Won and {Onaka}, Takashi and {Robitaille}, Jean-Fran{\c{c}}ois and {Soam}, Archana and {Tahani}, Mehrnoosh and {Tang}, Xindi and {Tamura}, Motohide and {Berry}, David and {Bastien}, Pierre and {Ching}, Tao-Chung and {Coud{\'e}}, Simon and {Kwon}, Woojin and {Wang}, Jia-Wei and {Hasegawa}, Tetsuo and {Lai}, Shih-Ping and {Qiu}, Keping},
        title = "{The JCMT BISTRO-2 Survey: The Magnetic Field in the Center of the Rosette Molecular Cloud}",
      journal = {\apj},
         year = 2021,
        month = may,
       volume = {913},
       number = {1},
          eid = {57},
        pages = {57},
          doi = {10.3847/1538-4357/abf3ca},
archivePrefix = {arXiv},
       eprint = {2104.00121},
 primaryClass = {astro-ph.GA},
       adsurl = {https://ui.adsabs.harvard.edu/abs/2021ApJ...913...57K}
}

@ARTICLE{2015MNRAS.451.4384T,
       author = {{Tritsis}, A. and {Panopoulou}, G.~V. and {Mouschovias}, T. Ch. and {Tassis}, K. and {Pavlidou}, V.},
        title = "{Magnetic field-gas density relation and observational implications revisited}",
      journal = {\mnras},
         year = 2015,
        month = aug,
       volume = {451},
       number = {4},
        pages = {4384-4396},
          doi = {10.1093/mnras/stv1133},
archivePrefix = {arXiv},
       eprint = {1505.05508},
 primaryClass = {astro-ph.GA},
       adsurl = {https://ui.adsabs.harvard.edu/abs/2015MNRAS.451.4384T}
}

@ARTICLE{2000MNRAS.311...85F,
       author = {{Fiege}, Jason D. and {Pudritz}, Ralph E.},
        title = "{Helical fields and filamentary molecular clouds - I}",
      journal = {\mnras},
         year = 2000,
        month = jan,
       volume = {311},
       number = {1},
        pages = {85-104},
          doi = {10.1046/j.1365-8711.2000.03066.x},
archivePrefix = {arXiv},
       eprint = {astro-ph/9901096},
 primaryClass = {astro-ph},
       adsurl = {https://ui.adsabs.harvard.edu/abs/2000MNRAS.311...85F}
}

@ARTICLE{2024MNRAS.528.1460R,
       author = {{Rawat}, Vineet and {Samal}, M.~R. and {Eswaraiah}, Chakali and {Wang}, Jia-Wei and {Elia}, Davide and {Panigrahy}, Sandhyarani and {Zavagno}, A. and {Yadav}, R.~K. and {Walker}, D.~L. and {Jose}, J. and {Ojha}, D.~K. and {Zhang}, C.~P. and {Dutta}, S.},
        title = "{Understanding the relative importance of magnetic field, gravity, and turbulence in star formation at the hub of the giant molecular cloud G148.24+00.41}",
      journal = {\mnras},
         year = 2024,
        month = feb,
       volume = {528},
       number = {2},
        pages = {1460-1475},
          doi = {10.1093/mnras/stae053},
archivePrefix = {arXiv},
       eprint = {2401.05310},
 primaryClass = {astro-ph.GA},
       adsurl = {https://ui.adsabs.harvard.edu/abs/2024MNRAS.528.1460R}
}

@ARTICLE{2016A&A...586A.135P,
       author = {{Planck Collaboration} and {Adam}, R. and {Ade}, P.~A.~R. and {Aghanim}, N. and {Alves}, M.~I.~R. and {Arnaud}, M. and {Arzoumanian}, D. and {Ashdown}, M. and {Aumont}, J. and {Baccigalupi}, C. and {Banday}, A.~J. and {Barreiro}, R.~B. and {Bartolo}, N. and {Battaner}, E. and {Benabed}, K. and {Benoit-L{\'e}vy}, A. and {Bernard}, J.-P. and {Bersanelli}, M. and {Bielewicz}, P. and {Bonaldi}, A. and {Bonavera}, L. and {Bond}, J.~R. and {Borrill}, J. and {Bouchet}, F.~R. and {Boulanger}, F. and {Bracco}, A. and {Burigana}, C. and {Butler}, R.~C. and {Calabrese}, E. and {Cardoso}, J.-F. and {Catalano}, A. and {Chamballu}, A. and {Chiang}, H.~C. and {Christensen}, P.~R. and {Colombi}, S. and {Colombo}, L.~P.~L. and {Combet}, C. and {Couchot}, F. and {Crill}, B.~P. and {Curto}, A. and {Cuttaia}, F. and {Danese}, L. and {Davies}, R.~D. and {Davis}, R.~J. and {de Bernardis}, P. and {de Rosa}, A. and {de Zotti}, G. and {Delabrouille}, J. and {Dickinson}, C. and {Diego}, J.~M. and {Dole}, H. and {Donzelli}, S. and {Dor{\'e}}, O. and {Douspis}, M. and {Ducout}, A. and {Dupac}, X. and {Efstathiou}, G. and {Elsner}, F. and {En{\ss}lin}, T.~A. and {Eriksen}, H.~K. and {Falgarone}, E. and {Ferri{\`e}re}, K. and {Finelli}, F. and {Forni}, O. and {Frailis}, M. and {Fraisse}, A.~A. and {Franceschi}, E. and {Frejsel}, A. and {Galeotta}, S. and {Galli}, S. and {Ganga}, K. and {Ghosh}, T. and {Giard}, M. and {Gjerl{\o}w}, E. and {Gonz{\'a}lez-Nuevo}, J. and {G{\'o}rski}, K.~M. and {Gregorio}, A. and {Gruppuso}, A. and {Guillet}, V. and {Hansen}, F.~K. and {Hanson}, D. and {Harrison}, D.~L. and {Henrot-Versill{\'e}}, S. and {Hern{\'a}ndez-Monteagudo}, C. and {Herranz}, D. and {Hildebrandt}, S.~R. and {Hivon}, E. and {Hobson}, M. and {Holmes}, W.~A. and {Hovest}, W. and {Huffenberger}, K.~M. and {Hurier}, G. and {Jaffe}, A.~H. and {Jaffe}, T.~R. and {Jones}, W.~C. and {Juvela}, M. and {Keih{\"a}nen}, E. and {Keskitalo}, R. and {Kisner}, T.~S. and {Kneissl}, R. and {Knoche}, J. and {Kunz}, M. and {Kurki-Suonio}, H. and {Lagache}, G. and {Lamarre}, J.-M. and {Lasenby}, A. and {Lattanzi}, M. and {Lawrence}, C.~R. and {Leonardi}, R. and {Levrier}, F. and {Liguori}, M. and {Lilje}, P.~B. and {Linden-V{\o}rnle}, M. and {L{\'o}pez-Caniego}, M. and {Lubin}, P.~M. and {Mac{\'\i}as-P{\'e}rez}, J.~F. and {Maffei}, B. and {Maino}, D. and {Mandolesi}, N. and {Maris}, M. and {Marshall}, D.~J. and {Martin}, P.~G. and {Mart{\'\i}nez-Gonz{\'a}lez}, E. and {Masi}, S. and {Matarrese}, S. and {Mazzotta}, P. and {Melchiorri}, A. and {Mendes}, L. and {Mennella}, A. and {Migliaccio}, M. and {Miville-Desch{\^e}nes}, M.-A. and {Moneti}, A. and {Montier}, L. and {Morgante}, G. and {Mortlock}, D. and {Munshi}, D. and {Murphy}, J.~A. and {Naselsky}, P. and {Natoli}, P. and {N{\o}rgaard-Nielsen}, H.~U. and {Noviello}, F. and {Novikov}, D. and {Novikov}, I. and {Oppermann}, N. and {Oxborrow}, C.~A. and {Pagano}, L. and {Pajot}, F. and {Paoletti}, D. and {Pasian}, F. and {Perdereau}, O. and {Perotto}, L. and {Perrotta}, F. and {Pettorino}, V. and {Piacentini}, F. and {Piat}, M. and {Plaszczynski}, S. and {Pointecouteau}, E. and {Polenta}, G. and {Ponthieu}, N. and {Popa}, L. and {Pratt}, G.~W. and {Prunet}, S. and {Puget}, J.-L. and {Rachen}, J.~P. and {Reach}, W.~T. and {Reinecke}, M. and {Remazeilles}, M. and {Renault}, C. and {Ristorcelli}, I. and {Rocha}, G. and {Roudier}, G. and {Rubi{\~n}o-Mart{\'\i}n}, J.~A. and {Rusholme}, B. and {Sandri}, M. and {Santos}, D. and {Savini}, G. and {Scott}, D. and {Soler}, J.~D. and {Spencer}, L.~D. and {Stolyarov}, V. and {Sudiwala}, R. and {Sunyaev}, R. and {Sutton}, D. and {Suur-Uski}, A.-S. and {Sygnet}, J.-F. and {Tauber}, J.~A. and {Terenzi}, L. and {Toffolatti}, L. and {Tomasi}, M. and {Tristram}, M. and {Tucci}, M. and {Umana}, G. and {Valenziano}, L. and {Valiviita}, J. and {Van Tent}, B. and {Vielva}, P. and {Villa}, F. and {Wade}, L.~A. and {Wandelt}, B.~D. and {Wehus}, I.~K.},
        title = "{Planck intermediate results. XXXII. The relative orientation between the magnetic field and structures traced by interstellar dust}",
      journal = {\aap},
         year = 2016,
        month = feb,
       volume = {586},
          eid = {A135},
        pages = {A135},
          doi = {10.1051/0004-6361/201425044},
archivePrefix = {arXiv},
       eprint = {1409.6728},
 primaryClass = {astro-ph.GA},
       adsurl = {https://ui.adsabs.harvard.edu/abs/2016A&A...586A.135P}
}

@ARTICLE{2016A&A...586A.136P,
       author = {{Planck Collaboration} and {Ade}, P.~A.~R. and {Aghanim}, N. and {Alves}, M.~I.~R. and {Arnaud}, M. and {Arzoumanian}, D. and {Aumont}, J. and {Baccigalupi}, C. and {Banday}, A.~J. and {Barreiro}, R.~B. and {Bartolo}, N. and {Battaner}, E. and {Benabed}, K. and {Benoit-L{\'e}vy}, A. and {Bernard}, J.-P. and {Bern{\'e}}, O. and {Bersanelli}, M. and {Bielewicz}, P. and {Bonaldi}, A. and {Bonavera}, L. and {Bond}, J.~R. and {Borrill}, J. and {Bouchet}, F.~R. and {Boulanger}, F. and {Bracco}, A. and {Burigana}, C. and {Calabrese}, E. and {Cardoso}, J.-F. and {Catalano}, A. and {Chamballu}, A. and {Chiang}, H.~C. and {Christensen}, P.~R. and {Clements}, D.~L. and {Colombi}, S. and {Colombo}, L.~P.~L. and {Combet}, C. and {Couchot}, F. and {Crill}, B.~P. and {Curto}, A. and {Cuttaia}, F. and {Danese}, L. and {Davies}, R.~D. and {Davis}, R.~J. and {de Bernardis}, P. and {de Rosa}, A. and {de Zotti}, G. and {Delabrouille}, J. and {Dickinson}, C. and {Diego}, J.~M. and {Donzelli}, S. and {Dor{\'e}}, O. and {Douspis}, M. and {Ducout}, A. and {Dupac}, X. and {Elsner}, F. and {En{\ss}lin}, T.~A. and {Eriksen}, H.~K. and {Falgarone}, E. and {Ferri{\`e}re}, K. and {Finelli}, F. and {Forni}, O. and {Frailis}, M. and {Fraisse}, A.~A. and {Franceschi}, E. and {Frejsel}, A. and {Galeotta}, S. and {Galli}, S. and {Ganga}, K. and {Ghosh}, T. and {Giard}, M. and {Giraud-H{\'e}raud}, Y. and {Gjerl{\o}w}, E. and {Gonz{\'a}lez-Nuevo}, J. and {G{\'o}rski}, K.~M. and {Gregorio}, A. and {Gruppuso}, A. and {Guillet}, V. and {Hansen}, F.~K. and {Hanson}, D. and {Harrison}, D.~L. and {Hern{\'a}ndez-Monteagudo}, C. and {Herranz}, D. and {Hildebrandt}, S.~R. and {Hivon}, E. and {Hobson}, M. and {Holmes}, W.~A. and {Huffenberger}, K.~M. and {Hurier}, G. and {Jaffe}, A.~H. and {Jaffe}, T.~R. and {Jones}, W.~C. and {Juvela}, M. and {Keskitalo}, R. and {Kisner}, T.~S. and {Knoche}, J. and {Kunz}, M. and {Kurki-Suonio}, H. and {Lagache}, G. and {Lamarre}, J.-M. and {Lasenby}, A. and {Lawrence}, C.~R. and {Leonardi}, R. and {Levrier}, F. and {Liguori}, M. and {Lilje}, P.~B. and {Linden-V{\o}rnle}, M. and {L{\'o}pez-Caniego}, M. and {Lubin}, P.~M. and {Mac{\'\i}as-P{\'e}rez}, J.~F. and {Maffei}, B. and {Mandolesi}, N. and {Mangilli}, A. and {Maris}, M. and {Martin}, P.~G. and {Mart{\'\i}nez-Gonz{\'a}lez}, E. and {Masi}, S. and {Matarrese}, S. and {Mazzotta}, P. and {Melchiorri}, A. and {Mendes}, L. and {Mennella}, A. and {Migliaccio}, M. and {Mitra}, S. and {Miville-Desch{\^e}nes}, M.-A. and {Moneti}, A. and {Montier}, L. and {Morgante}, G. and {Mortlock}, D. and {Munshi}, D. and {Murphy}, J.~A. and {Naselsky}, P. and {Nati}, F. and {Natoli}, P. and {N{\o}rgaard-Nielsen}, H.~U. and {Noviello}, F. and {Novikov}, D. and {Novikov}, I. and {Oppermann}, N. and {Pagano}, L. and {Pajot}, F. and {Paladini}, R. and {Paoletti}, D. and {Pasian}, F. and {Perrotta}, F. and {Pettorino}, V. and {Piacentini}, F. and {Piat}, M. and {Pierpaoli}, E. and {Pietrobon}, D. and {Plaszczynski}, S. and {Pointecouteau}, E. and {Polenta}, G. and {Pratt}, G.~W. and {Puget}, J.-L. and {Rachen}, J.~P. and {Rebolo}, R. and {Reinecke}, M. and {Remazeilles}, M. and {Renault}, C. and {Renzi}, A. and {Ricciardi}, S. and {Ristorcelli}, I. and {Rocha}, G. and {Rosset}, C. and {Rossetti}, M. and {Roudier}, G. and {Rubi{\~n}o-Mart{\'\i}n}, J.~A. and {Rusholme}, B. and {Sandri}, M. and {Savelainen}, M. and {Savini}, G. and {Scott}, D. and {Soler}, J.~D. and {Stolyarov}, V. and {Sutton}, D. and {Suur-Uski}, A.-S. and {Sygnet}, J.-F. and {Tauber}, J.~A. and {Terenzi}, L. and {Toffolatti}, L. and {Tomasi}, M. and {Tristram}, M. and {Tucci}, M. and {Valenziano}, L. and {Valiviita}, J. and {Van Tent}, B. and {Vielva}, P. and {Villa}, F. and {Wade}, L.~A. and {Wandelt}, B.~D. and {Yvon}, D. and {Zacchei}, A. and {Zonca}, A.},
        title = "{Planck intermediate results. XXXIII. Signature of the magnetic field geometry of interstellar filaments in dust polarization maps}",
      journal = {\aap},
         year = 2016,
        month = feb,
       volume = {586},
          eid = {A136},
        pages = {A136},
          doi = {10.1051/0004-6361/201425305},
archivePrefix = {arXiv},
       eprint = {1411.2271},
 primaryClass = {astro-ph.GA},
       adsurl = {https://ui.adsabs.harvard.edu/abs/2016A&A...586A.136P}
}

@ARTICLE{2017A&A...607A...2S,
       author = {{Soler}, J.~D. and {Hennebelle}, P.},
        title = "{What are we learning from the relative orientation between density structures and the magnetic field in molecular clouds?}",
      journal = {\aap},
         year = 2017,
        month = oct,
       volume = {607},
          eid = {A2},
        pages = {A2},
          doi = {10.1051/0004-6361/201731049},
archivePrefix = {arXiv},
       eprint = {1705.00477},
 primaryClass = {astro-ph.GA},
       adsurl = {https://ui.adsabs.harvard.edu/abs/2017A&A...607A...2S}
}

@ARTICLE{2016A&A...586A.138P,
       author = {{Planck Collaboration} and {Ade}, P.~A.~R. and {Aghanim}, N. and {Alves}, M.~I.~R. and {Arnaud}, M. and {Arzoumanian}, D. and {Ashdown}, M. and {Aumont}, J. and {Baccigalupi}, C. and {Banday}, A.~J. and {Barreiro}, R.~B. and {Bartolo}, N. and {Battaner}, E. and {Benabed}, K. and {Beno{\^\i}t}, A. and {Benoit-L{\'e}vy}, A. and {Bernard}, J.-P. and {Bersanelli}, M. and {Bielewicz}, P. and {Bock}, J.~J. and {Bonavera}, L. and {Bond}, J.~R. and {Borrill}, J. and {Bouchet}, F.~R. and {Boulanger}, F. and {Bracco}, A. and {Burigana}, C. and {Calabrese}, E. and {Cardoso}, J.-F. and {Catalano}, A. and {Chiang}, H.~C. and {Christensen}, P.~R. and {Colombo}, L.~P.~L. and {Combet}, C. and {Couchot}, F. and {Crill}, B.~P. and {Curto}, A. and {Cuttaia}, F. and {Danese}, L. and {Davies}, R.~D. and {Davis}, R.~J. and {de Bernardis}, P. and {de Rosa}, A. and {de Zotti}, G. and {Delabrouille}, J. and {Dickinson}, C. and {Diego}, J.~M. and {Dole}, H. and {Donzelli}, S. and {Dor{\'e}}, O. and {Douspis}, M. and {Ducout}, A. and {Dupac}, X. and {Efstathiou}, G. and {Elsner}, F. and {En{\ss}lin}, T.~A. and {Eriksen}, H.~K. and {Falceta-Gon{\c{c}}alves}, D. and {Falgarone}, E. and {Ferri{\`e}re}, K. and {Finelli}, F. and {Forni}, O. and {Frailis}, M. and {Fraisse}, A.~A. and {Franceschi}, E. and {Frejsel}, A. and {Galeotta}, S. and {Galli}, S. and {Ganga}, K. and {Ghosh}, T. and {Giard}, M. and {Gjerl{\o}w}, E. and {Gonz{\'a}lez-Nuevo}, J. and {G{\'o}rski}, K.~M. and {Gregorio}, A. and {Gruppuso}, A. and {Gudmundsson}, J.~E. and {Guillet}, V. and {Harrison}, D.~L. and {Helou}, G. and {Hennebelle}, P. and {Henrot-Versill{\'e}}, S. and {Hern{\'a}ndez-Monteagudo}, C. and {Herranz}, D. and {Hildebrandt}, S.~R. and {Hivon}, E. and {Holmes}, W.~A. and {Hornstrup}, A. and {Huffenberger}, K.~M. and {Hurier}, G. and {Jaffe}, A.~H. and {Jaffe}, T.~R. and {Jones}, W.~C. and {Juvela}, M. and {Keih{\"a}nen}, E. and {Keskitalo}, R. and {Kisner}, T.~S. and {Knoche}, J. and {Kunz}, M. and {Kurki-Suonio}, H. and {Lagache}, G. and {Lamarre}, J.-M. and {Lasenby}, A. and {Lattanzi}, M. and {Lawrence}, C.~R. and {Leonardi}, R. and {Levrier}, F. and {Liguori}, M. and {Lilje}, P.~B. and {Linden-V{\o}rnle}, M. and {L{\'o}pez-Caniego}, M. and {Lubin}, P.~M. and {Mac{\'\i}as-P{\'e}rez}, J.~F. and {Maino}, D. and {Mandolesi}, N. and {Mangilli}, A. and {Maris}, M. and {Martin}, P.~G. and {Mart{\'\i}nez-Gonz{\'a}lez}, E. and {Masi}, S. and {Matarrese}, S. and {Melchiorri}, A. and {Mendes}, L. and {Mennella}, A. and {Migliaccio}, M. and {Miville-Desch{\^e}nes}, M.-A. and {Moneti}, A. and {Montier}, L. and {Morgante}, G. and {Mortlock}, D. and {Munshi}, D. and {Murphy}, J.~A. and {Naselsky}, P. and {Nati}, F. and {Netterfield}, C.~B. and {Noviello}, F. and {Novikov}, D. and {Novikov}, I. and {Oppermann}, N. and {Oxborrow}, C.~A. and {Pagano}, L. and {Pajot}, F. and {Paladini}, R. and {Paoletti}, D. and {Pasian}, F. and {Perotto}, L. and {Pettorino}, V. and {Piacentini}, F. and {Piat}, M. and {Pierpaoli}, E. and {Pietrobon}, D. and {Plaszczynski}, S. and {Pointecouteau}, E. and {Polenta}, G. and {Ponthieu}, N. and {Pratt}, G.~W. and {Prunet}, S. and {Puget}, J.-L. and {Rachen}, J.~P. and {Reinecke}, M. and {Remazeilles}, M. and {Renault}, C. and {Renzi}, A. and {Ristorcelli}, I. and {Rocha}, G. and {Rossetti}, M. and {Roudier}, G. and {Rubi{\~n}o-Mart{\'\i}n}, J.~A. and {Rusholme}, B. and {Sandri}, M. and {Santos}, D. and {Savelainen}, M. and {Savini}, G. and {Scott}, D. and {Soler}, J.~D. and {Stolyarov}, V. and {Sudiwala}, R. and {Sutton}, D. and {Suur-Uski}, A.-S. and {Sygnet}, J.-F. and {Tauber}, J.~A. and {Terenzi}, L. and {Toffolatti}, L. and {Tomasi}, M. and {Tristram}, M. and {Tucci}, M. and {Umana}, G. and {Valenziano}, L. and {Valiviita}, J. and {Van Tent}, B. and {Vielva}, P. and {Villa}, F. and {Wade}, L.~A. and {Wandelt}, B.~D. and {Wehus}, I.~K. and {Ysard}, N. and {Yvon}, D. and {Zonca}, A.},
        title = "{Planck intermediate results. XXXV. Probing the role of the magnetic field in the formation of structure in molecular clouds}",
      journal = {\aap},
         year = 2016,
        month = feb,
       volume = {586},
          eid = {A138},
        pages = {A138},
          doi = {10.1051/0004-6361/201525896},
archivePrefix = {arXiv},
       eprint = {1502.04123},
 primaryClass = {astro-ph.GA},
       adsurl = {https://ui.adsabs.harvard.edu/abs/2016A&A...586A.138P}
}

@ARTICLE{2021ApJ...909..199O,
       author = {{Olguin}, Fernando A. and {Sanhueza}, Patricio and {Guzm{\'a}n}, Andr{\'e}s E. and {Lu}, Xing and {Saigo}, Kazuya and {Zhang}, Qizhou and {Silva}, Andrea and {Chen}, Huei-Ru Vivien and {Li}, Shanghuo and {Ohashi}, Satoshi and {Nakamura}, Fumitaka and {Sakai}, Takeshi and {Wu}, Benjamin},
        title = "{Digging into the Interior of Hot Cores with ALMA (DIHCA). I. Dissecting the High-mass Star-forming Core G335.579-0.292 MM1}",
      journal = {\apj},
         year = 2021,
        month = mar,
       volume = {909},
       number = {2},
          eid = {199},
        pages = {199},
          doi = {10.3847/1538-4357/abde3f},
archivePrefix = {arXiv},
       eprint = {2101.08284},
 primaryClass = {astro-ph.GA},
       adsurl = {https://ui.adsabs.harvard.edu/abs/2021ApJ...909..199O}
}

@ARTICLE{2026AJ....171...50H,
       author = {{Hwang}, Jihye and {Sanhueza}, Patricio and {Girart}, Josep Miquel and {Stephens}, Ian W. and {Beltr{\'a}n}, Maria T. and {Law}, Chi Yan and {Zhang}, Qizhou and {Liu}, Junhao and {Cort{\'e}s}, Paulo and {Olguin}, Fernando A. and {Koch}, Patrick M. and {Nakamura}, Fumitaka and {Saha}, Piyali and {Wang}, Jia-Wei and {Xu}, Fengwei and {Beuther}, Henrik and {Morii}, Kaho and {Fern{\'a}ndez L{\'o}pez}, Manuel and {Jiao}, Wenyu and {Kim}, Kee-Tae and {Li}, Shanghuo and {Zapata}, Luis A. and {Kim}, Jongsoo and {Choudhury}, Spandan and {Cheng}, Yu and {Pattle}, Kate and {Eswaraiah}, Chakali and {Sandhyarani}, Panigrahy and {Dewangan}, L.~K. and {Jadhav}, O.~R.},
        title = "{Magnetic Fields in Massive Star-forming Regions (MagMaR). VI. Magnetic Field Dragging in the Filamentary High-mass Star-forming Region G35.20─0.74N Due to Gravity}",
      journal = {\aj},
         year = 2026,
        month = jan,
       volume = {171},
       number = {1},
          eid = {50},
        pages = {50},
          doi = {10.3847/1538-3881/ae18c9},
archivePrefix = {arXiv},
       eprint = {2510.25078},
 primaryClass = {astro-ph.GA},
       adsurl = {https://ui.adsabs.harvard.edu/abs/2026AJ....171...50H}
}

\noindent\rule{\linewidth}{0.4pt}

\hspace{-5mm}$^{1}$Departamento de Astronom\'{i}a, Universidad de Concepci\'{o}n, Casilla 160-C, Concepci\'{o}n, Chile\\
\hspace{-1.3mm}$^{2}$Department of Astronomy, School of Science, The University of Tokyo, 7-3-1 Hongo, Bunkyo, Tokyo 113-0033, Japan\\
$^{3}$School of Astronomy and Space Science, Nanjing University, Nanjing, China\\
$^{4}$Key Laboratory of Modern Astronomy and Astrophysics (Nanjing University), Ministry of Education, Nanjing, China\\
$^{5}$Academia Sinica Institute of Astronomy and Astrophysics, No.1, Sec. 4., Roosevelt Road, Taipei 10617, Taiwan\\
$^{6}$National Astronomical Observatory of Japan, National Institutes of Natural Sciences, 2-21-1 Osawa, Mitaka, Tokyo 181-8588, Japan\\
$^{7}$Center for Gravitational Physics, Yukawa Institute for Theoretical Physics, Kyoto University, Kitashirakawa Oiwakecho, Sakyo-ku,Kyoto 606-8502, Japan\\
$^{8}$Department of Astronomy, University of Florida, PO Box 112055, USA\\

             Corresponding author: A. Koley \\
              
             \email{atanuphysics15@gmail.com}\\
            
\noindent\rule{\linewidth}{0.4pt}

\begin{appendix}

\onecolumn

\appendix

\section{Systematic velocity ($V_{\text{LSR}}$) of the region}\label{appendix_0}

We calculated the systematic velocity ($V_{\text{LSR}}$) using the following formula:

\begin{equation}\label{Eqn:A1}
   V_{\text{LSR}} = \frac{\sum_{i}^{} I_{\text{i}}{v}_{\text{i}}}{\sum_{i}^{} I_{\text{i}}}.
\end{equation}

\noindent Here $I_{\text{i}}$ and $v_{\text{i}}$ are the intensity and velocity of the $i^{\text{th}}$ channel. In Fig. \ref{Eqn:A1} we show the average spectra of the HN$^{13}$C (3$-$2) and H$^{13}$CO$^{+}$ (3$-$2) spectral lines. The $V_{\text{LSR}}$ obtained from these two lines are $-$47.0 km s$^{-1}$ and $-$46.2 km s$^{-1}$, respectively. The average $V_{\text{LSR}}$ obtained from the two spectral lines is $-$46.6 km s$^{-1}$.

\begin{figure*}
        \centering 
        \includegraphics[width=3.4in,height=2.8in,angle=0]{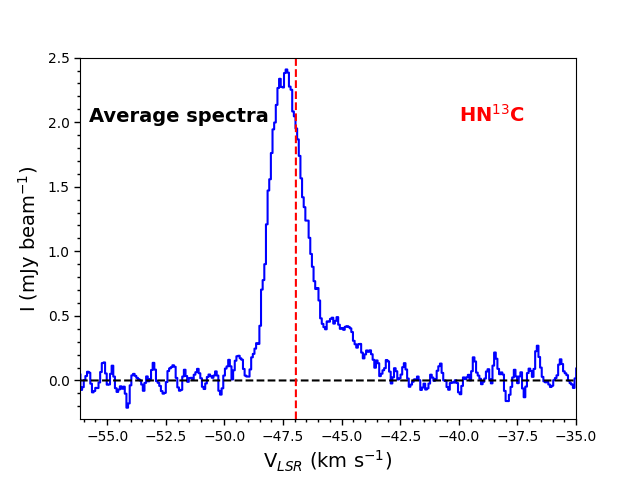}\includegraphics[width=3.4in,height=2.8in,angle=0]{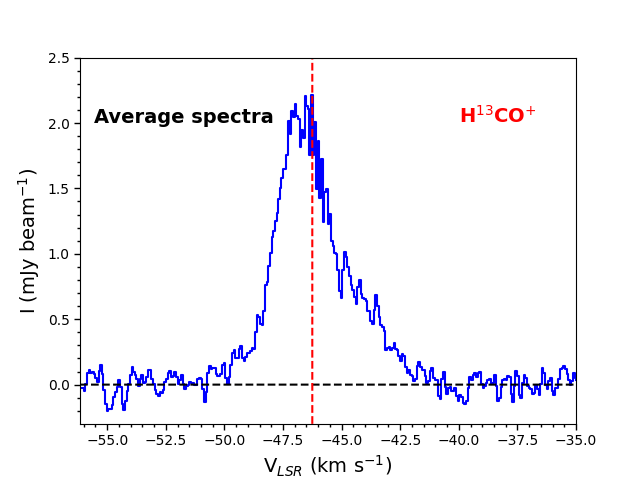}\\
        \caption{\textit{Left}: Average HN$^{13}$C (3$-$2) spectra (blue line) in the observed region of the G327.29 protocluster. The horizontal dashed black line is at 0.0 mJy beam$^{-1}$, and the vertical dashed red line is the systematic velocity ($V_{\text{sys}}$) of the region: $\sim$ $-$ 47.0 km s$^{-1}$. The $V_{\text{sys}}$ of this region is blueshifted from the entire G327.29 protocluster, which is at $\sim$ $-$ 45.0 km s$^{-1}$ \citep{2022A&A...662A...8M}.  \textit{Right}: Same but for the  H$^{13}$CO$^{+}$ (3$-$2) line. The $V_{\text{sys}}$ obtained from this line is $\sim$ $-$ 46.2 km s$^{-1}$.}
        \label{fig:figA1}
\end{figure*}


\section{Spatial distribution of the number of components and the velocity dispersion in HN$^{13}$C and H$^{13}$CO$^{+}$ spectral lines}\label{appendix_1}

\begin{figure*}
        \centering 
        \includegraphics[width=3.87in,height=3.4in,angle=0]{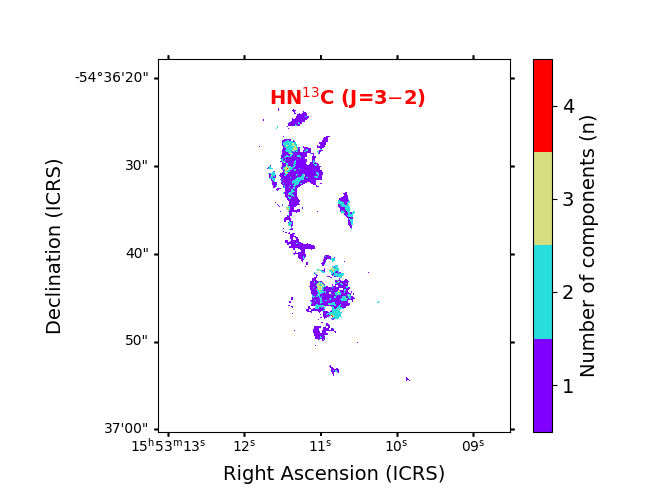}\includegraphics[width=3.6in,height=3.4in,angle=0]{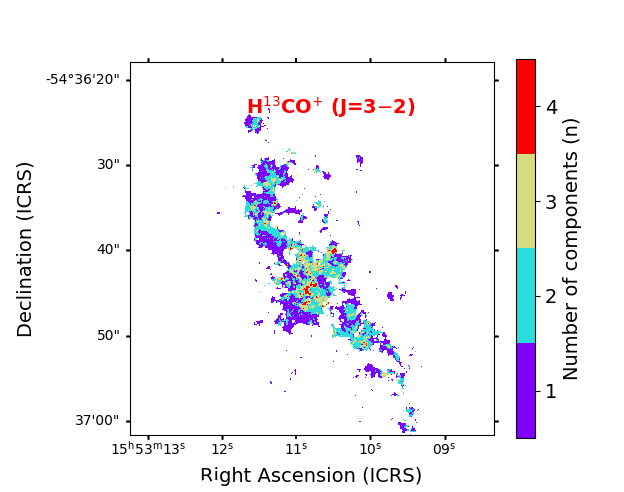}\\
        \caption{Distribution of the number of components of the HN$^{13}$C (J=3$-$2) and H$^{13}$CO$^{+}$ (J=3$-$2) lines. }
        \label{fig:fig2}
\end{figure*}

\begin{figure*}
        \centering 
        \includegraphics[width=3.8in,height=3.4in,angle=0]{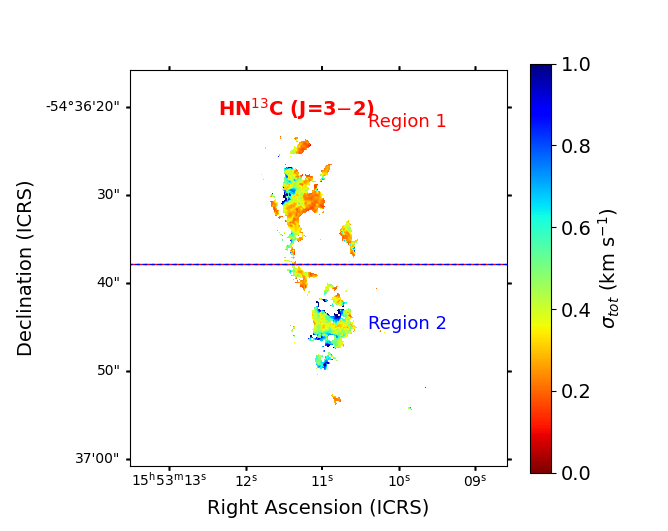}\includegraphics[width=3.8in,height=3.4in,angle=0]{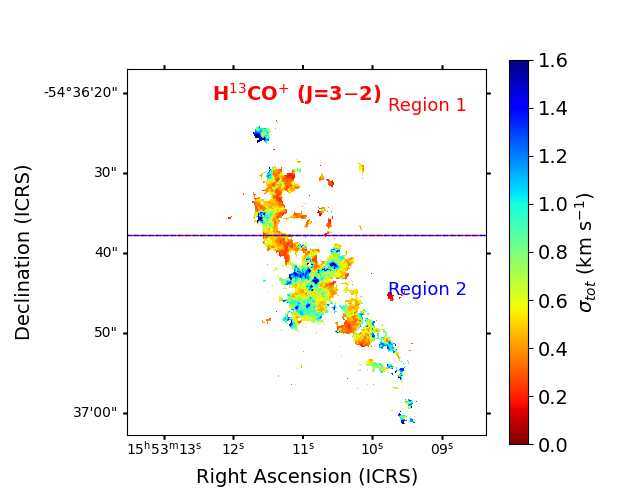}\\
        \caption{Velocity dispersion ($\sigma_{\text{tot}}$) maps of HN$^{13}$C (3$-$2) and H$^{13}$CO$^{+}$ (3$-$2) lines obtained from the pixel-wise fitting using the \texttt{Gausspy+} module. \texttt{Region 1} and \texttt{Region 2} are analysed independently to estimate the magnetic field.}
        \label{fig:fig3}
\end{figure*}

The spatial distributions of the number of components ($n$) for the HN$^{13}$C and H$^{13}$CO$^{+}$ spectral lines are shown in Fig.~\ref{fig:fig2}. These maps clearly indicate that multiple components are present in most pixels. This scenario is quite common in protocluster environments, as indicated by previous studies \citep{2024A&A...689A..74A,2025A&A...696A.202S,2025A&A...702A.133K,2025arXiv251003447S}. We also show the spatial distribution of the total velocity dispersion, $\sigma_{\rm tot}$, for the two spectral lines, HN$^{13}$C and H$^{13}$CO$^{+}$ in Fig. \ref{fig:fig3}. The entire mapped area is divided into two regions, labelled \texttt{Region~1} and \texttt{Region~2}. This division is necessary to estimate the magnetic-field strength independently in the two regions.

\section{Length ($L$) and radius ($R$) of the filament}\label{appendix_2_0}

\begin{figure*}[!ht]
        \centering 
        \includegraphics[width=2.5in,height=2.45in,angle=0]{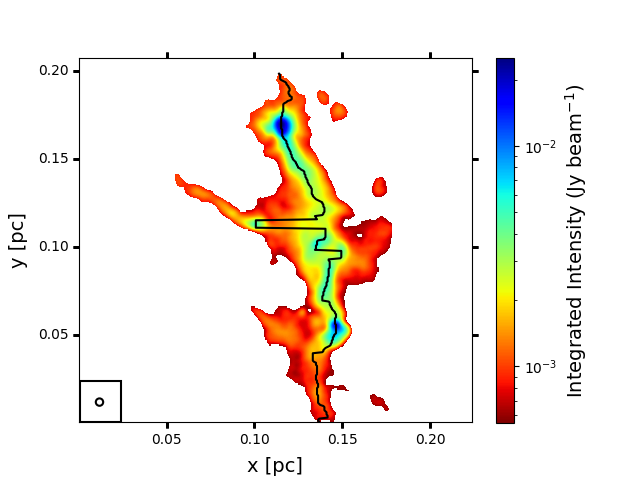}\includegraphics[width=2.5in,height=2.45in,angle=0]{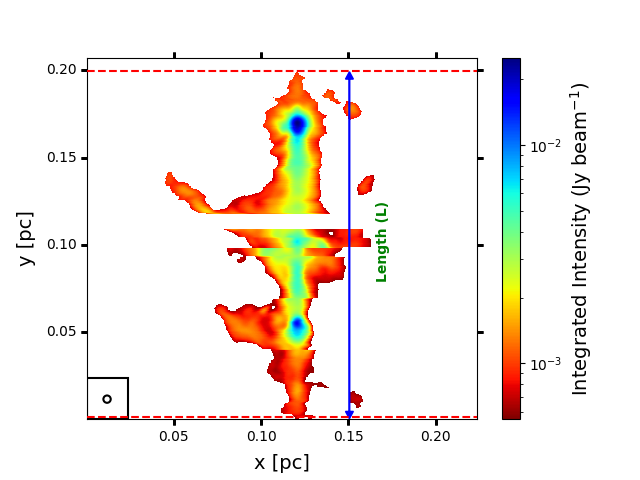}\includegraphics[width=2.5in,height=2.5in,angle=0]{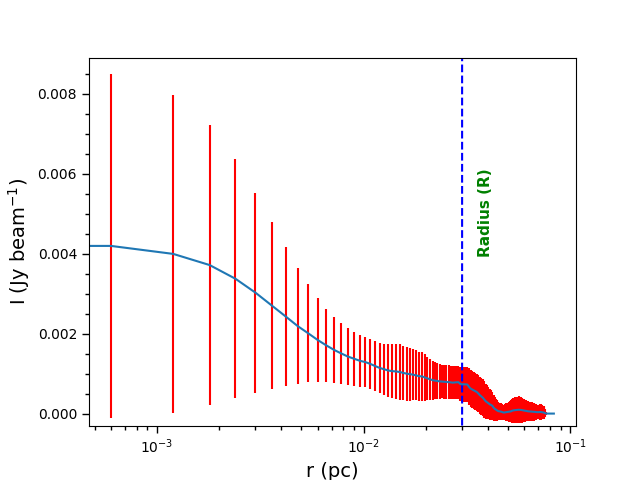}
        \caption{\textit{Left}: Observed filament in \texttt{Region 1}. The solid black line is the ridgeline of the filament. \textit{Middle}: Same filament after the ridgeline is vertically straightened. Due to certain jumps in the middle of the ridge line, those rows were masked. The length of the filament (L) is $\sim$ 0.20 pc. \textit{Right}: Radial profile of the integrated intensity (I) of the filament. The solid light sky-blue and red lines are the mean and dispersion ($\sigma$) values at each concentric cylindrical shell. We set the radius (R) to $\sim$ 0.03 pc because beyond that I decreases sharply and becomes flat. This is indicated by the vertical dashed blue line. This phenomenon is due to the mass distribution not being symmetrical on both sides of the ridgeline of the filament.}
        \label{fig:fig4}
\end{figure*}

Figure~\ref{fig:fig4} shows the filament and its ridgeline as well as the filament after the ridgeline is vertically straightened. To straighten the filament, we determined the pixel of maximum intensity in each row. We then shifted each row so that these peak-intensity pixels are aligned along a common column. We also excluded a few rows from the analysis where there is a significant jump in the positions of the maximum-intensity pixels compared to neighbouring rows. We then derived the filament length, $L = 0.20$~pc. The figure also shows the radial profile of the integrated intensity ($I$); the profile drops sharply beyond $\sim$0.03~pc and then flattens, defining a filament radius $R = 0.03$~pc. The uncertainties in $L$ and $R$ are taken as 20\% of their mean values, consistent with the adopted uncertainty in $d$. This is because we derived the physical length scale ($L$) and radius ($R$) from the corresponding angular sizes measured on the plane of the sky using $L = \theta_\text{L} d$ and $R = \theta_\text{R} d$, where $\theta_\text{L}$ and $\theta_\text{R}$ are the measured angular sizes along $L$ and $R$ and $d$ is the source distance. Since the uncertainty in the angular measurements is negligible compared to that in the distance, we propagated only the distance uncertainty when estimating the errors in $L$ and $R$. The adopted distance to the source is $2.5 \pm 0.5$~kpc, corresponding to a $20\%$ uncertainty. Consequently, the derived uncertainties in both $L$ and $R$ are also $20\%$.

\section{Virial parameters in this region}\label{appendix_4}

The virial parameters and its associated error for \texttt{Region 1} are

\begin{eqnarray}\label{eqn25}
\alpha_{\text{vir}} & = &\biggl (\frac{2E_{\text{k}} + E_{\text{B}}}{|E_{\text{G}}|} \biggl) \nonumber \\
                    & = & \biggl(\frac{3\sigma_{\text{tot}}^{2}L}{GM} +  \frac{B_\text{{tot}}^{2}R^{2}L^{2}}{8GM^{2}}\biggl),
\end{eqnarray}

\begin{eqnarray}\label{eqn26}
\Delta\alpha_{\text{vir}}&\approx&\Biggl[\biggl(\frac{3\sigma_{\text{tot}}^{2}L}{GM}\biggl)^{2}\biggl\{ \biggl(\frac{2\Delta\sigma_{\text{tot}}}{\sigma_{\text{tot}}}\biggl)^{2} ~+~ \biggl( \frac{\Delta S^{\text{1.2mm}}_{\text{int}}}{S^{\text{1.2mm}}_{\text{int}}}\biggl)^{2} ~+~  \biggl( \frac{\Delta d}{d}\biggl)^{2} ~+~ \nonumber \\ 
&  &  \biggl(\frac{\Delta B_{\text{1.2mm}}(T_{\text{d}},\nu)}{B_{\text{1.2mm}} (T_{\text{d}},\nu\Huge)}\biggl)^{2} \biggl\}  ~+~\biggl(\frac{B_\text{{tot}}^{2}R^{2}L^{2}}{8GM^{2}}\biggl)^{2}\biggl\{ \biggl(\frac{2\Delta \sigma_\text{{nth}}}{\sigma_\text{{nth}}}\biggl)^{2} + \nonumber \\
&  & \biggl(\frac{\Delta S^{\text{1.2mm}}_{\text{int}}}{S^{\text{1.2mm}}_{\text{int}}}\biggl)^{2} ~+~ \biggl( \frac{\Delta d}{d}\biggl)^{2} ~+~ \biggl(\frac{\Delta B_{\text{1.2mm}}(T_{\text{d}},\nu)}{B_{\text{1.2mm}} (T_{\text{d}},\nu\Huge)}\biggl)^{2}\biggl\} \Biggl]^{1/2}.
\end{eqnarray}

\noindent The virial parameter and its associated uncertainty for \texttt{Region 2} are 

\begin{eqnarray}\label{eqn30}
\alpha_{\text{vir}} & = &\biggl (\frac{2E_{\text{k}} + E_{\text{B}}}{|E_{\text{G}}|} \biggl) \nonumber \\
                    & = & \biggl(\frac{3\sigma_{\text{tot}}^{2}R}{GM} +  \frac{B_{\text{tot}}^{2}R^{4}}{6GM^{2}}\biggl),
\end{eqnarray}

\begin{eqnarray}\label{eqn31}
\Delta\alpha_{\text{vir}}&\approx&\Biggl[\biggl(\frac{3\sigma_{\text{tot}}^{2}R}{GM}\biggl)^{2}\biggl\{ \biggl(\frac{2\Delta\sigma_{\text{tot}}}{\sigma_{\text{tot}}}\biggl)^{2} +~\biggl( \frac{\Delta d}{d}\biggl)^{2} + ~\biggl( \frac{\Delta S^{\text{1.2mm}}_{\text{int}}}{S^{\text{1.2mm}}_{\text{int}}}\biggl)^{2} +   \nonumber \\
 &  & \biggl(\frac{\Delta B_{\text{1.2mm}} \Huge(T_{\text{d}},\nu)}{B_{\text{1.2mm}} (T_{\text{d}},\nu\Huge)}\biggl)^{2} \biggl\}~+~ \biggl(\frac{B_{\text{tot}}^{2}R^{4}}{6GM^{2}}\biggl)^{2}\biggl\{ \biggl(\frac{2\Delta \sigma_{\text{nth}}}{\sigma_{\text{nth}}}\biggl)^{2} +  \nonumber \\
 &  & \biggl( \frac{\Delta S^{\text{1.2mm}}_{\text{int}}}{S^{\text{1.2mm}}_{\text{int}}}\biggl)^{2} ~+~\biggl( \frac{\Delta d}{d}\biggl)^{2}~+~\biggl(\frac{\Delta B_{\text{1.2mm}} \Huge(T_{\text{d}},\nu)}{B_{\text{1.2mm}} (T_{\text{d}},\nu\Huge)}\biggl)^{2} \biggl\}\Biggl]^{1/2}. 
\end{eqnarray}

In Eqs. \ref{eqn33}--\ref{eqn39}, $R_\text{core}$ denotes the core radius, $S^{\text{1.2mm}}_{\text{int,core}}$ is the integrated intensity of the core, $\sigma_\text{tot,core}$ is the total velocity dispersion towards the cores derived from the two spectral lines, $M_\mathrm{core}$ is the core mass, $\rho_\text{core}$ is the core density, $B_\text{core}$ is the magnetic field strength in the core  obtained assuming a value of $p$ and $\rho_\text{sur}$, $B_\text{sur}$, and $\sigma_{\text{nth, sur}}$ are the surrounding density, the magnetic field, and the non-thermal velocity dispersion where cores are located. The surrounding region falls in either \texttt{Region 1} or \texttt{Region 2}. We used Eqs. \ref{eqn33} and \ref{eqn34} to calculate the virial parameters and their associated uncertainties for the cores without taking into account the effect of the magnetic field. When the magnetic field strength within the cores is assumed to be the same as that in the surrounding environment (i.e., $p=0.0$), we used Eqs. \ref{eqn35} and \ref{eqn36}. Finally, Eqs. \ref{eqn37}--\ref{eqn39} were used when we assumed that the magnetic field strength increases from the surrounding environment to the core scales, corresponding to $p=0.5$.

\begin{eqnarray}\label{eqn33}
\alpha_{\text{vir,turb}} = \Biggl(\frac{5\sigma_{\text{tot,core}}^{2}R_{\text{core}}}{GM_\text{core}}\Biggl)~,
\end{eqnarray}

\begin{eqnarray}\label{eqn34}
 \Delta\alpha_{\text{vir,turb}}&\approx&\Biggl[\biggl(\frac{5\sigma_{\text{tot,core}}^{2}R_{\text{core}}}{GM_\text{core}}\biggl)^{2}\biggl\{ \biggl(\frac{2\Delta\sigma_{\text{tot,core}}}{\sigma_{\text{tot,core}}}\biggl)^{2} +~\biggl( \frac{\Delta R_\text{core}}{R_\text{core}}\biggl)^{2} +   \nonumber \\
 &  &  + ~\biggl( \frac{\Delta S^{\text{1.2mm}}_{\text{int,core}}}{S^{\text{1.2mm}}_{\text{int,core}}}\biggl)^{2} ~+~ \biggl(\frac{\Delta B_{\text{1.2mm}} \Huge(T_{\text{core}},\nu)_\text{{}}}{B_{\text{1.2mm}} \Huge(T_{\text{core}},\nu\Huge)_\text{}}\biggl)^{2} \biggl\}\Biggl]^{1/2}~,    
\end{eqnarray}

\begin{eqnarray}\label{eqn35}
\alpha_{\text{vir,turb+mag}} = \Biggl(\frac{5\sigma_{\text{tot,core}}^{2}R_{\text{core}}}{GM_\text{core}}~+~\frac{5B_{\text{core}}^{2}R_{\text{core}}^{4}}{18GM_\text{core}^{2}}\Biggl)~,
\end{eqnarray}

\begin{eqnarray}\label{eqn36}
 \Delta\alpha_{\text{vir, turb+mag}}&\approx&\Biggl[\biggl(\frac{5\sigma_{\text{tot,core}}^{2}R_{\text{core}}}{GM_\text{core}}\biggl)^{2}\biggl\{ \biggl(\frac{2\Delta\sigma_{\text{tot,core}}}{\sigma_{\text{tot,core}}}\biggl)^{2} +~\biggl( \frac{\Delta R_\text{core}}{R_\text{core}}\biggl)^{2} +   \nonumber \\
 &  &  + ~\biggl( \frac{\Delta S^{\text{1.2mm}}_{\text{int,core}}}{S^{\text{1.2mm}}_{\text{int,core}}}\biggl)^{2} ~+~ \biggl(\frac{\Delta B_{\text{1.2mm}} \Huge(T_{\text{d}},\nu)_\text{{}}}{B_{\text{1.2mm}} \Huge(T_{\text{d}},\nu\Huge)_\text{}}\biggl)^{2} \biggl\}  \nonumber \\  &  &~+~ \biggl( \frac{5B_{\text{core}}^{2}R_{\text{core}}^{4}}{18GM_\text{core}^{2}}\biggl)^{2} \biggl\{ \biggl(\frac{2\Delta B_{\text{core}}}{B_{\text{core}}}\biggl)^{2}  
  + ~\biggl( \frac{\Delta S^{\text{1.2mm}}_{\text{int,core}}}{S^{\text{1.2mm}}_{\text{int,core}}}\biggl)^{2} \nonumber \\
  &  &  ~+~ \biggl(\frac{\Delta B_{\text{1.2mm}} \Huge(T_{\text{core}},\nu)_\text{{}}}{B_{\text{1.2mm}} \Huge(T_{\text{core}},\nu\Huge)_\text{}}\biggl)^{2} \biggl\}                                   \Biggl]^{1/2} ,    
\end{eqnarray}

\begin{eqnarray}\label{eqn37}
\alpha_{\text{vir,turb+mag}} = \Biggl(\frac{5\sigma_{\text{tot,core}}^{2}R_{\text{core}}}{GM_\text{core}}~+~\frac{5B_{\text{core}}^{2}R_{\text{core}}^{4}}{18GM_\text{core}^{2}}\Biggl)~,
\end{eqnarray}

\begin{eqnarray}\label{eqn38}
B_{\text{core}} =  B_\text{{sur}} \biggl(\frac{\rho_\text{core}}{\rho_\text{sur}} \biggl)^\frac{1}{2} = 1.25\sqrt{4\pi\rho_{\text{core}}}~\sigma_{\text{nth, sur}}~\biggl [ \frac{<B_\text{t}^{2}>}{<B_\text{pos}^2>}  \biggl ]^{-1/2}~ ,
\end{eqnarray}

\begin{eqnarray}\label{eqn39}
\Delta\alpha_{\text{vir,turb+mag}}&\approx&\Biggl[\biggl(\frac{5\sigma_{\text{tot,core}}^{2}R_{\text{core}}}{GM_\text{core}}\biggl)^{2}\biggl\{ \biggl(\frac{2\Delta\sigma_{\text{tot,core}}}{\sigma_{\text{tot,core}}}\biggl)^{2} +~\biggl( \frac{\Delta R_\text{core}}{R_\text{core}}\biggl)^{2} +   \nonumber \\
 &  &  + ~\biggl( \frac{\Delta S^{\text{1.2mm}}_{\text{int,core}}}{S^{\text{1.2mm}}_{\text{int,core}}}\biggl)^{2} ~+~ \biggl(\frac{\Delta B_{\text{1.2mm}} \Huge(T_{\text{core}},\nu)_\text{{}}}{B_{\text{1.2mm}} \Huge(T_{\text{core}},\nu\Huge)_\text{}}\biggl)^{2} \biggl\}  \nonumber \\  &  &~+~ \biggl( \frac{5B_{\text{core}}^{2}R_{\text{core}}^{4}}{18GM_\text{core}^{2}}\biggl)^{2} \biggl\{ \biggl(\frac{2\Delta \sigma_{\text{nth,sur}}}{\sigma_{\text{nth,sur}}}\biggl)^{2} \nonumber \\
  &  & + ~\biggl( \frac{\Delta S^{\text{1.2mm}}_{\text{int,core}}}{S^{\text{1.2mm}}_{\text{int,core}}}\biggl)^{2} 
   ~+~ \biggl(\frac{\Delta B_{\text{1.2mm}} \Huge(T_{\text{core}},\nu)_\text{{}}}{2~B_{\text{1.2mm}} \Huge(T_{\text{core}},\nu\Huge)_\text{}}\biggl)^{2} \nonumber \\
   &  & ~+~~\biggl( \frac{\Delta R_\text{core}}{R_\text{core}}\biggl)^{2}\biggl\}                                   \Biggl]^{1/2}.    
\end{eqnarray}

\section{Core extraction from 1.2 mm dust continuum emission}\label{appendix_2_0_0}

\begin{figure*}
        \centering 
        \includegraphics[width=6.4in,height=4.5in,angle=0]{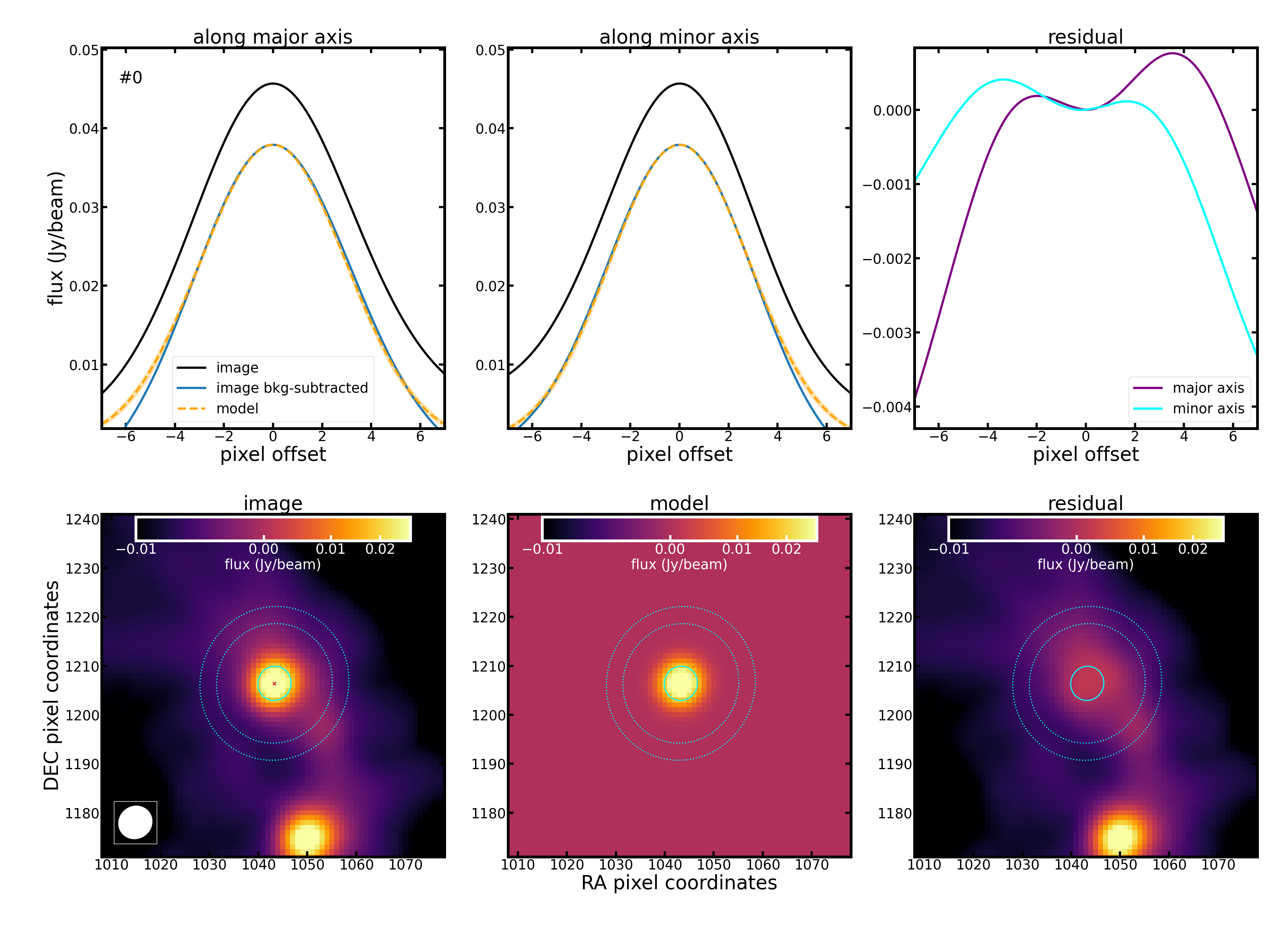}
        \caption{Example of 2D Gaussian modelling for the ALMA 1.2 mm flux measurement of cores in G327.29 protocluster. \textit{Top row}: 1D profile of the continuum image before (black) and after (blue) background subtraction and the model (dashed orange) along the major (left) and minor (middle) axes. The residual of the image profile subtracted by the model is presented in the right panel. \textit{\textit{Bottom row}}: Cutout of the original image (left), the model (centre), and the residual image (right). The solid~cyan ellipse at the centre indicates the FWHM of the 2D Gaussian model. The dashed cyan ellipses are the annulus where the local background value is evaluated. The size of the synthesized beam of the image is displayed in the lower-left corner. The rest of the cores are shown in \cite{Koley_zenodo}.}
        \label{fig:fig9}
\end{figure*}

An example of the 2D Gaussian modelling used to measure the 1.2 mm ALMA flux of the cores identified using \texttt{TGIF} \citep{tgif} and \texttt{astrodendro} \citep{2008ApJ...679.1338R} in the G327.29 protocluster is shown in Fig. \ref{fig:fig9}. The rest of the cores are shown in \cite{Koley_zenodo}.

\section{Core properties towards the observed region}\label{appendix_2}
The derived physical and observational properties of the identified cores, including their positions, sizes, flux densities, and other relevant parameters are summarized and listed in Table \ref{tab:table1}.

\begin{table*} 

        \caption{Core properties towards the observed region.}
        \begin{tabular}{ c c c c c c c c c c   }

        \hline
        
                Core &RA &  Dec &  $\theta_{\text{maj}}$ $\times$ $\theta_{\text{min}}$  &   PA  &  Size   &  $S_{\text{peak, core}}^\text{1.2mm}$ & $S_{\text{int, core}}^\text{1.2mm}$ &  $T_\text{core}$ & $M_\text{core}$  \\ [0.5 ex]

        id. &  [ICRS] &  [ICRS]   &  [$''$ $ \times$ $''$] &   [deg]  &   [au]  &  [mJy/beam] & [mJy] &  [K] & [M\textsubscript{\(\odot\)}]  \\ [0.5 ex]
                \hline  

        0 & 15:53:10.52   &  -54:36:39.83  & 0.35$\times$0.34$^{*}$  & 68.9  &  1724$\pm$344  & 37.9$\pm$0.27   &  55.7$\pm$1.7 & 28$\pm$7  & 4.2$\pm$2.1      \\ [0.5 ex]
                \hline  
        1 & 15:53:10.49  & -54:36:41.41   &  0.39$\times$0.35$^{*}$  &  56.7   &  1848$\pm$370   & 36.9$\pm$0.28  &  62.7$\pm$1.5  &   27$\pm$7  & 4.9$\pm$2.5      \\ [0.5 ex]
                \hline  
        2 & 15:53:10.90   & -54:36:46.16   &  0.38$\times$0.31  &    07.3   & 1716$\pm$344  &  25.2$\pm$0.28   &  74.2$\pm$2.7  &   27$\pm$7  & 5.6$\pm$2.9    \\ [0.5 ex]
                \hline  
        3 & 15:53:10.94   &-54:36:46.05    &  0.40$\times$0.35  &   00.0   & 1870$\pm$374   & 17.0$\pm$0.28  &  54.8$\pm$2.4  &   100$\pm$50  & 0.9$\pm$0.6    \\ [0.5 ex]
                \hline
        4 & 15:53:10.98   & -54:36:46.55    &  0.41$\times$0.28  &    48.8 & 1694$\pm$338  & 9.7$\pm$0.29  &  28.6$\pm$1.2  &   28$\pm$7  & 1.9$\pm$1.0      \\ [0.5 ex]
                \hline
        5 & 15:53:11.08   & -54:36:47.99   &  0.38$\times$0.22  &   00.0  & 1446$\pm$290  & 19.3$\pm$0.30  &  50.0$\pm$3.0  &   26$\pm$6  & 4.0$\pm$2.0       \\ [0.5 ex]
                \hline
        6 & 15:53:10.71   & -54:36:46.75   &  0.48$\times$0.13  &   46.5  & 1248$\pm$250  &  8.2$\pm$0.29 &  22.2$\pm$0.7  &   26$\pm$7 & 1.7$\pm$0.9       \\ [0.5 ex]
                \hline
        7 & 15:53:11.02   & -54:36:42.79   &  0.23$\times$0.21  &   54.4 &  796$\pm$160   & 13.4$\pm$0.27  &  24.2$\pm$0.8  &   25$\pm$6 & 2.1$\pm$1.1     \\ [0.5 ex]
                \hline
        8 & 15:53:11.19   & -54:36:48.69   &  0.35$\times$0.33$^{*}$  & 49.3 & 1700$\pm$340 & 18.4$\pm$0.32   &  26.4$\pm$1.0  &   25$\pm$6  & 2.2$\pm$1.1       \\ [0.5 ex]
                \hline
        9 & 15:53:11.59  & -54:36:25.08    &  0.43$\times$0.42  &  178.9 & 2124$\pm$424   &  20.8$\pm$0.32  &  77.2$\pm$5.2  &   27$\pm$7  & 5.6$\pm$2.9      \\ [0.5 ex]
                \hline
        10 & 15:53:11.29  & -54:36:34.60   &  0.24$\times$0.21  &  34.7 & 1122$\pm$224  &  11.5$\pm$0.23   &  24.1$\pm$1.5  &   28$\pm$7  & 1.8$\pm$0.9       \\ [0.5 ex]
                \hline
        11 & 15:53:11.73 & -54:36:29.78     &  0.48$\times$0.18  & 1.0  & 1470$\pm$294   &   3.6$\pm$0.27   &  10.3$\pm$0.3  &   25$\pm$6  & 0.8$\pm$0.4     \\ [0.5 ex]
                \hline
        12 & 15:53:09.40 &  -54:37:00.52  &  0.29$\times$0.18  & 29.2  & 1142$\pm$228  &  59.7$\pm$1.10   &  128.4$\pm$4.6  &   33$\pm$8  & 14.2$\pm$7.0       \\ [0.5 ex]
                \hline
        13 & 15:53:09.49 &  -54:37:01.12  &  0.33$\times$0.22  & 33.9 & 1348$\pm$270   &  54.4$\pm$1.04   &  132.0$\pm$4.7  &   29$\pm$7  & 12.9$\pm$6.4      \\ [0.5 ex]
                \hline
         14 & 15:53:09.47 & -54:37:00.39  &  0.66$\times$0.25  & 89.7  & 2032$\pm$406  &  45.9$\pm$0.96   &  180.6$\pm$9.7  &   100$\pm$50  & 2.9$\pm$1.9       \\ [0.5 ex]
                \hline
        15 & 15:53:09.46 & -54:36:58.83 &  0.29$\times$0.18  & 70.7  &  1142$\pm$228  &   10.90$\pm$0.80  &  23.4$\pm$0.7  &   32$\pm$8  & 1.4$\pm$0.7     \\ [0.5 ex]
                \hline
        16 & 15:53:09.23 &-54:36:59.13  &  0.20$\times$0.09  & 65.4  & 670$\pm$134  &  20.65$\pm$1.10   &  35.8$\pm$1.3  &   29$\pm$7  & 7.4$\pm$3.7       \\ [0.5 ex]
                \hline
        17 & 15:53:09.21 &-54:36:57.86 &  0.37$\times$0.34$^{*}$  & 59.5 &  1774$\pm$354  & 21.16$\pm$0.96    &  33.1$\pm$0.8  &   29$\pm$7  & 2.2$\pm$1.1       \\ [0.5 ex]
                \hline
        18 & 15:53:10.69 &-54:36:32.19 &  0.14$\times$0.08  & 38.3 &  530$\pm$106   &  9.14$\pm$0.25   &  14.7$\pm$0.5  &   27$\pm$7  & 1.8$\pm$0.9      \\ [0.5 ex]
                \hline
        19 & 15:53:10.45 &-54:36:50.29  &  0.22$\times$0.18  & 37.4  & 994$\pm$198   & 8.62$\pm$0.31    &  16.7$\pm$1.5  &   25$\pm$6  & 1.4$\pm$0.7    \\ [0.5 ex]
                \hline
        20 & 15:53:09.66 &-54:36:44.22  &  0.37$\times$0.34$^{*}$  & 48.8  & 1774$\pm$354  &  12.3$\pm$0.44   &  19.3$\pm$0.6  &   26$\pm$6  & 1.4$\pm$0.7     \\ [0.5 ex]
                \hline
        21 & 15:53:09.33 &-54:36:51.48  &  0.28$\times$0.25  &  30.8 & 1322$\pm$264    &   13.7$\pm$0.58  &  32.1$\pm$2.4  &   27$\pm$7  & 2.4$\pm$1.2    \\ [0.5 ex]
                \hline
        22 & 15:53:09.45 &-54:36:51.94  &  0.38$\times$0.33$^{*}$  & 60.7 &  1770$\pm$354 &  11.1$\pm$0.52   &  17.2$\pm$0.6  &   27$\pm$7  & 1.2$\pm$0.6    \\ [0.5 ex]
                \hline
        23 & 15:53:10.03 &-54:36:53.59  &  0.34$\times$0.32  & 25.1 &  1650$\pm$330  & 4.2$\pm$0.38    &  5.7$\pm$0.1  &   25$\pm$6  & 0.4$\pm$0.2    \\ [0.5 ex]
                \hline
        24 & 15:53:10.48 &-54:36:36.29  &  0.36$\times$0.18  &  0.0  & 1272$\pm$254  &  4.3$\pm$0.27   &  10.2$\pm$0.3  &   26$\pm$6  & 0.8$\pm$0.4   \\ [0.5 ex]
                \hline
        25 & 15:53:10.93 & -54:36:36.30  &  0.23$\times$0.09  & 81.7 & 720$\pm$144   & 1.9$\pm$0.24     &  3.4$\pm$0.0  &   26$\pm$6  & 0.3$\pm$0.1   \\ [0.5 ex]
                \hline
        26 & 15:53:08.87 &-54:36:59.65  &  0.26$\times$0.23  & 33.4 & 1222$\pm$244  &  18.3$\pm$1.9    &  40.6$\pm$0.1  &   34$\pm$8  & 2.3$\pm$1.1     \\ [0.5 ex]
        \hline
        27 & 15:53:12.16  & -54:36:12.74  &  0.38$\times$0.18  & 75.0 & 1308$\pm$262  &  32.9$\pm$3.6     &  81.2$\pm$4.0  &   25$\pm$5  & 8.3$\pm$4.0     \\ [0.5 ex]
                \hline
        \end{tabular}

        \vspace{2mm}
        \label {tab:table1}
    \tablefoot{Column 1: Core IDs. Columns 2 and 3: right ascension (RA) and declination (Dec) of the cores. Column 4: Deconvolved major ($\theta_{\text{maj}}$) and minor ($\theta_{\text{min}}$) axes of the cores (in arcsec unit). Column 5: Position angle (P.A.) of the cores. Column 6: Size of the cores.  Columns 7 and 8: Peak flux ($S_{\text{peak, core}}^\text{1.2mm}$) and integrated intensity ($S_{\text{int,core}}^\text{1.2mm}$) of the cores. Column 9: Temperature towards the cores ($T_{\text{core}}$) obtained from the works of \cite{2024A&A...687A.217D} and \cite{2024A&A...687A.163B}. Column 10: Mass of the cores ($M_\text{core}$). Asterisk (*) symbol in column 4  denotes the upper limit of the size of the cores, which is almost equal to the FWHM of the beam. For core id 27 there was no dust temperature measurements, consequently we take the temperature from the nearby pixels.}
\end{table*}

\section{Average spectra of HN$^{13}$C (3$-$2) and H$^{13}$CO$^{+}$ (3$-$2) towards 1.2 mm continuum cores}\label{appendix_3_0_0}

\begin{figure}[ht!]
        \centering 
        \includegraphics[width=3.4in,height=2.4in,angle=0]{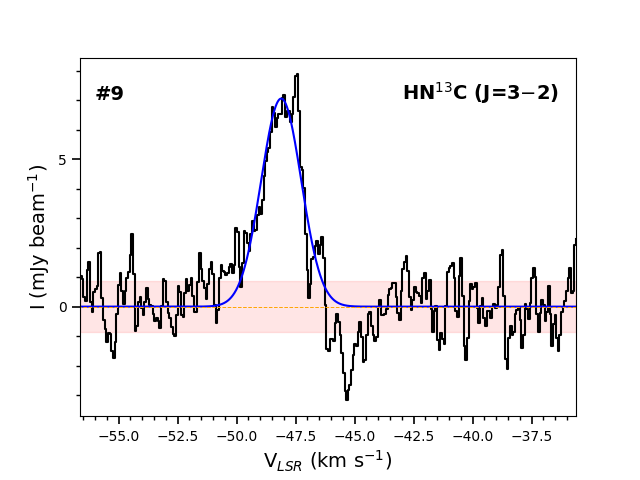}\includegraphics[width=3.4in,height=2.4in,angle=0]{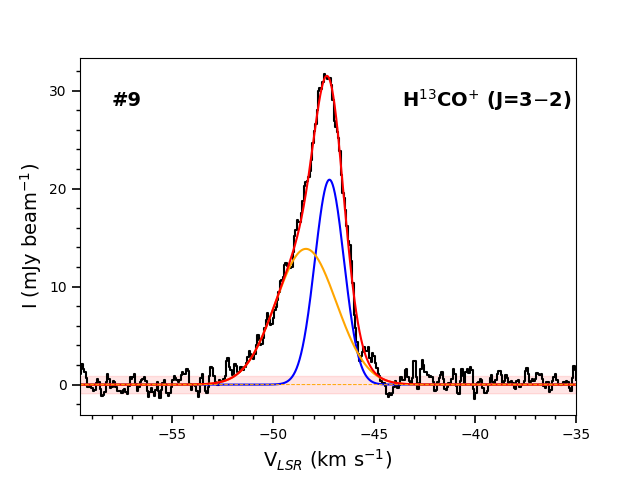}\\
        \caption{Average spectra of the HN$^{13}$C (J=3$-$2) and H$^{13}$CO$^{+}$ (J=3$-$2) lines towards core 9. The solid black hist line in each panel represents the observed spectrum, while the other lines indicate the decomposed components. The red-shaded area marks the $\pm$3$\sigma$ region, and the hashtag symbol in each panel denotes the core ID.  In a few cases, which are presented in \cite{Koley_zenodo}, the baseline was not properly subtracted due to complex continuum emission; however, this does not affect the overall analysis.}
        \label{fig:fig13_0}
\end{figure}

Figure \ref{fig:fig13_0} shows the averaged spectra towards core 9 for the HN$^{13}$C and H$^{13}$CO$^+$ ($3-2$) transitions. The corresponding spectra for the remaining cores are available in \cite{Koley_zenodo}.

\section{Properties of the line widths towards the dense cores}\label{appendix_3}
The measured properties of the line widths towards the identified dense cores are summarized in Table \ref{tab:table2}.

\begin{table} 
        \centering
        \caption{Properties of the line widths towards the dense cores.}
        \begin{tabular}{ c c c c c}

        \hline
         Core IDs & $\sigma_{\text{tot\_HN$^{13}$C}}$ &$\sigma_{\text{tot\_H$^{13}$CO$^{+}$}}$ & $\sigma_{\text{tot, avg}}$ & $\sigma_{\text{tot,core}}$ \\ [0.5 ex]
                  & (km s$^{-1}$) & (km s$^{-1}$) & (km s$^{-1}$)  & (km s$^{-1}$) \\ [0.5 ex]
         \hline

         0 & 0.41 & 0.92  & 0.67 (0.26) & 0.73 (0.26)  \\ [0.5 ex]
         1 & 0.64 & 0.89 & 0.77 (0.13)  & 0.82 (0.13)  \\ [0.5 ex]
         2 & 0.45 & 1.17 & 0.81  (0.36) & 0.86 (0.36)  \\ [0.5 ex]
         3 & 0.35 & 0.52 &  0.44 (0.09) & 0.72 (0.10)  \\ [0.5 ex]
         4 & 0.42 & 0.86 &  0.64 (0.22) & 0.71 (0.22)  \\ [0.5 ex] 
         5 & 0.42 & 0.61 &  0.52 (0.10) & 0.59 (0.10)  \\ [0.5 ex]
         6 & 0.35 & 0.40 &  0.38 (0.03) & 0.48 (0.03)  \\ [0.5 ex]
         7 & 0.35 & 1.19 &  0.77 (0.42) & 0.82 (0.42)  \\ [0.5 ex]
         8 & 0.66 & 1.00 &  0.83 (0.17) & 0.88 (0.17)  \\ [0.5 ex]
         9 & 0.85 & 1.24 &  1.05 (0.20) & 1.09 (0.20)  \\ [0.5 ex]
         10 & 0.31 & 0.53  & 0.42 (0.11)& 0.52 (0.11)  \\ [0.5 ex]
         11 & 0.44 & 0.35 &  0.40 (0.05)& 0.49 (0.05)  \\ [0.5 ex]
         12 &  --   & 0.88 & 0.88 (0.00)& 0.94 (0.00)  \\ [0.5 ex]
         13 &  --   & 0.94 & 0.94 (0.00)& 0.99 (0.00)  \\ [0.5 ex]
         14 &  --   & 1.24 & 1.24 (0.00)& 1.36 (0.01)  \\ [0.5 ex]
         15 &  --   & 0.71 & 0.71 (0.00)& 0.78 (0.00)  \\ [0.5 ex]
         16 &  --   & 0.82 & 0.82 (0.00)& 0.87 (0.00) \\ [0.5 ex]
         17 &  --   & 0.70 & 0.70 (0.00)& 0.76 (0.00) \\ [0.5 ex]
         18 & 0.30 & 0.29 & 0.30 (0.00) & 0.42 (0.00) \\ [0.5 ex]
         19 & 0.62 & 0.37 &  0.50 (0.13)& 0.57 (0.13) \\ [0.5 ex]
         20 & 0.33 & 0.37 & 0.35 (0.02) & 0.45 (0.02) \\ [0.5 ex]
         21 &  --   & 0.69 & 0.69 (0.00)& 0.75 (0.00)  \\ [0.5 ex]
         22 & 0.82 & 0.44 & 0.63 (0.19) & 0.70 (0.19) \\ [0.5 ex]
         23 & 0.34 & 0.54 & 0.44 (0.10) & 0.52 (0.11) \\ [0.5 ex]
         24 & 0.43 & 0.40 & 0.42 (0.01) & 0.51 (0.01) \\ [0.5 ex]
         25 & 0.47 & 0.23 & 0.35 (0.12) & 0.45 (0.12)\\ [0.5 ex]
         26 & 0.88 & 0.72 &  0.80  (0.08)& 0.87 (0.08)  \\ [0.5 ex]
         27 &  --   & -- &  --\\ [0.5 ex]
         
        \hline

        \end{tabular}

        \vspace{2mm}
        \label {tab:table2}

\tablefoot{Column 1: Core IDs. Columns 2 and 3: Average velocity dispersion of HN$^{13}$C ($\sigma_{\text{tot\_HN$^{13}$C}}$) and H$^{13}$CO$^{+}$ ($\sigma_{\text{tot\_H$^{13}$CO$^{+}$}}$) lines towards the dense cores.  Column 4: Average velocity dispersion ($\sigma_{\text{tot,avg}}$) obtained from two spectral lines. Column 5: Core velocity dispersion ($\sigma_{\text{tot,core}}$) including the thermal contribution. Values in parentheses represent the uncertainties of the corresponding quantities.}   
\end{table}

\section{Virial parameters of the cores without and with the inclusion of magnetic fields}\label{appendix_3_0}

The virial parameters of the cores, calculated both without and with the inclusion of magnetic support, are listed in Table  \ref{tab:table3}.

\begin{table*}
        \caption{Virial parameter ($\alpha_{\text{vir}}$) and the ratio of kinetic (2$E_{\text{k}}$) to magnetic energy ($E_{\text{B}}$) of the cores in the G327.29 protocluster.}
        \begin{tabular}{ c c c c c c }
   
        \hline
         Core IDs & $\alpha_{\text{vir, turb}}$ ($\Delta\alpha_{\text{vir, turb}}$) & $\alpha_{\text{vir, turb+mag}}$  ($\Delta\alpha_{\text{vir, turb+mag}}$) & $\alpha_{\text{vir, turb+mag}}$  ($\Delta\alpha_{\text{vir, turb+mag}}$) & 2$E_{\text{k}}$/$E_\text{{B}}$  &  2$E_{\text{k}}$/$E_\text{{B}}$    \\ [0.5 ex]
                  & (without $B_\text{{tot}}$)  & $ (B_{\text{tot}}\propto \rho^{0}$)  &  ($B_\text{{tot}}\propto \rho^{0.5}$) & ($B_{\text{tot}} \propto \rho^{0}$) & ($B_{\text{tot}} \propto \rho^{0.5}$) \\ [0.5 ex]
         \hline

         0 & 1.23 (0.99) & 1.32 (0.99) & 5.54 (3.69) & 13.67 & 0.29    \\ [0.5 ex]
         1 & 1.43 (0.70) & 1.51 (0.71) & 5.39 (2.35) & 17.87 & 0.36    \\ [0.5 ex]
         2 & 1.27 (1.17) & 1.32 (1.17) & 4.49 (2.91) & 25.40 & 0.39    \\ [0.5 ex]
         3 & 6.06 (3.83) & 8.64 (4.60) & 27.89 (19.00) & 2.35 & 0.28   \\ [0.5 ex]
         4 & 2.53 (1.82) & 2.92 (1.86) & 11.89 (7.94) & 6.49 & 0.27    \\ [0.5 ex] 
         5 & 0.71 (0.35) & 0.75 (0.35) & 4.50  (3.15) & 17.75 & 0.19   \\ [0.5 ex]
         6 & 0.95 (0.39) & 1.09 (0.41) & 8.66  (6.39) & 6.79  & 0.12   \\ [0.5 ex]
         7 & 1.43 (1.56) & 1.45 (1.56) & 5.41  (3.63) & 71.50 & 0.36   \\ [0.5 ex]
         8 & 3.36 (1.79) & 3.66 (1.81) & 11.48 (6.92) & 11.20 & 0.41   \\ [0.5 ex]
         9 & 2.53 (1.35) & 2.56 (1.35) & 3.94  (1.90) & 84.33 & 1.79   \\ [0.5 ex]
         10 & 0.95 (0.53) & 0.97 (0.53) & 3.26 (2.27) & 47.50 & 0.41   \\ [0.5 ex]
         11 & 2.48 (1.03)& 2.76  (1.08) & 9.31 (6.57) & 8.86 & 0.36    \\ [0.5 ex]
         12 & 0.40 (0.14) & 0.41 (0.14) & 1.24 (0.71) & 40.00 & 0.48   \\ [0.5 ex]
         13 & 0.58 (0.21) & 0.59 (0.21) & 1.67 (0.93) & 58.00 & 0.53   \\ [0.5 ex]
         14 & 7.29 (4.15) & 7.63 (4.17) & 14.65 (7.53) & 21.44 & 0.99  \\ [0.5 ex]
         15 & 2.79 (1.01) & 2.94 (1.02) & 11.36 (7.13) & 14.48 & 0.32  \\ [0.5 ex]
         16 & 0.39 (0.14) & 0.40 (0.14) & 1.34  (0.80) & 39.00 & 0.41  \\ [0.5 ex]
         17 & 2.62 (0.93) & 2.97 (0.98) & 11.09 (7.03) & 7.49  & 0.31  \\ [0.5 ex]
         18 & 0.29 (0.11) & 0.30 (0.11) & 1.39  (1.05) & 29.00 & 0.26  \\ [0.5 ex]
         19 & 1.30 (0.76) & 1.38 (0.77) & 8.75  (6.22) & 16.25 & 0.17  \\ [0.5 ex]
         20 & 1.44 (0.52) & 2.31 (0.92) & 14.75 (10.96) & 1.66 & 0.11  \\ [0.5 ex]
         21 & 1.74 (0.67) & 1.83 (0.68) & 7.53  (4.84) & 19.33 & 0.30  \\ [0.5 ex]
         22 & 4.06 (2.69) & 5.23 (2.89) & 19.56 (13.08) & 3.47 & 0.26  \\ [0.5 ex]
         23 & 6.27 (3.50) & 14.18 (7.84) & 49.60 (35.85) & 0.79 & 0.14   \\ [0.5 ex]
         24 & 2.33 (0.82) & 3.02 (1.03) & 8.23 (5.67)  &  3.38  & 0.39   \\ [0.5 ex]
         25 & 2.73 (1.74) & 2.85 (1.75) & 11.65 (8.64) &  22.75 & 0.31  \\ [0.5 ex]
         26 & 2.26 (0.88) & 2.28 (0.88) & 7.84 (4.67) &   113.00 & 0.41 \\ [0.5 ex]
         27 &   ---       &  ---        &  ---         &  --- & --- \\ [0.5 ex]
         
        \hline

        \end{tabular}

        \vspace{2mm}
        \label {tab:table3}

\tablefoot{Column 1: Core IDs. Column 2: Virial parameter ($\alpha_{\text{vir, turb}}$) of the cores without magnetic field   $B_{\text{tot}}$ $\propto$ $\rho^{0}$. Column 3: Virial parameter ($\alpha_{\text{vir, turb+mag}}$) of the cores when $B_{\text{tot}}$ $\propto$ $\rho^{0.0}$. Column 4: Same as column 3 but when 
 $B_{\text{tot}}$ $\propto$ $\rho^{0}$. Columns 5 and 6: Ratio of kinetic energy (2$E_{\text{k}}$) and magnetic energy ($E_{\text{B}}$) when   $B_{\text{tot}}$ $\propto$ $\rho^{0}$ and  $B_{\text{tot}}$ $\propto$ $\rho^{0.5}$.} 
\end{table*}

\end{appendix}

\end{document}